\documentclass[11pt,a4paper,english,twoside]{article}

\usepackage{a4wide}
\usepackage{amssymb, amsmath, amsthm}
\usepackage{graphicx}
\usepackage{subcaption}
\usepackage[all]{xy}
\usepackage[pdftex,hyperref,svgnames]{xcolor}
\usepackage[pdftex,colorlinks=true,
pdfstartview=FitV,
pdfnewwindow=true,
linktoc = page,
linkcolor= blue,
citecolor= red,
urlcolor= blue,
hyperindex=true,
hyperfigures=false]{hyperref}
\hypersetup{linktocpage}
\usepackage{dsfont}
\usepackage{empheq}
\usepackage{cite}
\usepackage{float}
\usepackage{cancel}
\usepackage{relsize}
\usepackage{soul}
\usepackage{enumerate}
\usepackage{enumitem}
\usepackage{hhline}
\usepackage{multirow}
\usepackage{xspace}
\usepackage{bbm}
\usepackage{pdfpages}
\usepackage{setspace}

\usepackage{framed}

\newcommand*\widefbox[1]{\fbox{\hspace{1em}#1\hspace{1em}}}

\newcommand{\nn}{\nonumber}

\def\del {\partial}
\def\d {{\rm d}}

\def\beq{\begin{equation}}
\def\eeq{\end{equation}}
\def\bea{\begin{eqnarray}}
\def\eea{\end{eqnarray}}

\begin{document}
\numberwithin{equation}{section}

\begin{titlepage}

\begin{center}

\phantom{DRAFT}

\vspace{1.2cm}

{\LARGE \bf{Quintessence: an analytical study,\vspace{0.4cm}\\with theoretical and observational applications}}\\

\vspace{2.2 cm} {\Large David Andriot}\\
\vspace{0.9 cm} {\small\slshape Laboratoire d’Annecy-le-Vieux de Physique Th\'eorique (LAPTh),\\
CNRS, Universit\'e Savoie Mont Blanc (USMB), UMR 5108,\\
9 Chemin de Bellevue, 74940 Annecy, France}\\
\vspace{0.5cm} {\upshape\ttfamily andriot@lapth.cnrs.fr}\\

\vspace{2.8cm}

{\bf Abstract}
\vspace{0.1cm}
\end{center}

\begin{quotation}
We focus on minimally coupled (multi)field quintessence models, of thawing type, and their realistic solutions. In a model-independent manner, we describe analytically these cosmological solutions throughout the universe history. Starting with a kination - radiation domination phase, we obtain an upper bound on the scalar potential to guarantee an early kination: $V(\varphi) \ll e^{-\sqrt{6} \varphi}$. Turning to the radiation - matter phase, we obtain analytic expressions for the scale factor $a(t)$ (not $t(a)$) and the scalar fields $\varphi^i(t)$ (usually neglected). These allow us to evaluate analytically the freezing of scalar fields, typically $\Delta \varphi \lesssim 10^{-2}$, as well as the transition moment of the dark energy equation of state parameter $w_{\varphi}$ from $+1$ to $-1$, with excellent agreement to the numerics. We comment on this freezing in view of string theory model building, and of some cosmological events. Turning to the latest phase of matter - dark energy domination, we show that the (multi)field displacement is sub-Planckian: $\Delta \varphi \leq 1$. We also provide for that phase analytic expressions for $\int (w_{\varphi}+1)\, \d N$ in terms of matter evolution; we relate those to observational targets that we propose. Using finally the CPL parametrisation, while discussing a phantom behaviour, we derive analytic bounds on $w_0$ and $w_a$.
\end{quotation}

\end{titlepage}

\newpage

\tableofcontents

\section{Introduction and results summary}\label{sec:intro}

For about a century, we have known that our universe is expanding. Less than thirty years ago, the present expansion was observed to be accelerating. This acceleration is due to a so-called dark energy, which behaves very differently than ordinary matter or radiation. {\sl What is the nature of dark energy?} This open problem is a key question of modern fundamental physics. A standard proposal is to describe dark energy as a constant energy density given by a cosmological constant $\Lambda >0$. This has led to a concordance cosmological model named $\Lambda$CDM, well-tested and in fairly good agreement with observations. Another option however is to have an energy density varying with time, giving a dynamical dark energy. One realisation of this possibility is through so-called quintessence models \cite{Ratra:1987rm, Peebles:1987ek, Wetterich:1994bg, Caldwell:1997ii} (see e.g.~\cite{Caldwell:2000wt, Caldwell:2005tm, Gupta:2011ip, Tsujikawa:2013fta} for early reviews and constraints), where the dynamics is that of scalar fields evolving in a scalar potential. Recent (and coming) observations, as well as theoretical arguments, have revived the possibility of explaining dark energy by quintessence. In this work, {\sl we perform a model-independent analysis of realistic quintessence scenarios, identifying interesting theoretical features for model building, as well as characteristic signatures that could possibly be observed, helping to distinguish such models from $\Lambda$CDM.}

On the observational side, the first results of the DESI collaboration have recently been released \cite{DESI:2024mwx}. Those are based on collected data until redshift $z=4$ with an unprecedent precision level. Consistently with the recent DES results \cite{DES:2024tys}, the DESI observational results are compatible with a dynamical dark energy model. More DESI results are announced for the coming years, and further experiments (Euclid, LSST/Vera Rubin) will provide independent measurements to similar precision. On the theoretical side, difficulties have been pointed-out in obtaining a universe with a positive cosmological constant (a.k.a.~a de Sitter spacetime) from string theory, a quantum gravity theory candidate to be a fundamental theory of Nature. These difficulties have been characterised in terms of the strong de Sitter conjecture of the swampland program \cite{Rudelius:2021oaz, Bedroya:2019snp}. While not claimed to be impossible, there is no known well-controlled string theory construction of such a solution up-to-date. As an alternative, there has been a recent interest in rather realising quintessence models from string theory (see \cite{Apers:2024ffe, Seo:2024fki, Gallego:2024gay, Seo:2024qzf, Andriot:2024jsh, Bhattacharya:2024hep, Alestas:2024gxe, Apers:2024dtn, Casas:2024oak} for a recent sample, and \cite{Andriot:2024jsh} for further references). While necessary ingredients are naturally present in string effective theories, finding the right dynamics and an appropriate scalar potential is not an easy task. But a lot of activity is currently dedicated to these matters, including cosmological model building and related investigations \cite{Hossain:2023lxs, Notari:2024rti, Shiu:2024sbe, Tonioni:2024huw, Wolf:2024stt, Boiza:2024azh, Smith:2024ibv}.\\

With these motivations in mind, we consider in this work the following type of (multi)field quintessence model, with a set of scalar fields $\{ \varphi^i \}$ minimally coupled to gravity, and without any direct coupling to matter and radiation described by ${\cal L}_{m,r}$
\beq
{\cal S} = \int \d^4 x\, \sqrt{|g_4|} \left( \frac{M_p^2}{2} {\cal R}_4 - \frac{1}{2} g_{ij}\, \del_{\mu} \varphi^i \del^{\mu} \varphi^j - V(\varphi^k) \  + {\cal L}_{m,r} \right) \ .\label{Sintro}
\eeq
We aim at obtaining general results, independent of the (positive definite) field space metric $g_{ij}(\varphi^k)$ and of the (positive) scalar potential $V(\varphi^k)$. The dynamics of the universe are described by solutions to this model. For those, we will restrict to an FLRW metric with scale factor $a(t)$, describing an expanding universe, and most of our results will be independent of the spatial curvature. We will also restrict ourselves to homogeneous and isotropic fields $\varphi^i(t)$. The complete formalism and the way to find solutions is described in Section \ref{sec:formalism}. We will consider the evolution of solutions either in terms of time $t$ or in terms of the number of e-folds $N \equiv \ln a$.

\begin{figure}[H]
\begin{center}
\begin{subfigure}[H]{0.48\textwidth}
\includegraphics[width=\textwidth]{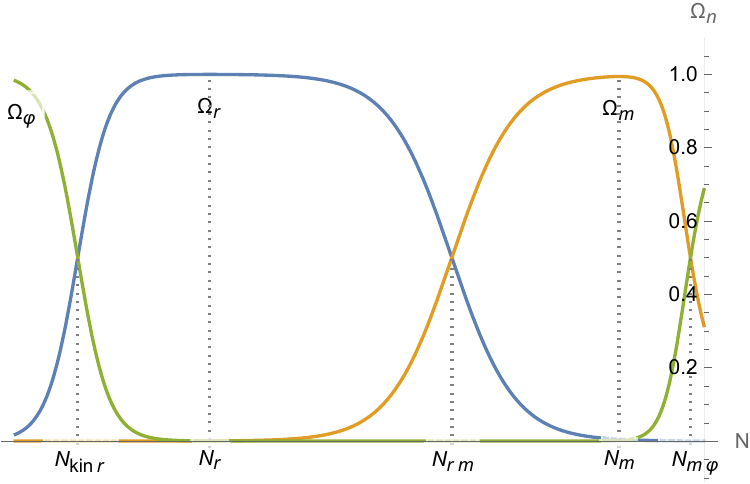}\caption{}\label{fig:ExpQuintONintro}
\end{subfigure}\quad
\begin{subfigure}[H]{0.48\textwidth}
\includegraphics[width=\textwidth]{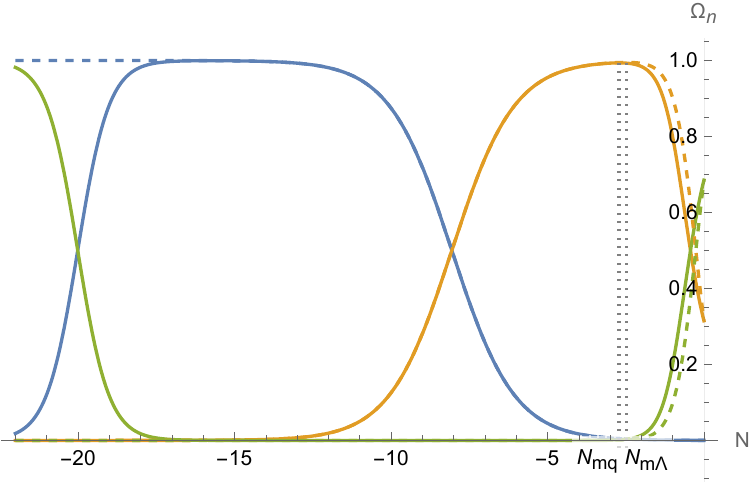}\caption{}\label{fig:ONsintro}
\end{subfigure}
\caption{Evolution of the $\Omega_{n}$ in terms of $N$ for a realistic solution of a quintessence model (plain lines), or of $\Lambda$CDM (dashed lines), as defined in the main text; $N=0$ corresponds here to today. The components $n$ are radiation ($r$), matter ($m$) and scalar ($\varphi$) where the latter plays the role of dark energy. The equality moments $N_{{\rm kin}r}, N_{rm}, N_{m\varphi}$ and the maxima moments $N_r, N_m$ are indicated. We see that the latter needs to be distinguished into $N_{m\, q}$ or $N_{m\, \Lambda}$, depending on the respective model.}\label{fig:ONintro}
\end{center}
\end{figure}

Before studying features and options offered by these models, a first necessary task is to obtain realistic solutions. By this we mean a solution describing a universe history which exhibits successively a radiation domination phase, a matter domination phase and today's universe content including dark energy. Those are captured by the respective energy density parameters $\Omega_r, \Omega_m, \Omega_{\varphi}$, where $0 \leq \Omega_n \leq 1$ gives the proportion of each component. The scalar component $\Omega_{\varphi}$ has contributions from the kinetic energy of the fields and the scalar potential, and plays the role of dark energy in the quintessence model. Today values should be close to the following fiducial values
\beq
\Omega_{r0}=0.0001\ ,\ \Omega_{m0}=0.3149\ ,\ \Omega_{\varphi0}=0.6850 \ , \label{fidOintro}
\eeq
indicating a dominant dark energy component, that provides acceleration. We consider in Section \ref{sec:examples} three different models, and for each of them we obtain a realistic solution. The first one is $\Lambda$CDM, which can be viewed as a particular case of \eqref{Sintro}, and the other two are single field quintessence models, with respectively an exponential or a hilltop scalar potential. While these two single field models are presented for illustration, most of our analytic results will be derived in multifield situations. For quintessence models, obtaining a realistic solution (especially with a radiation domination phase) requires a tuning of the initial conditions, which we achieve. The evolution of the $\Omega_n$ is then similar from one model to the other, and we display it in Figure \ref{fig:ONintro}. We indicate in Figure \ref{fig:ExpQuintONintro} that different phases in this universe history can be delimited by moments, expressed in terms of $N$, that characterise equality ($N_{{\rm kin}r}, N_{rm}, N_{m\varphi}$) or maxima ($N_r, N_m$) of the $\Omega_n$. We discuss these phases and specific moments in great details in Section \ref{sec:phases}, and provide various related analytic expressions. Given some values today $\Omega_{n0}$, it can be seen in Figure \ref{fig:ONsintro} that some of these moments are necessarily different between $\Lambda$CDM and quintessence, leading us to introduce extra notations: for instance, we label the moment for the maximum of $\Omega_m$ as $N_{m\, \Lambda}$ or $N_{m\, q}$ respectively. Similarly for the equality moment, we distinguish $N_{m\varphi\, \Lambda}$ and $N_{m\varphi\, q}$.
A specificity of the solutions to be considered is that they exhibit an initial phase where $\Omega_{\varphi}$ is dominating, as in Figure \ref{fig:ONintro}. It does not happen for $\Lambda$CDM, which can be viewed as a limiting case where $N_{{\rm kin}r}= N_r=- \infty$. As will be made clear, this initial phase is only due to the kinetic energy contribution, while the scalar potential is negligible. We then refer to a kination phase. For our universe, such a phase is hypothetical, because the above history is not expected to be complete. Indeed, one suspects for instance an early inflation, followed by some reheating process, which should then be patched at some point to the history described by Figure \ref{fig:ONintro}. This patching could leave room for kination, but it does not have to be and could join directly radiation domination. In this work we will study cosmological solutions that include (at least part of) such an initial kination phase.

\begin{figure}[H]
\begin{center}
\begin{subfigure}[H]{0.48\textwidth}
\includegraphics[width=\textwidth]{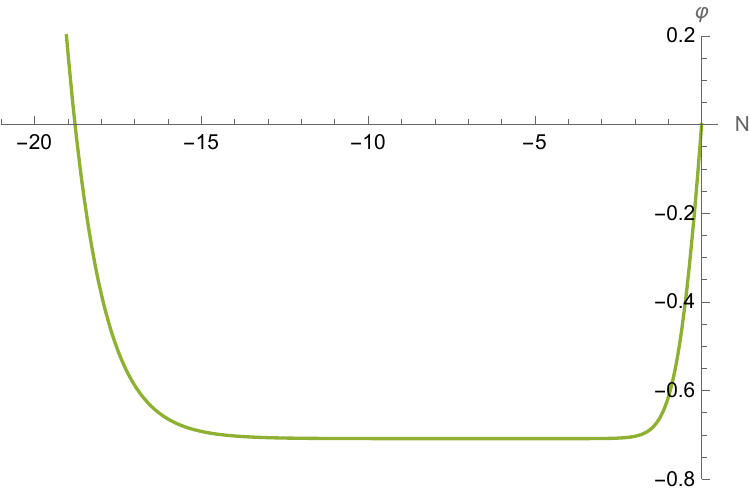}\caption{$\varphi(N)$}\label{fig:ExpQuintpNintro}
\end{subfigure}\quad
\begin{subfigure}[H]{0.48\textwidth}
\includegraphics[width=\textwidth]{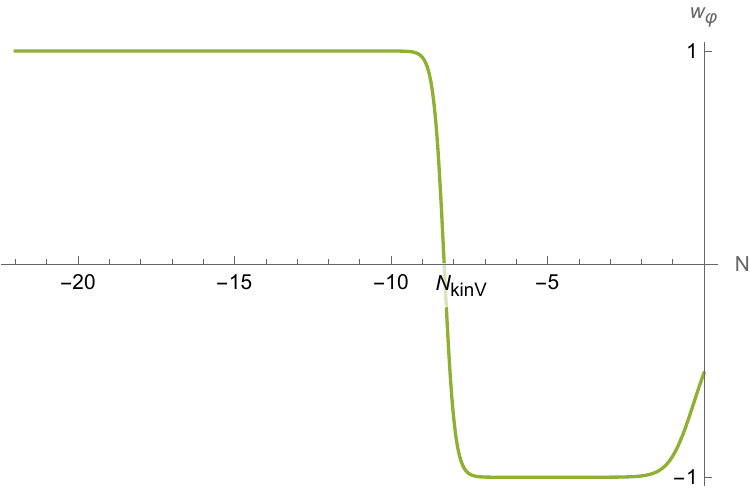}\caption{$w_{\varphi}(N)$}\label{fig:ExpQuintwNintro}
\end{subfigure}
\caption{Evolution of a scalar field $\varphi(N)-\varphi(0)$ and of the scalar equation of state parameter $w_{\varphi}(N)$ for a realistic solution of a quintessence model, as defined in the main text; we recall that for $\Lambda$CDM, one has the constant $w_{\varphi}=-1$. $N=0$ corresponds here to today. These curves display special features of interest in this work, namely a freezing of the field, as well as a moment $N_{{\rm kin}V}$ at which $w_{\varphi}=0$.} \label{fig:ExpQuintpwNintro}
\end{center}
\end{figure}

Having identified realistic solutions in the quintessence models, we notice, at least in our examples, special features that we will analyse in depth. One is the fact that the scalar fields get ``frozen'', namely seemingly constant, during most of radiation and matter domination, as illustrated in Figure \ref{fig:ExpQuintpNintro}. In recent times, fields start changing again: how much is the recent field displacement will also be a question of interest.
A second feature is the evolution of the scalar (dark energy) equation of state parameter $w_{\varphi}$, given by the following combination of the kinetic energy $\rho_{\rm kin}$ and the scalar potential
\beq
w_{\varphi} = \frac{\rho_{\rm kin} - V }{\rho_{\rm kin} + V} \ ,\qquad \rho_{\rm kin} = \frac{1}{2} g_{ij} \del_t \varphi^i \del_t \varphi^j \ .
\eeq
The evolution, depicted in Figure \ref{fig:ExpQuintwNintro}, indicates the initial domination of kinetic energy ($w_{\varphi}=1$) as in kination, followed by a domination of the scalar potential ($w_{\varphi}=-1$) as for a cosmological constant, and a final rise towards the value today. This recent evolution is the prime target of DESI observations and will be of interest here; in addition the transition moment at which $w_{\varphi}=0$ will be determined analytically. Note that the latest part of this evolution is that of thawing quintessence models, which are included in this study, and differs from that of freezing or tracker models \cite{Caldwell:2005tm}.

While we have expressed so far the evolution of solutions in terms of $N$, the dependence in time $t$ is also interesting. In particular, we note in our examples in Section \ref{sec:examples} that for a given Hubble parameter today $H_0$, the age of the universe is shorter in quintessence solutions than in $\Lambda$CDM.\\

Given these realistic cosmological solutions of quintessence models with their typical features, we turn in Section \ref{sec:analytic} to the core of this work: {\sl providing an analytical description of these solutions and their characteristics.} To that end, we separate the study in three phases: kination - radiation ($-\infty \leq N \leq N_r$), radiation - matter ($N_r \leq N \leq N_m$) and matter - dark energy ($N_m \leq N \leq N_{\rm today}$). Analytic solutions, namely $(a(t),\varphi^i(t))$, are discussed for each of these phases in Section \ref{sec:single}, \ref{sec:kinradanalytic}, \ref{sec:radmatat}, \ref{sec:anafrozenfield} and \ref{sec:lastphaseanalytic}. Prime results are obtained for the radiation - matter phase: first, we show how the well-known expression of $t(a)$ during this phase can be analytically inverted into $a(t)$, giving the explicit expressions \eqref{atsolradmat}. Second, while scalar field contributions are typically neglected during this phase (see Figure \ref{fig:ONintro} where $\Omega_{\varphi} \ll \Omega_r+\Omega_m$), we obtain an analytic expression for $\varphi^i(t)$ in \eqref{phit1} and \eqref{phit2}, and for $\varphi^i(N)$ in \eqref{PhiN1} and \eqref{PhiN2}. Those will be used to characterise the features noticed in the solutions, such as those of Figure \ref{fig:ExpQuintpwNintro}. Related expressions are given below in \eqref{Deltaphifrozenintro} and \eqref{NkinVequalityintro}.

The analytic solutions obtained provide a better understanding of the {\sl physics of the scalar fields}; that of the scale factor is nothing but a continuous growth at various rates. Provided some initial kinetic energy and a negligible potential, the fields start by rolling up or down the potential: this is kination. The Hubble friction being huge in the early universe, the fields are quickly slowed down. During radiation domination, the kinetic energy continues to get highly reduced, so much that fields appear frozen. This goes on until the potential forces become non-negligible: this happens at some point during the radiation - matter phase. Then the scalar fields get accelerated and their kinetic energy starts rising again. It remains very small until after matter domination, when finally Hubble friction becomes small enough to allow a proper field displacement. Then, the fields roll down the potential, while dark energy is getting dominant. What happens exactly today and in the future is model dependent, but the whole dynamics just described for the past is (almost) not.

We prove several results related to these dynamics. We discuss in Section \ref{sec:kincond} under which conditions the quintessence model allows for an initial kination phase. A necessary condition for the potential to be negligible enough is given in Lemma \eqref{lemmakin} to be
\beq
V(\varphi) \ll e^{-\sqrt{6}\, \varphi} \ ,
\eeq
where the early times correspond to a large negative field.

Turning to the radiation - matter phase, we use the analytic expressions to obtain in \eqref{PhiN1} - \eqref{PhiN2} the field variations during this period in a multifield setting, boiling down in the single field case to
\beq
\Delta \varphi = \sqrt{6}\ e^{N_{{\rm kin}r}-N_r} - \frac{2}{9} \left.\frac{\del_{\varphi}V}{V}\right|_{0rm} e^{4( N_{m\, \Lambda} - N_{m\, q} )} \ e^{3 (N_{m\, q} - N_{m\varphi\, \Lambda})} \label{Deltaphifrozenintro}\ .
\eeq
This analytic expression, when evaluated on the examples, gives a very good match with the numerics. It shows in particular how frozen the fields are, namely $\Delta \varphi \lesssim 10^{-2}$. We further use the analytic solutions to get an expression \eqref{NkinVequality} for $N_{{\rm kin}V}$, the moment at which $w_{\varphi}=0$, characterising its transition from $+1$ to $-1$: it is given by
\beq
N_{{\rm kin}V} =  -\frac{2}{3}( N_{m\, \Lambda} - N_{m\, q} )  + \frac{1}{6} (2 N_{{\rm kin}r} + N_{rm} + 3 N_{m\varphi\, \Lambda} ) \label{NkinVequalityintro}\ .
\eeq
When tested on examples, we find a perfect match to the numerical values. The main contributions to $N_{{\rm kin}V}$ are the equality moments $N_{{\rm kin}r}$ and $N_{rm}$.

The final matter - dark energy phase is more difficult to tackle analytically, since its latest dynamics is model-dependent, as discussed in Section \ref{sec:lastphaseanalytic}. We then adopt a different strategy, that is to consider integral of relevant quantities over this phase, as well as considering average simple parametrisation of some quantities. This allows us in Section \ref{sec:integral} to derive model-independent upper bounds to the field distance (multifield displacement) $\Delta \varphi$. In particular we show with \eqref{deltaphiupper2}, that
\beq
\Delta \varphi \leq 1 \ ,
\eeq
namely that the field distance in the matter - dark energy phase is sub-Planckian, as e.g.~in Figure \ref{fig:ExpQuintpNintro}. This is important in order to avoid possible quantum gravity corrections to the effective theory used, as argued e.g.~through the refined swampland distance conjecture (see e.g.~\cite{Palti:2019pca}).

We also focus during this phase on the recent variation of $w_{\varphi}$. We first provide in \eqref{relareaw} an expression for the integral of $w_{\varphi} + 1$ during this period, which captures the difference to a cosmological constant
\beq
\int_{N_{m\, q}}^0  (w_{\varphi} +1)\, \d N = \frac{4}{3}\, ( N_{m\, \Lambda} - N_{m\, q} ) \label{relareawintro}\ .
\eeq
We also give and comment on an analogous relation to the equality times $N_{m\varphi\, \Lambda} - N_{m\varphi\, q}$ in \eqref{relareaequality}. In Section \ref{sec:phantom}, we discuss the $w_0 w_a$ parametrisation of $w_{\varphi}$ used in DESI results \cite{DESI:2024mwx}, and comment on the ``observed'' phantom behaviour ($w_{\varphi} \leq -1$). In a Lemma in \eqref{lemmaw}, we relate such a behaviour within this parametrisation to the inequality $w_0 +w_a <-1$. From the latter and the integral expression \eqref{relareaw}, we conclude with a lower bound on $w_0$ \eqref{w0lowerbound} and an upper bound on $w_a$ \eqref{waupperbound}. We summarize our results (without the latter) as follows
\beq
\exists\ a > 0 \ {\rm s.t.} \ \ w_{\varphi} < -1 \quad \Leftrightarrow \quad w_0 +w_a < -1 \quad \Leftrightarrow \quad w_0 > -1 + \frac{4}{3} \left( N_{m\, \Lambda} - N_{m\, q} \right) \ .\label{w0waintro}
\eeq

We believe that these results are interesting for theoretical model building or for identifying new observational targets and constraints. We comment on these applications in detail in the Outlook in Section \ref{sec:outlook}.

\section{Quintessence: (multifield) formalism}\label{sec:formalism}

In this section, we present the 4-dimensional (4d) cosmological model considered, the equations to be solved and properties of solutions of interest. We then provide several necessary reformulations of the equation system.

\subsection{Cosmological model and equations}\label{sec:setting}

In this work we consider the following 4d cosmological model
\beq
{\cal S} = \int \d^4 x\, \sqrt{|g_4|} \left( \frac{M_p^2}{2} {\cal R}_4 - \frac{1}{2} g_{ij}\, \del_{\mu} \varphi^i \del^{\mu} \varphi^j - V(\varphi^k) \  + {\cal L}_{m,r} \right) \ ,
\eeq
where $M_p$ is the 4d reduced Planck mass, $\{ \varphi^i \}$ is a set of scalar fields (labeled by an index $i$) with scalar potential $V(\varphi^k)$ and field space metric $g_{ij}(\varphi^k)$, and ${\cal L}_{m,r}$ describes the physics of (relativistic) radiation and (non-relativistic) matter content. This model describes interacting scalar fields, minimally coupled to gravity, without any direct coupling to matter and radiation.

We will focus on solutions to this model, which admit a FLRW metric, namely
\beq
\d s_4^2 = - \d t^2 + a^2(t) \, \left( \frac{\d r^2}{1- k r^2} + r^2 \d \Omega_2^2 \right) \ ,
\eeq
with $a(t)$ the scale factor. The spatial curvature is captured by the constant $k$; we keep it generic for now. We restrict to homogeneous and isotropic scalar fields, i.e.~$\varphi^i(t)$. With such an ansatz, the equations of motion, namely the two Friedmann equations and the scalar fields equations of motion, are given by
\beq
\boxed{\hspace{1em} F_1 =0 \ ,\quad F_2=0 \ ,\quad E^i=0 \hspace{1em}}
\eeq
where
\bea
&& F_1 = 3 H^2 - \frac{1}{M_p^2} \sum_n \rho_n \ ,\quad F_2= \dot{H} + \frac{1}{2\, M_p^2} \sum_n (1+w_n)\rho_n \ ,\\
&& E^i= \ddot{\varphi}^i +\Gamma^i_{jk}\, \dot{\varphi}^j \dot{\varphi}^k + \ 3 H \dot{\varphi}^i + g^{ij} \del_{\varphi^j} V \ .
\eea
For a function $f$, we denote $\dot{f} \equiv \del_t f$. The Hubble parameter is $H=\dot{a}/a$, and we restrict to $a(t)>0$ for all times except in the limit to the origin. $\Gamma^i_{jk}$ is the Christoffel symbol for the field space metric $g_{ij}$. The energy densities $\rho_n$ are those of perfect fluids with pressure $p_n$, to which we associate equation of state parameters $w_n = p_n / \rho_n$. We consider in $F_1, F_2$ the components listed in Table \ref{tab:rhos}, namely radiation, matter, curvature and scalar field. The scalar component $\rho_{\varphi}$ will stand for dark energy.

In the following, we set $M_p=1$. Without loss of generality, we normalise $a(t_0)=1$ for a given $t_0>0$; this time will often correspond to today. We denote by an index ${}_0$ quantities at this time $t_0$. In particular, $\rho_{r0}$ and $\rho_{m0}$ are constant.

\begin{table}[H]
\begin{center}
\begin{tabular}{|l|c||c|c|c|}
\hline
&&&&\\[-8pt]
component & $n$ & $\rho_n$ & $w_n$ & $\rho_a$ \\[4pt]
\hline
&&&&\\[-8pt]
radiation & $r$ & $\rho_{r0}\, a^{-4}$ & $\frac{1}{3}$ & \\[4pt]
\hhline{----~}
&&&&\\[-8pt]
matter & $m$ & $\rho_{m0}\, a^{-3}$ & $0$ & $\rho_{a0}\, a^{-3(1+w_a)}$ \\[4pt]
\hhline{----~}
&&&&\\[-8pt]
curvature & $k$ & $-3k\, a^{-2}$ & $-\frac{1}{3}$ & \\[4pt]
\hline
&&&&\multicolumn{1}{|c}{}\\[-8pt]
scalar & $\varphi$ & $\frac{1}{2} g_{ij} \dot{\varphi}^i \dot{\varphi}^j  + V$ & $\frac{\frac{1}{2} g_{ij} \dot{\varphi}^i \dot{\varphi}^j  - V}{\frac{1}{2} g_{ij} \dot{\varphi}^i \dot{\varphi}^j  + V}$ & \multicolumn{1}{|c}{}\\[6pt]
\hhline{----~}
\end{tabular}
\end{center}\caption{Energy density and equation of state parameter of each component entering the cosmological model.}\label{tab:rhos}
\end{table}

For future convenience, we introduce notations for the separate scalar components that are the kinetic energy and the potential, as indicated in Table \ref{tab:rhokinV}. Since $\rho_{\varphi} = \rho_{\rm kin} + \rho_V$, and $w_{\varphi} \rho_{\varphi} = w_{\rm kin} \rho_{\rm kin} + w_V \rho_V$, the separate scalar components can equivalently be used in the various equations.
\begin{table}[H]
\begin{center}
\begin{tabular}{|l|c||c|c|}
\hline
&&&\\[-8pt]
component & $n$ & $\rho_n$ & $w_n$ \\[4pt]
\hline
&&&\\[-8pt]
kinetic energy & ${\rm kin}$ & $\frac{1}{2} g_{ij} \dot{\varphi}^i \dot{\varphi}^j$ & $1$ \\[4pt]
\hhline{----}
&&&\\[-8pt]
potential & $V$ & $V$ & $-1$ \\[4pt]
\hhline{----}
\end{tabular}
\end{center}\caption{Energy density and equation of state parameter for the separate scalar components.}\label{tab:rhokinV}
\end{table}

Given this setting, let us make few comments on the solutions to be considered. To start with, we will consider a positive definite $g_{ij}$, and $V \geq 0$. We thus get $\rho_{\varphi} \geq 0$. For $\rho_{\varphi} > 0$, we deduce $-1 \leq w_{\varphi} \leq 1$: this implies that a phantom behaviour $w_{\varphi} < -1$ is not possible in such a model.

In addition, for a flat universe, as will mostly be considered (as well as for an open one, $k=-1$), we get $\rho_n\geq 0$. Then, since $1+w_n \geq 0$, we deduce from $F_2=0$ that $\dot{H} \leq 0$. We will always have $\rho_r \neq 0$ or $\rho_m \neq 0$, so we deduce that $\dot{H}<0$ in our solutions. In addition, we will restrict ourselves to expanding universes, i.e.~$\dot{a}>0$, so $H > 0$. Therefore, in the solutions considered, $H$ will always be larger in the early universe.

Finally, it is straightforward to show that
\beq
\dot{\rho}_{\varphi} = \dot{\varphi}^i g_{ij} (E^j - 3 H  \dot{\varphi}^j) \ .\label{rhodotE}
\eeq
Therefore, in a solution with some field speed (satisfying $E^j=0$ and with $H>0$), we deduce that $\dot{\rho}_{\varphi} < 0$. In other words, Hubble friction, corresponding to the term $- 3 H  \dot{\varphi}^j$, makes $\rho_{\varphi}$ decrease with time. In addition, using $E^j=0$ gives that
\beq
\dot{\rho}_{\varphi} + 3 H \rho_{\varphi} = -3H w_{\varphi}  \rho_{\varphi} \ , \label{rhodot}
\eeq
which is the continuity equation for the dark energy or scalar field component. These results will be useful later.\\

Since $H \neq0$, we can introduce the energy density parameter $\Omega_n \equiv \rho_n / (3 H^2)$ for each component. The equations of motion then get rewritten in the convenient form
\beq
\boxed{\hspace{1em} f_1 =0 \ ,\quad f_2=0 \ ,\quad e^i=0 \hspace{1em}}
\eeq
where
\beq
f_1 = 1 -\sum_n \Omega_n \ ,\quad f_2 = \frac{\dot{H}}{3 H^2} + \frac{1}{2} \sum_n (1+w_n)\Omega_n \ ,\quad e^i= \frac{E^i}{3 H^2} \ .
\eeq
With $\rho_n \geq 0$, we deduce $\Omega_n \geq 0$, so $f_1=0$ imposes $\Omega_n \leq 1$: those parameters thus indicate the proportion of the component $n$ in the universe at a given time.

Last but not least, a useful expression for the energy density parameters can be obtained from $F_1 /(3 H_0^2) = 0$:
\beq
\frac{H^2}{H_0^2} = \Omega_{r0} \, a^{-4} + \Omega_{m0} \, a^{-3} + \Omega_{k0} \, a^{-2} + \Omega_{\varphi0}\, \frac{\rho_{\varphi}}{\rho_{\varphi0}} \ . \label{HH0}
\eeq
One deduces the following expressions
\bea
&& \Omega_r= \frac{\Omega_{r0}\, a^{-4}}{\Omega_{r0} \, a^{-4} + \Omega_{m0} \, a^{-3} + \Omega_{k0} \, a^{-2} + \Omega_{\varphi0}\, \frac{\rho_{\varphi}}{\rho_{\varphi0}}} \ ,\nn\\
&& \Omega_m= \frac{\Omega_{m0}\, a^{-3}}{\Omega_{r0} \, a^{-4} + \Omega_{m0} \, a^{-3} + \Omega_{k0} \, a^{-2} + \Omega_{\varphi0}\, \frac{\rho_{\varphi}}{\rho_{\varphi0}}}  \ , \label{Omegas}\\
&& \Omega_{\varphi}= \frac{\Omega_{\varphi0}\, \frac{\rho_{\varphi}}{\rho_{\varphi0}}}{\Omega_{r0} \, a^{-4} + \Omega_{m0} \, a^{-3} + \Omega_{k0} \, a^{-2} + \Omega_{\varphi0}\, \frac{\rho_{\varphi}}{\rho_{\varphi0}}} \ ,\nn
\eea
and similarly for $\Omega_k$. As mentioned above, we have here $\Omega_n \geq 0$; the expressions above then make it obvious that $\Omega_n \leq 1$, and that $\Omega_n$ gives the proportion of each component.

Let us also introduce the effective equation of state parameter $w_{{\rm eff}}= \sum_n w_n \, \Omega_n$. We can then show
\beq
\frac{F_1 + 2 F_2}{3 H^2} = \frac{2}{3 H^2} \frac{\ddot{a}}{a} + \frac{1}{3} + w_{{\rm eff}} \ ,
\eeq
giving the well-known condition for an accelerating solution
\beq
\ddot{a} > 0 \ \Leftrightarrow \ w_{{\rm eff}} < - \frac{1}{3} \ . \label{acc}
\eeq

With this formalism at hand, we can now look for solutions. This requires some rewriting of the equations that we now turn to.

\subsection{Equation system and reformulations}\label{sec:eqreform}

The three equations to solve form a differential equation system on the variables $a(t), \varphi(t)$, depending on the parameters $\rho_{r0}, \rho_{m0}, k$ and $g_{ij},V$. Having specified these parameters, as well as initial conditions on $a(t), \varphi(t), \dot{\varphi}(t)$, one can obtain a solution.

A useful property to solve the system is the following. The three equations above (with components of Table \ref{tab:rhos}) obey the following relation
\beq
\dot{F_1} = -\dot{\varphi}^i g_{ij}\, E^j + 6 H \, F_2 \ , \label{relEq}
\eeq
indicating that it is not necessary to solve all 3 equations.\footnote{\eqref{relEq} is consistent with \eqref{rhodotE} combined with the general continuity equation, given as $6H F_2 - \dot{F}_1 = \sum \dot{\rho}_n + 3 H \sum (1+w_n) \rho_n$. Note that the continuity equation is also obeyed by each component of Table \ref{tab:rhos} individually, but not by those of Table \ref{tab:rhokinV}.} Similarly, one obtains
\beq
\dot{f_1} + 2\frac{\dot{H}}{H} f_1 =  -\dot{\varphi}^i g_{ij}\, e^j + 6 H \, f_2 \ .
\eeq
In the following we will then specify, depending on the case and example, which equations get explicitly solved.\\

Several reformulations of the system prove useful, depending in particular on whether one wants the solution to evolve in terms of time (actually in terms of $H_0\, t$), or rather in terms the number of e-folds, $N \equiv \ln a$.

For the former, we consider $F_1=0, \ E^i=0$, which are sufficient to solve the system thanks to \eqref{relEq} with $H\neq0$. We rewrite them as follows
\begin{subequations}
\begin{empheq}[box=\widefbox]{align}
& a' = \sqrt{\Omega_{r0} \, a^{-2} + \Omega_{m0} \, a^{-1} + \Omega_{k0} + \frac{a^2}{3} \left(\frac{1}{2} g_{ij} {\varphi^i}' {\varphi^j}' + \tilde{V} \right)} \label{eqF1H0t}\\
& {\varphi^i}'' + \Gamma^i_{jk}\, {\varphi^j}' {\varphi^k}' + 3\, \frac{a'}{a} {\varphi^i}' + g^{ij}\del_{\varphi^i} \tilde{V} =0\ , \label{eqEiH0t}
\end{empheq}
\end{subequations}
where $\tilde{V} \equiv \frac{V}{H_0^2}$ and for a function $f$, $f' \equiv \dot{f}/H_0$. We also recall that we focus on solutions with an expanding universe, i.e.~$\dot{a}> 0$, and $a'>0$. The system now depends on the parameters $\Omega_{r0}, \Omega_{m0}, \Omega_{k0}$ and $g_{ij},\tilde{V}$. A solution is given by $a,\varphi^i$ that evolve in terms of $H_0\, t$. Initial conditions are $a(t_0)=1$, and $\varphi^i(t_0), {\varphi^i}'(t_0)$, to be specified. Using the definitions of $\Omega_{\varphi0},w_{\varphi0}$, one obtains the relations
\beq
g_{ij} {\varphi^i}' {\varphi^j}'(t_0) = 3 \Omega_{\varphi0} (1+w_{\varphi0}) \ , \quad \tilde{V}(t_0) = \frac{3}{2} \Omega_{\varphi0} (1-w_{\varphi0}) \ .\label{relphiOw}
\eeq
In case of a single, canonical (i.e.~with $g_{ij}=\delta_{ij}$) scalar field, and with the possibility of inverting the function $\tilde{V}(\varphi)$, it is sufficient to specify $\Omega_{\varphi0}, w_{\varphi0}$ to get the field initial conditions. Note that the sign of the speed should still be fixed, and we will typically take ${\varphi^i}'(t_0)>0$. Finally, one can also take $\Omega_{\varphi0} \equiv 1-\Omega_{r0}- \Omega_{m0}- \Omega_{k0}$, because of $f_1=0$.

To summarize, to get a solution evolving in terms the time $H_0\, t$, with a single canonical field, it is sufficient to solve \eqref{eqF1H0t} and \eqref{eqEiH0t}, with the data
\beq
\Omega_{r0}, \Omega_{m0}, \Omega_{k0},\quad  \tilde{V}, \quad {\rm and}\  w_{\varphi0} \ .
\eeq
Throughout this work, we will take the following fiducial values, in agreement with observational constraints from the CMB on the flat $\Lambda$CDM model \cite{Planck:2018vyg}, and very close to those observationally obtained with quintessence models \cite{DESI:2024mwx}
\beq
\Omega_{r0}=0.0001\ ,\ \Omega_{m0}=0.3149\ ,\ \Omega_{k0}=0 \ ,\ \Omega_{\varphi0}=1-\Omega_{r0}- \Omega_{m0}- \Omega_{k0}=0.6850 \ . \label{fidO}
\eeq
This will leave us to specify $\tilde{V}, w_{\varphi0}$. The latter two parameters are thus crucial as they determine the whole physics of the solution.\\

We turn to reformulations where solutions evolve in terms of $N=\ln a$, the number of e-folds. For a function $f$, $\del_N f= \dot{f}/H = f' \times H_0/H$. It is then straightforward to rewrite equations $F_1=0,F_2=0,E^i=0$ respectively as follows
\begin{subequations}
\begin{empheq}[box=\widefbox]{align}
& \tilde{H} - \sqrt{\Omega_{r0} \, e^{-4 N} + \Omega_{m0} \, e^{-3 N} + \Omega_{k0} \, e^{-2 N} + \frac{1}{3} \left(\frac{\tilde{H}^2}{2} g_{ij} \del_N \varphi^i \del_N \varphi^j + \tilde{V} \right)} =0  \label{eqF1N} \\
&  2 \tilde{H} \del_N \tilde{H} + 4 \Omega_{r0} \, e^{-4 N} + 3 \Omega_{m0} \, e^{-3 N} + 2 \Omega_{k0} \, e^{-2 N} + \tilde{H}^2 g_{ij} \del_N \varphi^i \del_N \varphi^j = 0 \label{eqF2N} \\
& \tilde{H}^2 \del_N^2 \varphi^i + \tilde{H} \del_N \tilde{H} \del_N \varphi^i + \tilde{H}^2 \Gamma^i_{jk} \del_N \varphi^j \del_N \varphi^k +  3 \tilde{H}^2 \del_N \varphi^i + g^{ij} \del_{\varphi^j} \tilde{V} =0 \label{eqEiN}
\end{empheq}
\end{subequations}
where $\tilde{V} \equiv \frac{V}{H_0^2}$, $\tilde{H} \equiv \frac{H}{H_0}$. One of the equations is redundant in view of \eqref{relEq}; we will come back to this point. The system depends on the parameters $\Omega_{r0}, \Omega_{m0}, \Omega_{k0}$ and $g_{ij},\tilde{V}$, and a solution is now given by the functions $\tilde{H},\varphi^i$ that depend on $N$. The initial conditions are set at $t_0$, namely $a=1$ or $N=0$, for which we get $\tilde{H}(0)=1$. We are once again left with specifying those for $\varphi^i, \del_N \varphi^i$; since ${\varphi^i}' = \tilde{H} \del_N \varphi^i$, we get ${\varphi^i}'(0) = \del_N \varphi^i(0)$. We then get from above the relations
\beq
g_{ij} \del_N\varphi^i \del_N \varphi^j (0) = 3 \Omega_{\varphi0} (1+w_{\varphi0}) \ , \quad \tilde{V}(0) = \frac{3}{2} \Omega_{\varphi0} (1-w_{\varphi0}) \ . \label{kinViniN}
\eeq
As before, for a single canonical field, and an invertible potential function, we can trade the field initial conditions for the data $\Omega_{\varphi0}, w_{\varphi0}$, with the (typically positive) sign of the field speed. Taking again the fiducial values \eqref{fidO}, we are left to specify $\tilde{V},w_{\varphi0}$ to get a solution.

If we solve $E^i=0, F_2=0$, the relation \eqref{relEq} implies that $F_1$ is a constant. Evaluating that constant at $t_0$ using the initial conditions, we get the left-hand side of \eqref{eqF1N} to be
\beq
1 - \sqrt{\Omega_{r0}  + \Omega_{m0} + \Omega_{k0} + \Omega_{\varphi0}}
\eeq
which vanishes with the fiducial values \eqref{fidO}. We conclude that $F_1=0$ at any time, as it should. In this way, it is sufficient to solve equations $E^i=0, F_2=0$, i.e.~\eqref{eqF2N}, \eqref{eqEiN}.\\

A last reformulation of the system can be obtained in terms of the variables
\beq
x= \sqrt{\Omega_{{\rm kin}}} \ ,\ y= \sqrt{\Omega_V} \ ,\ z= \sqrt{\Omega_k} \ ,\ u=\sqrt{\Omega_r} \ , \label{xyzuO}
\eeq
evolving in terms of $N$. The three equations can be used to form a system of first order differential equations. With a single canonical scalar field, an extra variable, $-\del_{\varphi} V /V$, can be introduced. In the case where this variable is constant, it can be considered as a parameter. The system then becomes a closed first order differential system, that can be studied using the methods of dynamical systems. $-\del_{\varphi} V /V$ being constant amounts to have an exponential potential; we refer to \cite[(2.17)]{Andriot:2024jsh} for the corresponding system, and subsequent dynamical system analysis. Beyond the exponential potential however, this formulation is less interesting and we will rather use the above formulation, solving \eqref{eqF2N}, \eqref{eqEiN}.\\

Having discussed the equation system to be solved, we turn in the next section to solution examples.

\section{Numerical solution examples, and domination phases}\label{sec:examples}

In this section, we numerically obtain examples of solutions to the system of equations detailed in the previous section. We do so restricting to a flat universe, i.e.~$k=0$ or $\Omega_{k0}=0$, and to a single canonical scalar field. We use the fiducial values \eqref{fidO} for $\Omega_{r0}, \Omega_{m0}, \Omega_{\varphi0}$. As explained in Section \ref{sec:eqreform}, we are then left to specify $\tilde{V}$ and $w_{\varphi0}$. We will consider three different scalar potentials: a (cosmological) constant, falling back to $\Lambda$CDM model with $w_{\varphi}=-1$, an exponential and a hilltop potential. The last two are quintessence models.

For $\Lambda$CDM, the solution is then completely fixed, but for the exponential and hilltop quintessence models, different solutions are obtained for different values of $w_{\varphi0}$. In the following, we will fine-tune this value in order to get a solution with a radiation domination phase, that starts around $N=-20$. This is motivated by the time of BBN, which has the corresponding redshift $z_{{\rm BBN}} \sim e^{20}$ \cite{Andriot:2024jsh}; an earlier value for the start of the radiation phase is possible, by a further tuning of $w_{\varphi0}$. Without such a fine-tuned value of $w_{\varphi0}$, most of the solutions obtained do not have a radiation domination phase; those are thus unrealistic (see e.g.~\cite[Sec. 3.1]{Andriot:2024jsh}).

\subsection{$\Lambda$CDM}\label{sec:LCDM}

We start by considering the case of a (positive) cosmological constant: the potential is a constant function $V= \Lambda >0$. We further fix $w_{\varphi}=-1$ (at any time), which amounts to take constant scalar fields, $\dot{\varphi}^i=0$. This setup corresponds to the so-called $\Lambda$CDM model; the solution is completely fixed, as explained above. The equations $E^i=0$ are automatically satisfied. Thanks to \eqref{relEq}, solving $F_1=0$ is sufficient to have the complete solution.

\begin{figure}[H]
\centering
\includegraphics[width=0.6\textwidth]{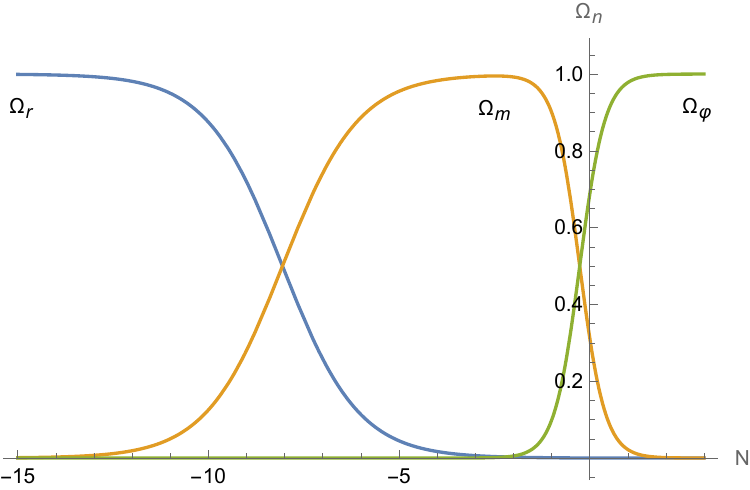}
\caption{Evolution of the $\Omega_{n}$ for $\Lambda$CDM.}\label{fig:LCDMON}
\end{figure}

Let us first note that since $\rho_{\varphi}=\Lambda$ is constant, the expressions \eqref{Omegas} of the $\Omega_n$ simplify:
\bea
&& \Omega_r= \frac{\Omega_{r0} \, a^{-4}}{\Omega_{r0} \, a^{-4} + \Omega_{m0} \, a^{-3} + \Omega_{\varphi0}}  \ ,\nn\\
&& \Omega_m= \frac{\Omega_{m0} \, a^{-3}}{\Omega_{r0} \, a^{-4} + \Omega_{m0} \, a^{-3} + \Omega_{\varphi0}}  \ ,\\
&& \Omega_{\varphi}= \frac{\Omega_{\varphi0}}{\Omega_{r0} \, a^{-4} + \Omega_{m0} \, a^{-3} + \Omega_{\varphi0}} \ .\nn
\eea
Those only depend on $a=e^N$, it is then straightforward to depict in Figure \ref{fig:LCDMON} their evolution in terms of the number of e-folds $N$. We read from Figure \ref{fig:LCDMON} the well-known successive phases of radiation, matter and dark energy domination; we will come back to those in Section \ref{sec:phases}.

\begin{figure}[H]
\centering
\includegraphics[width=0.6\textwidth]{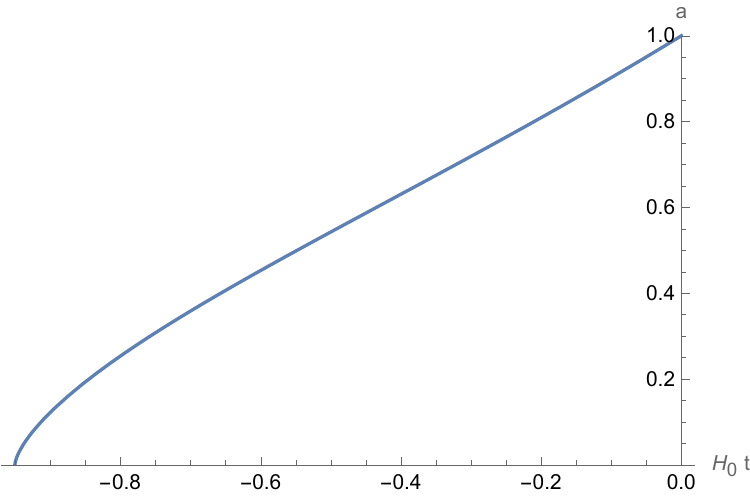}
\caption{$a(t)$ in terms of $H_0\, t$ with the normalisation $a(t_0)=1$. Time is shifted such that $t_0=0$ corresponds to the universe today. We verify the expansion, namely the growth of $a(t)$, and observe that in the past, $a(t)=0$ at $H_0\, t=-0.9506$.}\label{fig:LCDMaT}
\end{figure}

To get the evolution in terms of time, we still need to solve $F_1=0$ and get $a(t)$. As in \eqref{eqF1H0t}, this equation can be written as
\beq
\frac{\d a}{\sqrt{\Omega_{r0}\, a^{-2} + \Omega_{m0}\, a^{-1} + \Omega_{\varphi0}\, a^2}} = H_0\, \d t \ .
\eeq
We solve this equation numerically, with the normalisation $a(t_0)=1$, where $t_0$ represents here the time today. The solution is depicted in Figure \ref{fig:LCDMaT}.
We see in Figure \ref{fig:LCDMaT} the expansion of the universe, and observe the ``beginning'' or initial singularity, $a(t)=0$, at $H_0\, (t_0-t)=0.9506$. From this and a value of $H_0$, the ``Hubble constant'' today, one deduces the age of the universe. $H_0$ is nowadays subject to the well-known Hubble tension; if we stick for illustration to the values of \cite{Planck:2018vyg} and take $H_0=67.4\, {\rm km}.{\rm s}^{-1}.{\rm Mpc}^{-1}$, we obtain\footnote{We recall that $1\, {\rm km}.{\rm s}^{-1}.{\rm Mpc}^{-1} \approx 1.022 \cdot 10^{-3}\, (10^9 {\rm yrs})^{-1}$, giving here $\frac{1}{H_0} = 14.517 \cdot 10^9\, {\rm yrs}$.}
\beq
\text{Age of the universe}\approx 13.80 \cdot 10^9\, {\rm yrs} \ .\label{ageLCDM}
\eeq

\begin{figure}[H]
\centering
\includegraphics[width=0.48\textwidth]{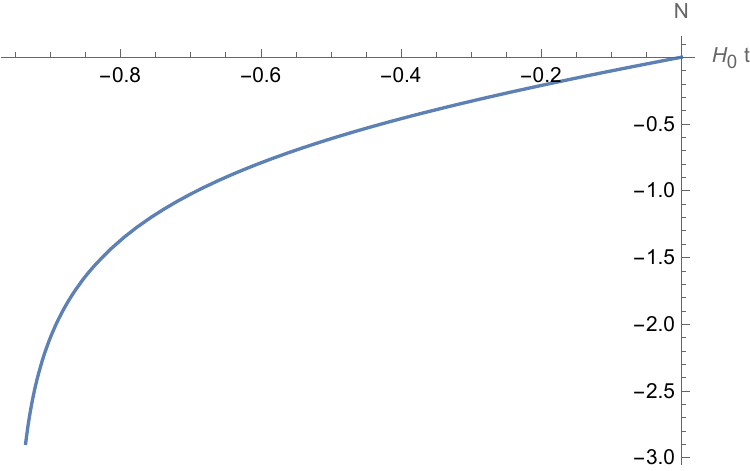}
\caption{$N(t)=\ln a(t)$ in terms of $H_0\, t$ for $\Lambda$CDM, with $t_0=0$.}\label{fig:LCDMNt}
\end{figure}

\begin{figure}[H]
\begin{center}
\begin{subfigure}[H]{0.48\textwidth}
\includegraphics[width=\textwidth]{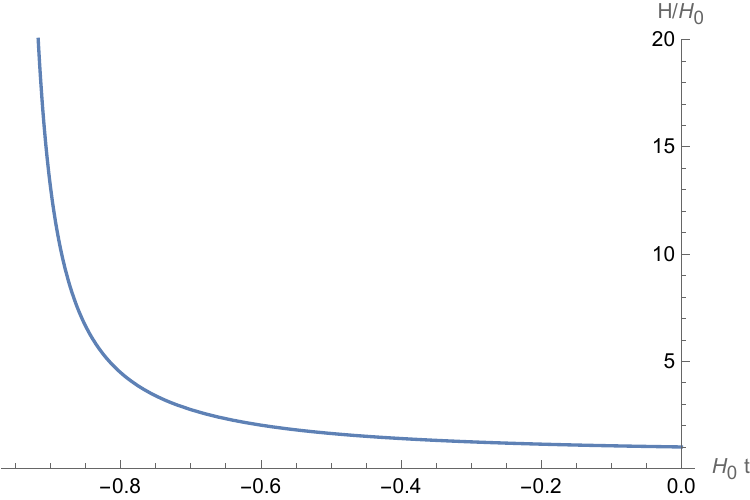}\caption{$H(t)$}\label{fig:LCDMHt}
\end{subfigure}\quad
\begin{subfigure}[H]{0.48\textwidth}
\includegraphics[width=\textwidth]{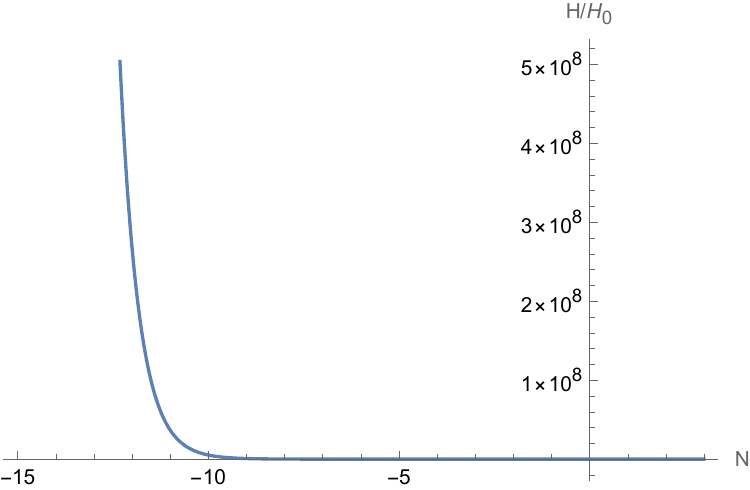}\caption{$H(N)$}\label{fig:LCDMHN}
\end{subfigure}
\caption{$H(t)/H_0$ and $H(N)/H_0$ for $\Lambda$CDM.} \label{fig:LCDMH}
\end{center}
\end{figure}

We finally depict in Figure \ref{fig:LCDMNt} and \ref{fig:LCDMH} the evolution of further quantities, and comment on them.
The number of e-folds evolving in terms of time in Figure \ref{fig:LCDMNt} allows to realise that most of the lifetime of the universe is captured approximately by the last 3 e-folds. This also shows the relevance of one or the other dependence, $t$ or $N$, according to what period is studied.

From Figure \ref{fig:LCDMH}, we learn that $H/H_0$ takes huge values in the early universe. This quantity decreases as expected (see Section \ref{sec:setting}), reaching in this model a value close to a constant today; the fact it asymptotes to a constant for $\Lambda$CDM can only be seen via the analytic solution, discussed in Section \ref{sec:lastphaseanalytic}. Using some values of $N$ that will become relevant in Section \ref{sec:phases}, we get the following evaluation of the Hubble parameter
\bea
&& N=-8.055:\quad H/H_0= 1.402 \cdot 10^5 \\
&& N= -2.483:\quad H/H_0= 23.32
\eea
Most of these observations will remain true for the quintessence models and their cosmological solutions, that we now turn to.

\subsection{Exponential quintessence}\label{sec:expquint}

We restrict to a single canonical field $\varphi$ and consider an exponential potential
\beq
V(\varphi) = V_0 \, e^{-\lambda\, \varphi} \ ,\quad \lambda,V_0 >0 \ .
\eeq
This model has been studied extensively in \cite{Andriot:2024jsh} (see also references therein). We introduce a constant field value $\varphi_*$, such that $e^{\lambda\, \varphi_*} \equiv \frac{V_0}{H_0^2}$, i.e.~$\tilde{V} = e^{-\lambda(\varphi - \varphi_*)}$. We can then redefine the field $\varphi \rightarrow \varphi + \varphi_*$ to absorb this constant; the value of the field will always be known up to a constant. The potential is then only characterised by $\lambda$: we get $\tilde{V} = e^{-\lambda\, \varphi }$. For this example, we pick $\lambda =\sqrt{3}$, but a lower value could equally be chosen: our goal is here to illustrate qualitative features of the solution.

\begin{figure}[H]
\begin{center}
\begin{subfigure}[H]{0.48\textwidth}
\includegraphics[width=\textwidth]{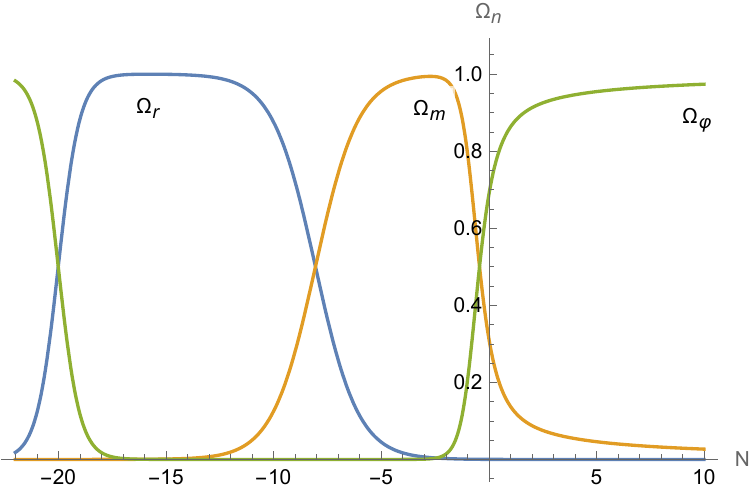}\caption{}\label{fig:ExpQuintON}
\end{subfigure}\quad
\begin{subfigure}[H]{0.48\textwidth}
\includegraphics[width=\textwidth]{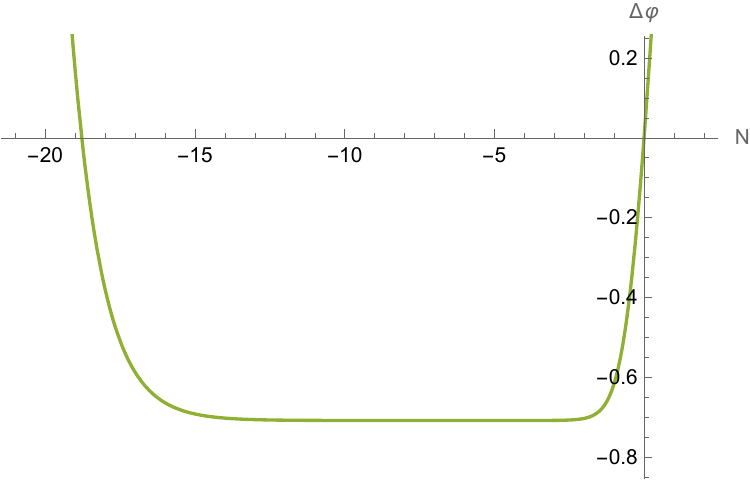}\caption{}\label{fig:ExpQuintpN}
\end{subfigure}
\caption{$\Omega_n(N)$ and $\Delta \varphi(N) = \varphi(N) - \varphi_0$ in terms of $N$, for exponential quintessence with $\lambda=\sqrt{3}$ and $w_{\varphi0}=-0.51073604885$ as explained in the main text. The scalar field is extracted from the solution as $\varphi = -\frac{1}{\lambda} \ln \left( \frac{y^2}{\Omega_m \, e^{3\, N}} \right) + {\rm constant}$, where we recall $y^2 = \Omega_V = V/(3H^2)$.} \label{fig:ExpQuintN}
\end{center}
\end{figure}

As explained at the beginning of Section \ref{sec:examples}, we are left to fix $w_{\varphi0}$ to get a solution: we take $w_{\varphi0}=-0.51073604885$ allowing the radiation phase to start around $N=-20$. To get the solution in terms of $N$, we use, as explained in Section \ref{sec:eqreform}, the dynamical system formulation of the equations in terms of variables $(x,y,z,u)$, and solve it numerically. This gives direct access to the $\Omega_n (N)$, as indicated in \eqref{xyzuO}. We depict the solution in Figure \ref{fig:ExpQuintN}. We recall the notation for the field value today: $\varphi_0=\varphi(N=0)$.
We see in Figure \ref{fig:ExpQuintON} the different domination phases, that we will comment on in Section \ref{sec:phases}. Turning to the field evolution, we see in Figure \ref{fig:ExpQuintpN} that its value seems to get frozen, essentially between radiation and matter domination: we will study this phenomenon in Section \ref{sec:anafrozenfield} and \ref{sec:anafieldapplication}.

\begin{figure}[H]
\begin{center}
\begin{subfigure}[H]{0.48\textwidth}
\includegraphics[width=\textwidth]{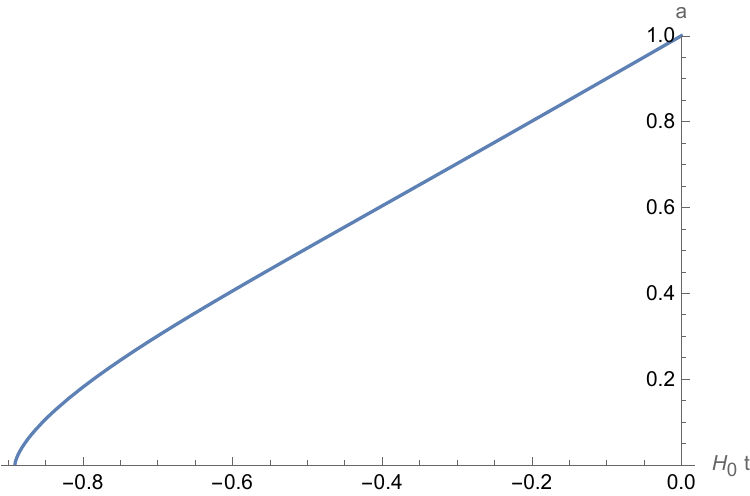}\caption{$a(t)$}\label{fig:ExpQuinta}
\end{subfigure}\quad
\begin{subfigure}[H]{0.48\textwidth}
\includegraphics[width=\textwidth]{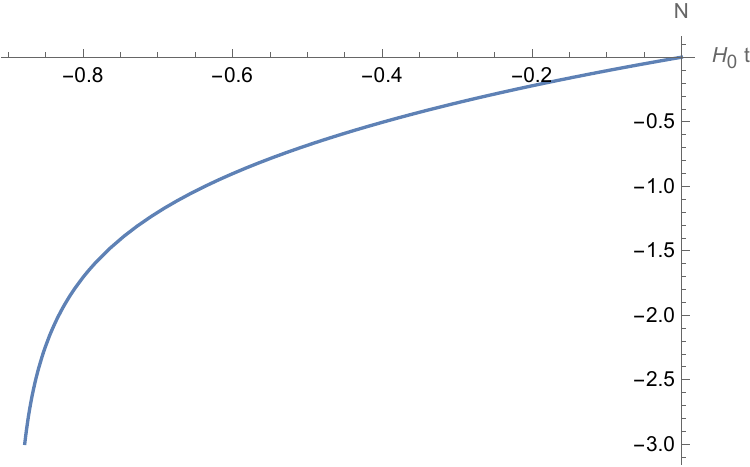}\caption{$N(t)$}\label{fig:ExpQuintNt}
\end{subfigure}\\
\begin{subfigure}[H]{0.48\textwidth}
\includegraphics[width=\textwidth]{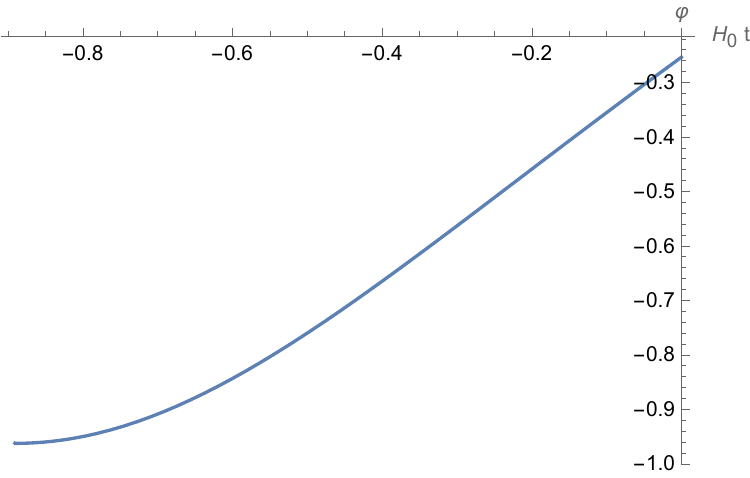}\caption{$\varphi(t)$}\label{fig:ExpQuintp}
\end{subfigure}\quad
\begin{subfigure}[H]{0.48\textwidth}
\includegraphics[width=\textwidth]{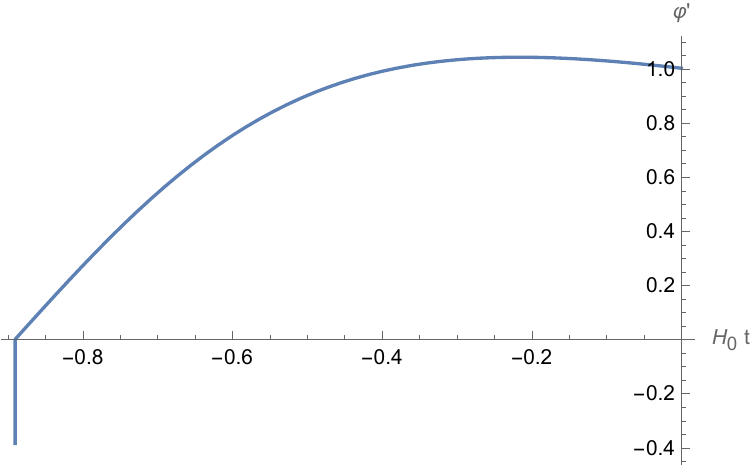}\caption{$\varphi'(t)$}\label{fig:ExpQuintppt}
\end{subfigure}
\caption{Solution $a(t), N(t), \varphi(t), \varphi'(t)$ in terms of $H_0\, t$ (with $t_0=0$), for exponential quintessence with $\lambda=\sqrt{3}$ and $w_{\varphi0}=-0.51073604885$. Having $\varphi'(t_0) \approx 1$ is accidental.} \label{fig:ExpQuintt}
\end{center}
\end{figure}

In order to get the solution in terms of time, we follow the discussion of Section \ref{sec:eqreform} and solve the reformulated equations \eqref{eqF1H0t} and \eqref{eqEiH0t}. We recall from the discussion around \eqref{relphiOw} that the initial conditions are given here by
\beq
\varphi'(t_0) = \sqrt{3 \Omega_{\varphi0} (1+w_{\varphi0})} \ ,  \quad \varphi_0 = -\frac{1}{\lambda} \ln \left(\frac{3}{2} \Omega_{\varphi0} (1-w_{\varphi0})\right)\ ,
\eeq
where as mentioned, we consider a solution with $\varphi'(t_0)>0$; it was noticed in \cite{Andriot:2024jsh} that this sign allows to get a solution with past matter domination. We then obtain the solution numerically, and display it in Figure \ref{fig:ExpQuintt}.
$a(t)$ and $N(t)$ in Figure \ref{fig:ExpQuintt} are very similar to those of the $\Lambda$CDM solution depicted in Figure \ref{fig:LCDMaT} and \ref{fig:LCDMNt}: we verify again the universe expansion, and we see that most of the lifetime of the universe is captured by 3 e-folds. The main difference is that $a(t)=0$ is now reached at $H_0\, (t_0-t)=0.8908$, giving a priori a slightly shorter age of the universe, although strictly speaking, $H_0$ should be reevaluated within this model.

We turn to the scalar field in Figure \ref{fig:ExpQuintt}, not present for $\Lambda$CDM. We observe an initial phase in $\varphi'(t)$ with a non-zero kinetic energy (here with $\varphi'(t)<0$):\footnote{\label{foot:sign}The sign of $\varphi'(t)$ in this very early universe is fixed here by the tuning of $w_{\varphi0}$. A tiny change in that value can give a different solution with the opposite sign of $\varphi'(t)$. This other solution is otherwise almost indistinguishable, since most quantities do not depend on this sign: see e.g.~\cite[Fig. 8, 17]{Andriot:2024jsh}.} this corresponds to an initial kination phase that we will comment on in Section \ref{sec:phases}, and that can be seen already in Figure \ref{fig:ExpQuintON}. We do not see this phase in $\varphi(t)$, due to numerical precision issues in the very initial moments. The freezing of the field, noticed in Figure \ref{fig:ExpQuintpN}, is barely visible in Figure \ref{fig:ExpQuintp}: we notice a constant value of $\varphi(t)$ in the early universe, around the time when $\varphi'(t)=0$.\\

\begin{figure}[H]
\begin{center}
\begin{subfigure}[H]{0.48\textwidth}
\includegraphics[width=\textwidth]{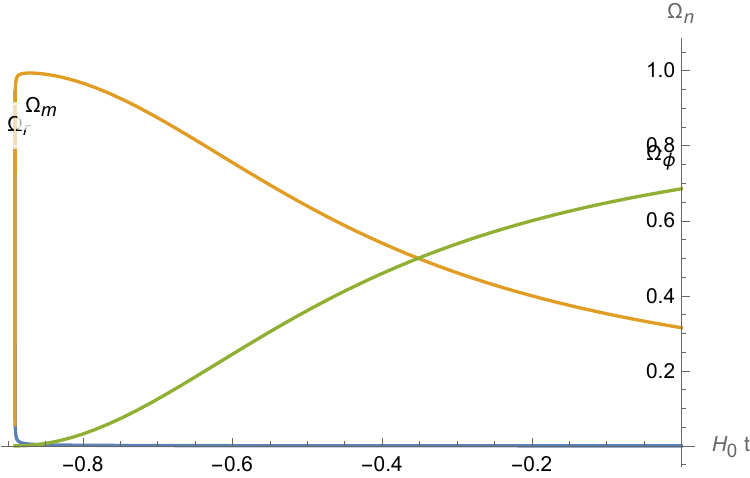}\caption{}\label{fig:ExpQuintOt}
\end{subfigure}\quad
\begin{subfigure}[H]{0.48\textwidth}
\includegraphics[width=\textwidth]{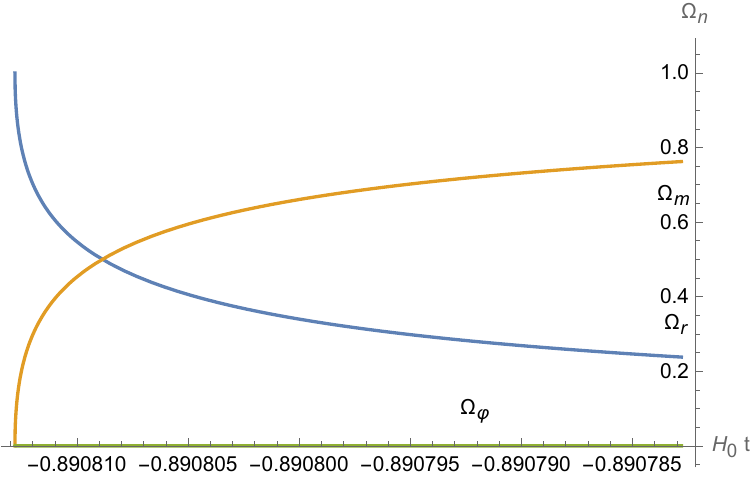}\caption{}\label{fig:ExpQuintO0}
\end{subfigure}
\caption{$\Omega_n(t)$ in terms of $H_0\,t$ for the exponential quintessence example. The radiation domination can barely be seen in Figure \ref{fig:ExpQuintOt}, but is more noticeable when zooming on the early times in Figure \ref{fig:ExpQuintO0}. Due to precision issues in the initial moments, we do not see $\Omega_r$ going down and $\Omega_{\varphi}$ rising in the initial kination phase.} \label{fig:ExpQuintOtt}
\end{center}
\end{figure}

We provide in addition the evolution of the $\Omega_n$ in terms of time in Figure \ref{fig:ExpQuintOtt}. Most of their evolution in terms of $N$, displayed in Figure \ref{fig:ExpQuintON}, can be seen here, but the evolution in terms of time still appears less suited.

\begin{figure}[H]
\begin{center}
\begin{subfigure}[H]{0.48\textwidth}
\includegraphics[width=\textwidth]{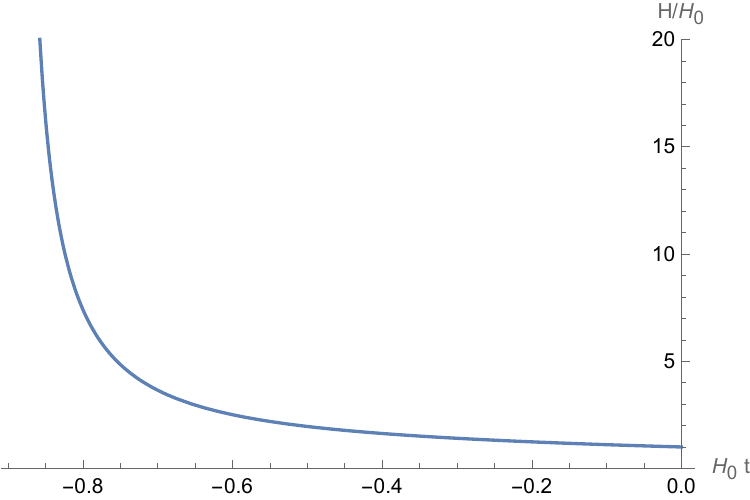}\caption{$\frac{H}{H_0}(t)$}\label{fig:ExpQuintHt}
\end{subfigure}\quad
\begin{subfigure}[H]{0.48\textwidth}
\includegraphics[width=\textwidth]{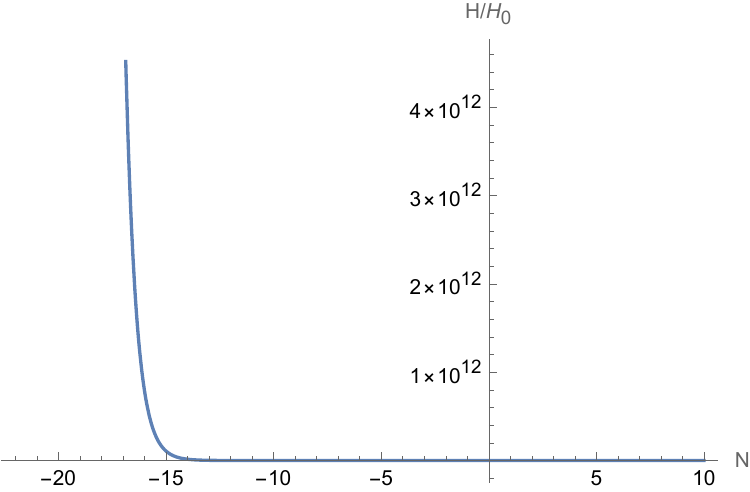}\caption{$\frac{H}{H_0}(N)$}\label{fig:ExpQuintHN}
\end{subfigure}\\
\begin{subfigure}[H]{0.48\textwidth}
\includegraphics[width=\textwidth]{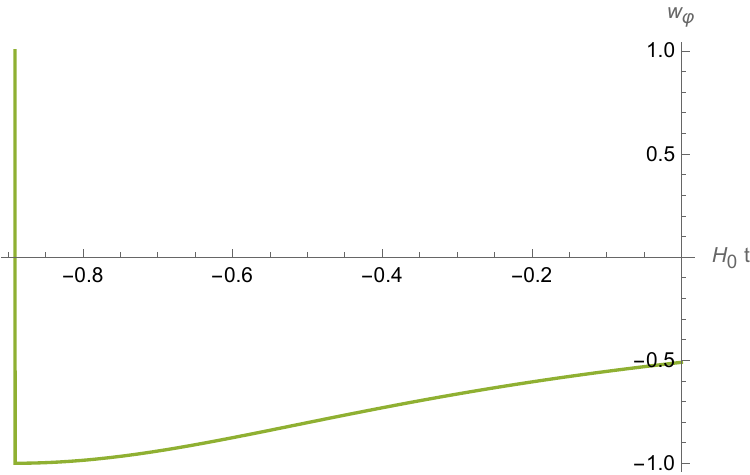}\caption{$w_{\varphi}(t)$}\label{fig:ExpQuintwt}
\end{subfigure}\quad
\begin{subfigure}[H]{0.48\textwidth}
\includegraphics[width=\textwidth]{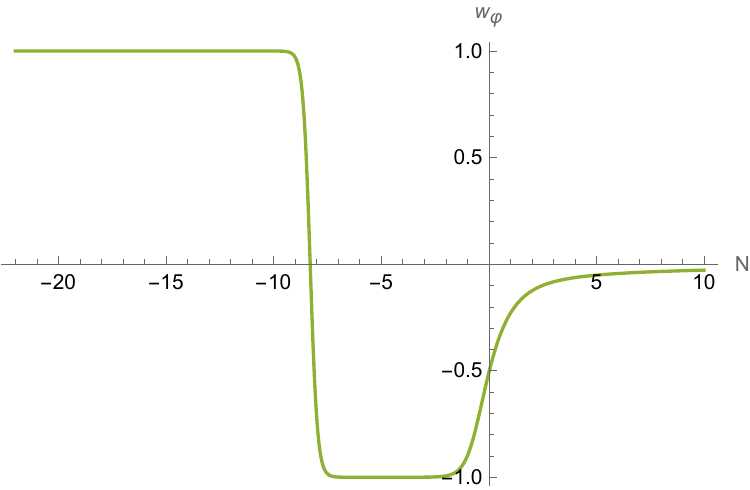}\caption{$w_{\varphi}(N)$}\label{fig:ExpQuintwN}
\end{subfigure}
\caption{$H/H_0$ and $w_{\varphi}$ evolving in terms time and in terms of e-folds, for the exponential quintessence example. $H/H_0$ can be accessed as $(H/H_0)^2=a^{-3} \Omega_{m0}/\Omega_m$.} \label{fig:ExpQuintwH}
\end{center}
\end{figure}
The evolution of further quantities in terms of time and e-folds is given in Figure \ref{fig:ExpQuintwH}. There is not much qualitative difference for $H$ compared to that of $\Lambda$CDM, depicted in Figure \ref{fig:LCDMH}: it takes huge values in the early universe, and decreases. Last but not least, we depict $w_{\varphi}$ in Figure \ref{fig:ExpQuintwH}, that was not discussed for $\Lambda$CDM (there, $w_{\varphi}=-1$): its evolution deserves some comments.

The evolution of $w_{\varphi}$ is remarkable: it starts at $+1$, in agreement with an initial kination phase (where kinetic energy dominates). It then sharply drops to $-1$, indicating a domination of the potential, analogous to a cosmological constant. It finally raises up to the value today, $w_{\varphi0}$, and further evolves in the future. We will come back in great detail in Section \ref{sec:anafrozenfield} and \ref{sec:anafieldapplication} to this evolution, which turns out to be generic.

\subsection{Hilltop quintessence}\label{sec:hilltopquint}

We finally consider the hilltop potential, given by
\beq
V(\varphi) =V_0 \left( 1 - \frac{\kappa^2}{2} \varphi^2 \right) \ , \ V_0,\kappa>0 \ ,
\eeq
discussed e.g.~in \cite{Shlivko:2024llw}. We restrict implicitly to the field range for which $V>0$. The potential now depends on two parameters, for which we take
\beq
\tilde{V}_0=\frac{V_0}{H_0^2}=5 \ ,\quad \kappa=\frac{1}{2} \ . \label{Hilltopparam}
\eeq
Note that $\tilde{V}/\tilde{V}_0 \leq 1$. In particular, let us recall from \eqref{kinViniN} that the value today is given by $\tilde{V}(0) = \frac{3}{2} \Omega_{\varphi0} (1-w_{\varphi0})$. We deduce that $\tilde{V}_0 > \frac{3}{2} \Omega_{\varphi0} (1 - w_{\varphi0})$. Using in addition \eqref{acc}, assuming solutions that describe an accelerating universe today, we have $\Omega_{\varphi0}  w_{\varphi0} < -\frac{1}{3} (1 + \Omega_{r0}) < -\frac{1}{3}$, giving $-w_{\varphi0}>\frac{1}{3}$. We conclude
\beq
\tilde{V}_0 > 2 \Omega_{\varphi0} \approx 1.3700 \ .
\eeq
In other words, $V_0> H_0^2 M_p^2$, and the value chosen above for $\tilde{V}_0$ cannot be lowered much.

\begin{figure}[H]
\begin{center}
\begin{subfigure}[H]{0.48\textwidth}
\includegraphics[width=\textwidth]{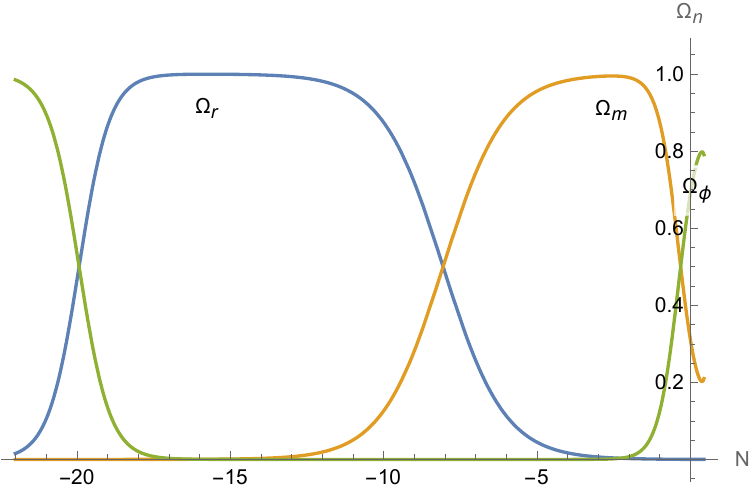}\caption{$\Omega_n(N)$}\label{fig:HilltopON}
\end{subfigure}\quad
\begin{subfigure}[H]{0.48\textwidth}
\includegraphics[width=\textwidth]{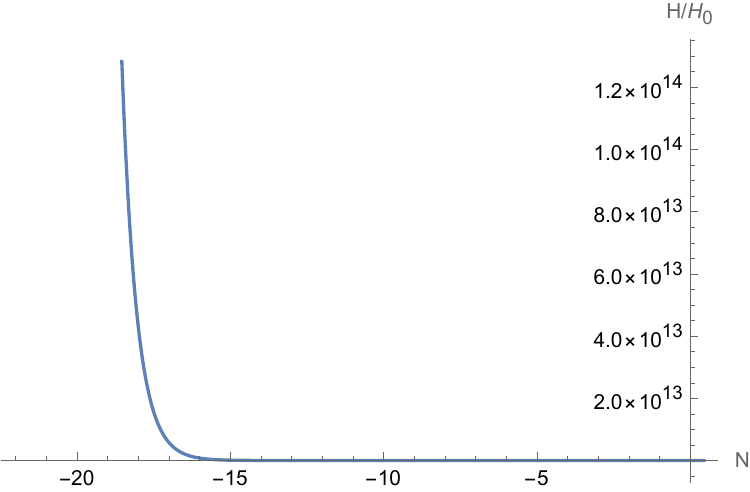}\caption{$\frac{H}{H_0}(N)$}\label{fig:HilltopHN}
\end{subfigure}\\
\begin{subfigure}[H]{0.48\textwidth}
\includegraphics[width=\textwidth]{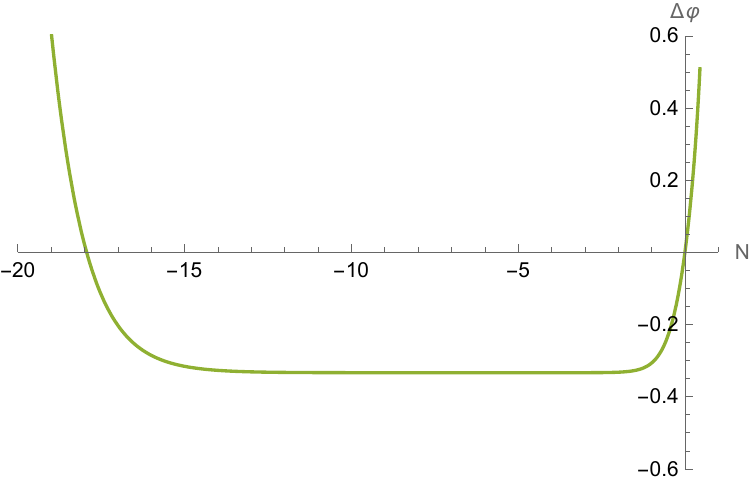}\caption{$\Delta \varphi(N) = \varphi(N) - \varphi(0)$}\label{fig:HilltopDPhiN}
\end{subfigure}\quad
\begin{subfigure}[H]{0.48\textwidth}
\includegraphics[width=\textwidth]{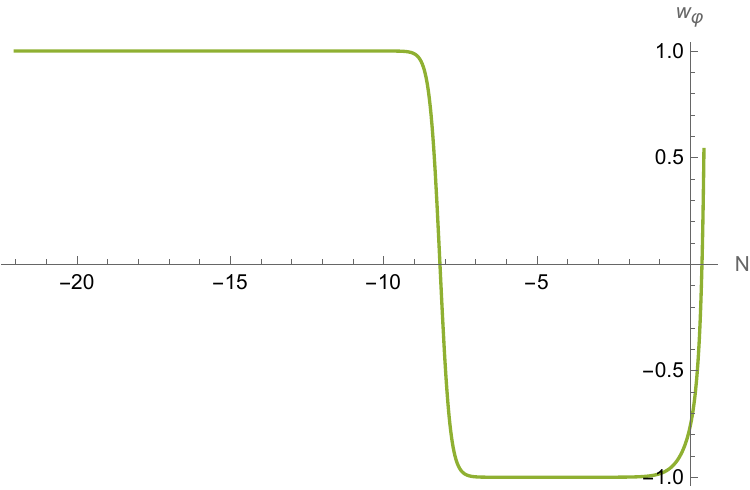}\caption{$w_{\varphi}(N)$}\label{fig:HilltopwN}
\end{subfigure}
\caption{Solution and relevant quantities evolving in terms of $N$, for the hilltop quintessence example with parameters \eqref{Hilltopparam} and $w_{\varphi0} = -0.76201230846$.} \label{fig:Hilltop}
\end{center}
\end{figure}

As explained at the beginning of Section \ref{sec:examples}, we are left to fix $w_{\varphi0} = -0.76201230846$. This value allows the radiation domination to start around $N=-20$. To get the solution evolving in terms of $N$, we follow Section \ref{sec:eqreform} and solve \eqref{eqF2N}, \eqref{eqEiN}, giving $\tilde{H}(N),\varphi(N)$. We do so with the following initial conditions, read from \eqref{kinViniN}
\beq
\tilde{H}(0)=1 \ ,\ \varphi(0) = \sqrt{\left(1 - \frac{3}{2} \Omega_{\varphi0} (1 - w_{\varphi0}) \frac{1}{\tilde{V}_0}\right) \frac{2}{\kappa^2} }\ ,\ \varphi'(0) = \sqrt{3\Omega_{\varphi0}(1 + w_{\varphi0})} \ .
\eeq
We display in Figure \ref{fig:Hilltop} the solution and further quantities. Qualitatively, the behaviour is the same as for exponential quintessence.

Finally, we can also obtain the solution in terms of time, as explained in Section \ref{sec:eqreform}. Features are similar to the above examples. We obtain that $a(t)=0$ for $H_0 (t_0-t) = 0.9313$, giving as for exponential quintessence an a priori shorter age of the universe compared to $\Lambda$CDM; once again, $H_0$ should however be reevaluated observationally for this model.

\subsection{Domination phases, equality times and maxima}\label{sec:phases}

In the three examples presented above, we see some patterns. To start with, they all exhibit successive domination phases, from the past until today (future is not discussed in this work): radiation, matter and dark energy domination phases. By domination, we refer to the fact that one component has the highest value of $\Omega_n$. This is a priori different than $\Omega_n > 0.5$, but in practice, since only two components are non-negligible simultaneously, the two criteria are almost equivalent.

\begin{figure}[H]
\begin{center}
\begin{subfigure}[H]{0.48\textwidth}
\includegraphics[width=\textwidth]{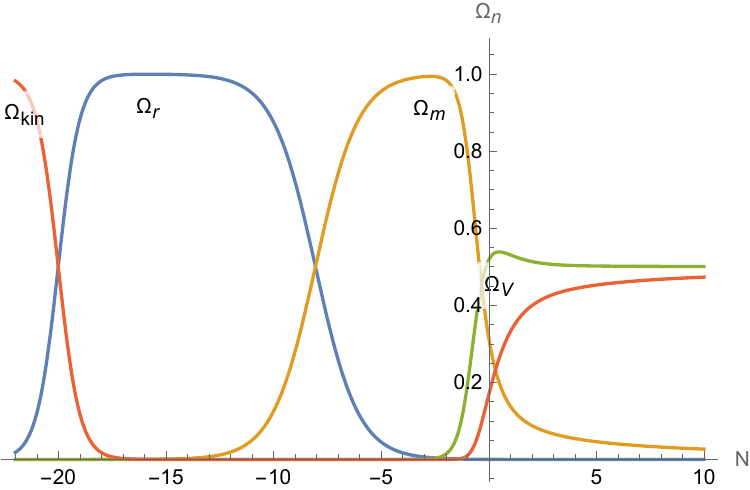}\caption{exponential}\label{fig:ExpQuintONkinV}
\end{subfigure}\quad
\begin{subfigure}[H]{0.48\textwidth}
\includegraphics[width=\textwidth]{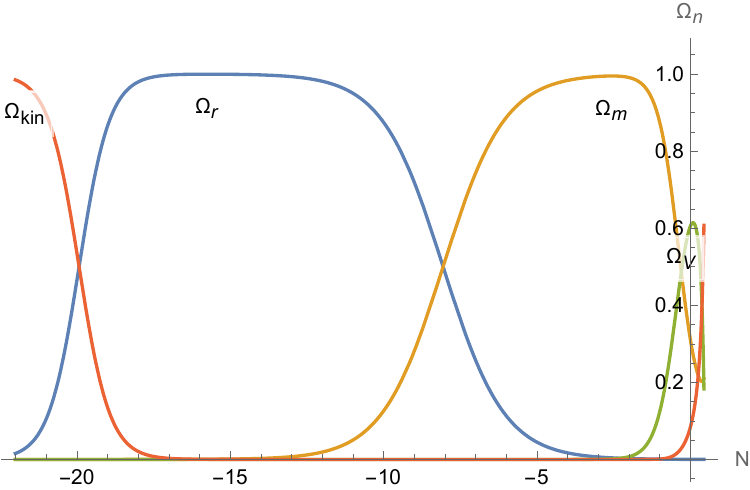}\caption{hilltop}\label{fig:HilltopONkinV}
\end{subfigure}
\caption{$\Omega_n(N)$ for our two quintessence examples, distinguishing $\Omega_{{\rm kin}}$ (red) and $\Omega_V$ (green) contributions.} \label{fig:ONkinV}
\end{center}
\end{figure}

The two quintessence examples are supplemented by an initial phase where $\Omega_{\varphi}$ dominates. While we could also call the latter dark energy domination, it turns out that the kinetic energy is by far the main contributor to this initial phase, while the potential is very negligible: we verify this explicitly in Figure \ref{fig:ONkinV}. We will study in Section \ref{sec:analytic} under which conditions this is true. For now, we make this approximation, which amounts to set $V=0$ in this first phase. In that case, one refers to the initial kination phase, and trades $\Omega_{\varphi}$ for $\Omega_{{\rm kin}}$. By definition, $\Lambda$CDM cannot exhibit such a phase; there one extends the radiation domination phase to $N= -\infty$.

With these definitions, a domination phase starts and end at equality times, namely when the two successively highest $\Omega_n$ are equal. In terms of e-folds, we use the following notations for these equality moments
\bea
\Omega_{{\rm kin}} = \Omega_r:&&\qquad N= N_{{\rm kin} r} \nn\\
\Omega_r = \Omega_m:&&\qquad N= N_{rm} \nn\\
\Omega_m = \Omega_{\varphi}:&&\qquad N= N_{m \varphi} \nn
\eea
where for $\Lambda$CDM, $N_{{\rm kin} r} = - \infty$. The successive domination phases are defined as
\bea
\text{kination}: &&\qquad -\infty < N < N_{{\rm kin} r} \nn\\
\text{radiation}: &&\qquad N_{{\rm kin} r} < N < N_{rm} \nn\\
\text{matter}: &&\qquad N_{rm} < N < N_{m \varphi} \nn\\
\text{dark energy}: &&\qquad N_{m \varphi} < N < 0 \nn
\eea
Let us now compute these equality times. We do so using the expressions \eqref{Omegas} for the $\Omega_n$, together with the information, justified in Section \ref{sec:kinradphase}, that $\rho_{{\rm kin}} = \rho_{{\rm kin}0}\, a^{-6}$ during the kination phase. It is straightforward to obtain the following expressions
\beq
\boxed{\hspace{1em} N_{{\rm kin} r} = \frac{1}{2} \ln \frac{\Omega_{{\rm kin}0}}{\Omega_{r0}} \ ,\quad N_{rm} = \ln \frac{\Omega_{r0}}{\Omega_{m0}} \hspace{1em}} \quad  N_{m \varphi} = \frac{1}{3} \ln \left( \frac{\Omega_{m0}}{\Omega_{\varphi0}} \frac{\rho_{\varphi0}}{\rho_{\varphi\, m \varphi}} \right) \ , \label{Nequality}
\eeq
where $\rho_{\varphi\, m \varphi}$ refers to the scalar field energy density at the last equality time. A priori, we do not know this value. Note though that for $\Lambda$CDM, $\rho_{\varphi\, m \varphi}=\rho_{\varphi0}$. So we introduce the notations
\beq
\boxed{\hspace{1em} N_{m \varphi \, \Lambda} = \frac{1}{3} \ln \frac{\Omega_{m0}}{\Omega_{\varphi0}} \ , \quad  N_{m \varphi \, q} = \frac{1}{3} \ln \left( \frac{\Omega_{m0}}{\Omega_{\varphi0}} \frac{\rho_{\varphi0}}{\rho_{\varphi\, m \varphi}} \right) \hspace{1em}}  \label{Nmp}
\eeq
giving the expression for a quintessence model
\beq
\frac{\rho_{\varphi\, m \varphi}}{\rho_{\varphi0}} = e^{3 ( N_{m \varphi \, \Lambda} - N_{m \varphi \, q}) } \ . \label{rhoratioequality}
\eeq
As discussed around \eqref{rhodot}, Hubble friction imposes a decreasing $\rho_{\varphi}$. We deduce that
\beq
N_{m \varphi \, q} < N_{m \varphi \, \Lambda}  \ ,
\eeq
which will be verified numerically in Table \ref{tab:Neq}.

Using the fiducial values \eqref{fidO}, we obtain from the above expressions
\beq
N_{rm} \approx -8.0548 \ ,\quad N_{m \varphi \, \Lambda} \approx -0.2591 \ .
\eeq
In addition, for the quintessence examples, we have tuned the initial conditions ($w_{\varphi0}$) to obtain $N_{{\rm kin} r} \approx -20$, as explained at the beginning of Section \ref{sec:examples}. In Table \ref{tab:Neq}, we list the values numerically obtained for the three examples.
\begin{table}[H]
\begin{center}
\begin{tabular}{|l||c|c|c|}
\hhline{~---}
\multicolumn{1}{c|}{} &&&\\[-8pt]
\multicolumn{1}{c|}{} & $\Lambda$CDM & Exp. quint. & Hill. quint. \\[4pt]
\hline
&&&\\[-8pt]
$N_{{\rm kin} r}$ & $-\infty$ & $-20.008$ & $-19.923$ \\[4pt]
\hline
&&&\\[-8pt]
$N_{rm}$ & $-8.0548$ & $-8.0548$ & $-8.0548$ \\[4pt]
\hline
&&&\\[-8pt]
$N_{m \varphi}$ & $-0.2591$ & $-0.4295$ & $-0.3091$ \\[4pt]
\hline
\end{tabular}
\end{center}\caption{Equality times separating the different domination phases for the solution considered in each model. The values rely on the fiducial values \eqref{fidO} for the $\Omega_{n0}$, and the tuned value of $w_{\varphi0}$ that fixes $N_{{\rm kin} r}$.}\label{tab:Neq}
\end{table}
The value of $\Omega_{{\rm kin}0}$ is not known. We read from above that $\Omega_{{\rm kin}0} = \Omega_{r0} \times e^{2\, N_{{\rm kin} r} }$. For $N_{{\rm kin} r} \approx -20$ and the fiducial value \eqref{fidO}, this gives $\Omega_{{\rm kin}0} \approx 10^{-22}$. This number is only meaningful in the past, even though normalised today. Indeed, the kinetic energy later evolves differently, eventually contributing non-trivially to $\Omega_{\varphi0}$.\\

\begin{figure}[H]
\centering
\includegraphics[width=0.6\textwidth]{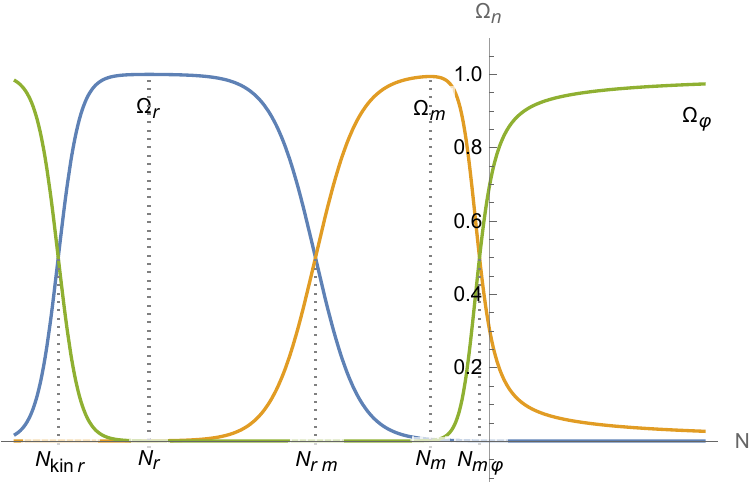}
\caption{Equality and maxima moments in a quintessence model, here for the exponential quintessence example.}\label{fig:ExpQuintONs}
\end{figure}

Other remarkable moments in the history of the universe are the maxima of the $\Omega_n$: the maximum of $\Omega_m$, present for all models, reached at $N_m$, and that of $\Omega_r$, for the quintessence models, reached at $N_r$. The fact a maximum is reached is related to the fact that other dominations appear before and after; as a consequence, the value of the maximum can be related to the equality times before and after, as we will see. We illustrate in Figure \ref{fig:ExpQuintONs} the various moments discussed in this section.
We define phases thanks to these maxima moments
\bea
\text{kination - radiation}: &&\qquad -\infty < N < N_r \nn\\
\text{radiation - matter}: &&\qquad N_{r} < N < N_{m} \label{phases2}\\
\text{matter - dark energy}: &&\qquad N_{m} < N < 0 \nn
\eea
and those will be the topic of the next section.

These maxima can be computed. Let us start with radiation. We use the expression \eqref{Omegas} for $\Omega_r$. We assume that before $N_r$, $\Omega_{\varphi}$ is only given by the kinetic energy, as in the kination phase. While this will be justified later analytically, we can already see in the above examples, in Figure \ref{fig:ExpQuintwN} and \ref{fig:HilltopwN}, that $w_{\varphi} \approx +1$ during radiation domination and much after the maximum of $\Omega_r$: this is consistent with the domination of kinetic energy over the potential. This gives the expression
\beq
\Omega_r = \frac{\Omega_{r0}\, a^{-4}}{\Omega_{{\rm kin}0}\, a^{-6} +\Omega_{r0}\, a^{-4} +\Omega_{m0}\, a^{-3} } \ .
\eeq
Extremizing it gives
\beq
\boxed{\hspace{1em} N_{r} = \frac{1}{3} \ln \frac{2 \Omega_{{\rm kin}0}}{\Omega_{m0}} \ ,\qquad \Omega_{r\, {\rm max}} = \left( 1+ \frac{3}{2} \frac{\Omega_{m0}}{\Omega_{r0}} \left( 2 \frac{\Omega_{{\rm kin}0}}{\Omega_{m0}} \right)^{\frac{1}{3}}  \right)^{-1} \hspace{1em}}
\eeq
Consistently, these expressions apply to $\Lambda$CDM with $\Omega_{{\rm kin}0}=0$, $N_r = - \infty$, $\Omega_{r\, {\rm max}} =1$. As mentioned, we can rewrite the above in terms of the equality times as
\beq
3\, N_r = \ln 2 + 2 N_{{\rm kin}r} + N_{rm} \ ,\qquad \Omega_{r\, {\rm max}}^{-1} - 1 =  3 \times 2^{-\frac{2}{3}}\ e^{- \frac{2}{3} |N_{{\rm kin} r} - N_{rm} |} \ .
\eeq
We see that the longer radiation domination lasts, the closer to 1 $\Omega_{r\, {\rm max}}$  is. Also, with the rewriting $3\, |N_r-N_{rm}| = -\ln 2 + 2 |N_{{\rm kin}r} - N_{rm}|$, we see that the longer radiation domination lasts, the earlier the maximum is. Finally, we can look at the other $\Omega_n$ at $N_r$: using \eqref{Omegas}, we derive the relation
\beq
r\ {\rm max}:\quad \frac{\Omega_m}{\Omega_{{\rm kin}}} = 2 \ .\label{OmOp2}
\eeq
We will see in Table \ref{tab:max} an excellent numerical agreement.

We turn to matter. The expression \eqref{Omegas} for $\Omega_m$ is
\beq
\Omega_m = \frac{\Omega_{m0}\, a^{-3}}{\Omega_{r0}\, a^{-4} + \Omega_{m0}\, a^{-3} + \Omega_{\varphi0}\, \frac{\rho_{\varphi}}{\rho_{\varphi0}}} \ .\label{Omgen}
\eeq
We can extremize it, but this requires the variation of $\rho_{\varphi}$, and is thus more involved than the radiation case. Getting to the quantity $a \del_a \rho_{\varphi} = H^{-1} \dot{\rho}_{\varphi} $, we can use \eqref{rhodot}. We obtain the following expressions
\bea
N_{m\, q} &=& \frac{1}{4} \ln \left( \frac{\Omega_{r0}}{3 \Omega_{\varphi0}} \times \frac{\rho_{\varphi0}}{\rho_{\varphi\, m}(-w_{\varphi\, m})} \right) \ ,\\
\Omega_{m\, {\rm max}\, q} &=& \left( 1 + \frac{\Omega_{\varphi0}}{\Omega_{m0}} \left(1-3 w_{\varphi\, m}\right) \frac{\rho_{\varphi\, m}}{\rho_{\varphi0}} \left( \frac{\Omega_{r0}}{3 \Omega_{\varphi0}} \times \frac{\rho_{\varphi0}}{\rho_{\varphi\, m}(-w_{\varphi\, m})}  \right)^{\frac{3}{4}}
 \right)^{-1} \ ,
\eea
where $\rho_{\varphi\, m}, w_{\varphi\, m}$ denote the values of these quantities at this point, and we take $w_{\varphi\, m}<0$. The above expressions simplify for $\Lambda$CDM, where the variation of $\rho_{\varphi\, m}$ can be avoided. There, one has $ \rho_{\varphi\, m} = \rho_{\varphi0}$, $w_{\varphi\, m}=-1$, giving
\beq
\boxed{\hspace{1em} N_{m\, \Lambda} = \frac{1}{4} \ln \frac{\Omega_{r0}}{3 \Omega_{\varphi0}} \ , \qquad \Omega_{m\, {\rm max}\, \Lambda} = \left( 1 + 4\frac{\Omega_{\varphi0}}{\Omega_{m0}} \left( \frac{\Omega_{r0}}{3 \Omega_{\varphi0}} \right)^{\frac{3}{4}}
 \right)^{-1} \hspace{1em}}
\eeq
For $\Lambda$CDM, we can rewrite the above expressions in terms of equality times
\beq
4 N_{m\, \Lambda} = -\ln 3 + N_{rm} + 3 N_{m\varphi\, \Lambda} \ , \qquad \Omega_{m\, {\rm max}\, \Lambda}^{-1} -1 = 4 \times 3^{-\frac{3}{4}}\ e^{-\frac{3}{4} |N_{rm}-N_{m\varphi\, \Lambda}| } \ .
\eeq
We see again that the longer the matter domination phase, the closer to 1 $\Omega_{m\, {\rm max}}$ is. And we get $4 |N_{m\, \Lambda} - N_{m\varphi\, \Lambda}| = \ln 3 + |N_{rm} - N_{m\varphi\, \Lambda}|$, showing again that the longer the matter domination phase is, the earlier the maximum is.

As for the equality times, the above notations allow us to get the expression
\beq
\boxed{\hspace{1em} \frac{\rho_{\varphi0}}{\rho_{\varphi\, m}(-w_{\varphi\, m})} = e^{4( N_{m\, q} - N_{m\, \Lambda} )} \hspace{1em}} \label{rhoratio}
\eeq
giving an interesting estimate of the variation of the ratio of $\rho_{\varphi}$. We recall from \eqref{rhodot} that Hubble friction forces $\rho_{\varphi}$ to decrease. Taking in addition that $w_{\varphi\, m} \approx -1$ as in our examples, we deduce that
\beq
N_{m\, q}  < N_{m\, \Lambda} \ ,
\eeq
a point that will be well verified numerically in Table \ref{tab:max}. Using the expression \eqref{rhoratio} leads to the following quintessence expression
\beq
\Omega_{m\, {\rm max}\, q} = \left( 1 + \frac{1+3(-w_{\varphi\, m})}{(-w_{\varphi\, m})} \frac{\Omega_{\varphi0}}{\Omega_{m0}} e^{3 N_{m\, q}}  e^{4( N_{m\, \Lambda} - N_{m\, q} )}  \right)^{-1} \ .
\eeq
Finally, we look at the other $\Omega_n$ at $N_{m\, q}$. Using \eqref{Omegas} and \eqref{rhoratio}, it is straightforward to derive the relation
\beq
m\ {\rm max}:\quad \frac{\Omega_r}{\Omega_{\varphi}} = 3 (-w_{\varphi\, m}) \approx 3 \ ,\label{OrOp3}
\eeq
which will also get an excellent numerical agreement.

Using the fiducial values \eqref{Omegas} for the $\Omega_n$, we can either compute or get numerically the various quantities at the maxima. We give them in Table \ref{tab:max}.
\begin{table}[H]
\begin{center}
\begin{tabular}{|l|l||c|c|c|}
\hhline{~~---}
\multicolumn{2}{c|}{} &&&\\[-8pt]
\multicolumn{2}{c|}{} & $\Lambda$CDM & Exp. quint. & Hill. quint. \\[4pt]
\hline
&&&&\\[-8pt]
 & $N_r$ & $-\infty$ & $-15.793$ & $-15.737$ \\[4pt]
\hhline{~----}
&&&&\\[-8pt]
$r\, {\rm max}$ & $\Omega_{r}$ & $1$ & $0.9993$ & $0.9993$ \\[4pt]
\hhline{~----}
&&&&\\[-8pt]
 & $\Omega_{m}$ & $0$ & $0.00044$ & $0.00046$ \\[4pt]
\hhline{~----}
&&&&\\[-8pt]
 & $\Omega_{{\rm kin}}$ & $0$ & $0.00022$ & $0.00023$ \\[4pt]
\hline
&&&&\\[-8pt]
 & $N_m$ & $-2.483$ & $-2.719$ & $-2.549$ \\[4pt]
\hhline{~----}
&&&&\\[-8pt]
$m\, {\rm max}$ & $\Omega_{r}$ & $0.0038$ & $0.0048$ & $0.0040$ \\[4pt]
\hhline{~----}
&&&&\\[-8pt]
 & $\Omega_{m}$ & $0.9950$ & $0.9936$ & $0.9946$ \\[4pt]
\hhline{~----}
&&&&\\[-8pt]
 & $\Omega_{\varphi}$ & $0.0013$ & $0.0016$ & $0.0013$ \\[4pt]
\hline
\end{tabular}
\end{center}\caption{Relevant quantities at radiation and matter maxima, for each example.}\label{tab:max}
\end{table}

Having defined the various phases according to equality or maxima moments, and having evaluated those, we are now ready to study each phase separately.

\section{Analytic solutions and properties}\label{sec:analytic}

Having presented in Section \ref{sec:formalism} the equations to solve, and in Section \ref{sec:examples} several examples of numerical solutions and their features, we tackle in this section the question of getting analytic solutions and proving some of their properties. We will address this question by focusing separately on the different phases of the universe history, listed in \eqref{phases2}.

\subsection{Single component solutions}\label{sec:single}

To start with, several well-known analytic solutions (to $f_1=f_2=e^i=0$) are obtained by considering a single fluid component in the universe. We will see that they correspond to limits of the complete solutions within the corresponding component domination phase. In the dynamical system study of the exponential quintessence \cite{Andriot:2024jsh}, these single component solutions correspond to fixed points; by extension here, we denote them as $P_n$ for the $n$-component. The solutions are the following
\bea
P_{{\rm kin}}^{\pm ... \pm}:\quad && a(t)= \left(3\, \rho_{{\rm kin}0} \right)^{\frac{1}{6}} \,  t^{\frac{1}{3}} \ , \quad \varphi^i(t) = \varphi^i_{0{\rm kin}} \pm \sqrt{\frac{2}{3}} c_k^i \, \ln t \ ,\quad \rho_{r,m0}=k=V =0 \ ,\ g_{ij}=\delta_{ij} \ ,\nn\\
&& {\rm with} \ \sum_i (c_k^i)^2 = 1 \ , \ c_k^i\geq0 \ ,\nn\\
P_r:\quad && a(t)= \left(\frac{4}{3}\, \rho_{r0} \right)^{\frac{1}{4}} \, t^{\frac{1}{2}} \ , \quad \varphi^i(t) = \varphi^i_{0r} \ ,\quad \rho_{m0}=k=V =0 \ ,\nn\\
P_m:\quad && a(t)= \left(\frac{3}{4}\, \rho_{m0} \right)^{\frac{1}{3}} \, t^{\frac{2}{3}} \ , \quad \varphi^i(t) = \varphi^i_{0m} \ ,\quad \rho_{r0}=k=V =0 \ ,\label{Pnsol}\\
P_k:\quad && a(t)=  t \ , \quad \varphi^i(t) = \varphi^i_{0k} \ ,\quad \rho_{r0}=\rho_{m0}=V =0  \ ,\ k=-1\ ,\nn\\
P_{\Lambda}:\quad && a(t) = a_{\Lambda}\, e^{ \sqrt{\frac{\rho_{\Lambda}}{3}}\ t} \ , \quad \varphi^i(t) = \varphi^i_{0\Lambda} \ ,\quad \rho_{r0}=\rho_{m0}=k=0 \ , \ \rho_{\Lambda} = V = \Lambda = 3 H_0^2 \ ,\nn
\eea
where for each solution $P_n$ one has $\Omega_n=1$. In each solution, $\varphi^i_{0n}$ is a free constant. $\rho_{{\rm kin}0}, \rho_{\Lambda}, a_{\Lambda}$ are positive free constants (as are $\rho_{r0}, \rho_{m0}$); $c_k^i$ are also free up to their orthonormalisation condition. In $P_{{\rm kin}}^{\pm ... \pm}$, the $\pm$ in position $i$ denotes the $\pm$ sign of field $\varphi^i$. In all solutions, the time $t$ can be shifted by a constant; we ignore the latter for simplicity, and take $t>0$.

The first four solutions scale factors can be repackaged as
\beq
a(t)=\left(\frac{3(1+w_n)^2}{4}\, \rho_{n0} \right)^{\frac{1}{3(1+w_n)}} \, t^{\frac{2}{3(1+w_n)}} \ ,
\eeq
which is the general solution to $f_2=0$ (with single component) for a constant $w_n > -1$, again up to a shift in time. The normalisation is expressed in terms of the free constant $\rho_{n0}$, such that
\beq
\rho_n = \rho_{n0}\ a^{-3(1+w_n)} \ ,
\eeq
which is the solution to $\Omega_n=1 \Leftrightarrow f_1=0$.

$P_{\Lambda}$ is different since one has $w_{\Lambda}=-1$. It is the de Sitter solution (in flat space). To obtain it, one restricts themselves to $H>0$, which we do here with an expanding universe; one then gets for that solution a constant $H =H_0$. We also recall that we take $M_p=1$.\\

In realistic cosmologies, $P_r$ and $P_m$ should be good approximations of complete solutions around their respective domination: we see for instance in Table \ref{tab:max} that $\Omega_n \approx 1$ and $\Omega_{n'\neq n} \ll 1$ close to these maxima.\footnote{Since complete quintessence solutions evolve away from these dominations before and after the maxima, we deduce that as fixed points, $P_r$ and $P_m$ cannot be stable but only saddle points, and cosmological solutions pass only close by. The latter is consistent with the dynamical system analysis of \cite{Andriot:2024jsh}.} Similarly, $P_{{\rm kin}}^{\pm ... \pm}$ should be the approximation of the complete quintessence solutions at early times, when kinetic energy dominates ($\Omega_{{\rm kin}} \approx 1$), and $P_{\Lambda}$ should play the same role for the asymptotic future in $\Lambda$CDM. We will verify these claims in the following when getting more complete analytic solutions.

\subsection{Kination - radiation phase}\label{sec:kinradphase}

\subsubsection{Analytic solution $(a(t),\varphi^i(t))$}\label{sec:kinradanalytic}

We are interested in describing analytically this initial phase that lasts between $N =-\infty$ and $N_r$, as defined in \eqref{phases2}. As argued above, $P_{{\rm kin}}^{\pm ... \pm}$ and $P_r$ are expected to provide asymptotic expressions of the complete solution at small and large times respectively. Note that for each of them separately, one has $\rho_{{\rm kin}} \propto a^{-6}$ and $\rho_r \propto a^{-4}$, so in an expanding universe, kination is expected to dominate before radiation. We are thus interested in finding the interpolating solution between $P_{{\rm kin}}^{\pm ... \pm}$ and $P_r$. To that end, we neglect the other components, which amounts to set $V=\rho_{m0}=k=0$. The scaling in $a$ of $\rho_m, \rho_k$ is such that it is legitimate to neglect them during this phase; we will discuss in the next subsubsection to what extent neglecting $V$ is justified, but we can already verify the validity of this approximation for all three components in Figure \ref{fig:ONkinV} for our two quintessence examples. Last but not least, for most of the resolution, we do not need to restrict ourselves to canonically normalized fields, i.e.~we can leave $g_{ij}$ free; only when needing an expression for $\varphi^i$, we will have to specify the field space metric.\\

To find the solution, it is sufficient to solve $E^i=0$ and $F_1=0$ thanks to the relation \eqref{relEq}. We start with the former: with $V=0$ and \eqref{rhodotE}, $E^i=0$ implies
\beq
\dot{\rho}_{{\rm kin}} = - 3 H \dot{\varphi}^i g_{ij}  \dot{\varphi}^j = -6 H \rho_{{\rm kin}} \ \Rightarrow \ \boxed{\hspace{1em} \rho_{{\rm kin}} = \rho_{{\rm kin}0}\ a^{-6} \hspace{1em}} \label{kincont}
\eeq
We integrated and solved that equation, with the integration constant $\rho_{{\rm kin}0}$, matching to the standard notation for $a=1$. This solution for $\rho_{{\rm kin}}$ is sufficient to solve $F_1=0$, as we will do in the following.

However, we have not yet solved $E^i=0$. To that end, we would need the explicit $g_{ij}$. Let us only give here the solution for canonically normalised fields, $g_{ij}=\delta_{ij}$. In that case, $E^i=0$ with $V=0$ can be integrated into
\beq
\dot{\varphi}^i =\pm \sqrt{2 \, \rho_{{\rm kin}0}}\, c_k^i \ a^{-3} \ ,\quad c_k^i \geq 0 \ ,
\eeq
where the sign and normalisation chosen for the integration constant $c_k^i$ are introduced for convenience. Indeed, normalising further the $c_k^i$ as follows (which can be done without loss of generality), the constant $\rho_{{\rm kin}0}$ just introduced matches the above one, and we get the same expression for $\rho_{{\rm kin}}$
\beq
\sum_i (c_k^i)^2 = 1 \ \Rightarrow \ \rho_{{\rm kin}} = \rho_{{\rm kin}0}\ a^{-6} \ .
\eeq
A complete expression for $\varphi^i(t)$ requires to find $a(t)$: we turn to this task by tackling $F_1=0$.

That equation can now be brought to the form
\beq
\frac{a^2 \d a}{\sqrt{\frac{1}{3}\left( \rho_{{\rm kin}0} + \rho_{r0} \, a^2 \right)}} = \d t \ \Leftrightarrow \ \frac{x^2\, \d x}{\sqrt{1+x^2}} = \frac{\rho_{r0}^{\frac{3}{2}}}{\sqrt{3}\, \rho_{{\rm kin}0}} \, \d t \quad {\rm with} \ x= \sqrt{\frac{\rho_{r0}}{\rho_{{\rm kin}0}}} \, a \ .
\eeq
With an integration by parts and using
\beq
\int \sqrt{1+x^2} \, \d x = \frac{1}{2}\, x \sqrt{1+x^2} + \frac{1}{2}\, \ln \left(x+ \sqrt{1+x^2} \right) \ ,
\eeq
up to a constant, we eventually obtain
\beq
\boxed{\hspace{1em} \frac{\rho_{r0}^{\frac{3}{2}}}{\sqrt{3}\, \rho_{{\rm kin}0}} \, ( t -t_*) = \frac{1}{2}\, x \sqrt{1+x^2} - \frac{1}{2}\, \ln \left(x+ \sqrt{1+x^2} \right) \hspace{1em}}
\eeq
where $t_*$ captures possible integration constants. Setting $a(t=0)=0$ fixes $t_*=0$.\\

We can verify the asymptotics of this solution. For large $a$, one gets $\sqrt{\frac{\rho_{r0}}{3}} \, t = \frac{1}{2} a^2$, from which we recover $P_r$. In addition, for $g_{ij}=\delta_{ij}$, using $\dot{\varphi}^i =\pm \sqrt{2 \, \rho_{{\rm kin}0}}\, c_k^i \, a^{-3} \sim t^{-3/2}$, we obtain $\varphi^i = {\rm constant} + {\cal O}(t^{-1/2})$: in this limit of large $a$ and large $t$, we recover that $\varphi^i$ is constant as in $P_r$. For small $a$, developing to third order in $x$, we get $\frac{1}{3} x^3$, meaning $\sqrt{3\, \rho_{{\rm kin}0}} \, t = a^3 $ corresponding to the $P_{{\rm kin}}^{\pm ... \pm}$ solution. We also get $\dot{\varphi}^i =\pm \sqrt{2 \, \rho_{{\rm kin}0}}\, c_k^i \, a^{-3} = \pm \sqrt{\frac{2}{3}}\, c_k^i \, t^{-1}$, which integrates to the solution of $P_{{\rm kin}}^{\pm ... \pm}$ for $\varphi^i(t)$. Our solution therefore interpolates, as expected, between $P_{{\rm kin}}^{\pm ... \pm}$ and $P_r$.

One may wonder whether the above solution can be inverted into an expression for $a(t)$, beyond the asymptotics. This is in general not obvious. However, note that for $x>0$, one has
\beq
x \sqrt{1+x^2} = (x-1) \sqrt{1+x^2} + \sqrt{1+x^2} >  x-1 + \sqrt{1+x^2} > \ln \left(x+ \sqrt{1+x^2} \right) \ ,
\eeq
in agreement with $t-t_*>0$. As soon as the $\ln$ term becomes negligible, for large $x$ or $a$, we can invert the expression and obtain
\beq
{\rm Large} \ t:\quad a(t)= \left(\sqrt{\frac{\rho_{{\rm kin}0}^2}{4\rho_{r0}^2} + \frac{4}{3} \rho_{r0}\, t^2} - \frac{\rho_{{\rm kin}0}}{2\rho_{r0}} \right)^{\frac{1}{2}} \ .
\eeq
From this, one recovers $P_r$ in the large $t$ limit, but one also gets the solution before it.

\subsubsection{Is kination necessarily happening?}\label{sec:kincond}

One may ask whether an initial kination phase necessarily happens in the cosmological solution, as observed in our quintessence examples (see Figure \ref{fig:ONkinV}). The answer is affirmative, provided conditions on $V$ (that make it negligible) are satisfied. Indeed, we have seen above that if $V$ is negligible, then $E^i=0$ automatically leads to the solution $\rho_{{\rm kin}} = \rho_{{\rm kin}0}\, a^{-6}$. Let us emphasize that $V$ being negligible compared to the kinetic energy is sufficient for this, as shown with \eqref{rhodotE} and \eqref{kincont}: this solution is then obtained from the kinetic energy continuity equation. Strictly speaking, that solution being non-trivial also requires $\rho_{{\rm kin}0} \neq 0$; in other words, we also need initial conditions allowing for non-zero field speed at some early time, even if that speed is very small. Then, the equations impose that $\rho_{{\rm kin}}$ grows towards the past, and at some point dominates over radiation, matter and curvature: {\sl kination necessarily happens in the early universe, given some early non-zero (even small) field speed, and conditions on $V$, making it negligible, are obeyed.}

Let us clarify that it is reasonable to expect a different early universe, with e.g.~inflation and reheating. It is thus understood that solutions considered here should be cut at some point in the past to be patched to such possibly more realistic early scenario. If the cut is done during radiation domination, most of what we discuss here is meaningless. For the sake of the discussion, we assume here that the cut is done at an early enough time, allowing to have some initial kination phase.

The analytic solution found during kination - radiation domination required to neglect $V, \rho_m,k$, compared to $\rho_{{\rm kin}}, \rho_r$. The two densities $\rho_m, \rho_k$ are naturally neglected given their scaling in $a$, that gets subdominant during an early kination - radiation phase. Having $V$ negligible during this phase is the condition that we will discuss below. In \cite[Fig. 7]{Andriot:2024jsh}, one can find examples where even in absence of a radiation domination phase (due to different initial conditions), kination can still take place. The argument given at the beginning of this subsubsection holds indeed independently of radiation.\\

In the following, we restrict to a single, canonically normalised field, even though we believe that our results could, to some extent, apply to a multifield case. We also assume that there is some non-zero kinetic energy at some early time, as discussed above; we then take that $\dot{\varphi}(t)>0$, and will discuss the symmetric (negative) case below. We then prove the following lemma
\beq
\text{{\bf Lemma.}} \quad \quad \boxed{\hspace{1em} \text{{\it Early kination phase}}\quad \Rightarrow \quad V(\varphi) \ll e^{-\sqrt{6}\, \varphi} \ \text{{\it at early times}} \hspace{1em}} \label{lemmakin}
\eeq
where the early times will correspond here to a large, negative $\varphi(t)$, since the early times speed is positive. We believe the converse statement is true but do not prove it here. As side remarks, the early times refer here to the initial moments (where in the limit, quantities become infinite), so the lemma does not apply to a transient, finite duration kination phase. The idea here is simply that by definition, kinetic energy dominates the potential in these early moments, and we give the potential bound beyond which this cannot happen anymore. We also recall that such an early time kination is happening in our realistic quintessence solutions (see Figure \ref{fig:ONkinV}).

\begin{proof}
If we have a kination phase, then the potential can be neglected against kinetic energy, and as argued above, with a non-zero kinetic energy at some early time, one gets from the equations of motion (with $E^i=0$ and \eqref{kincont}) that the kinetic energy is $\rho_{{\rm kin}} = \rho_{{\rm kin}0}\, a^{-6}$. In addition, if this happens at sufficiently early times, then the kinetic energy dominates over radiation, matter or curvature, so that the solution to the full system is given by $P_{{\rm kin}}^{\pm}$. For simplicity of the discussion, we selected $\dot{\varphi}(t)>0$, i.e.~$P_{{\rm kin}}^{+}$. We read from this solution that
\beq
\rho_{{\rm kin}} = \frac{1}{3}\, t^{-2} \ ,\ \varphi(t) \sim +\sqrt{\frac{2}{3}} \, \ln t \sim - \sqrt{\frac{1}{6}} \, \ln \rho_{{\rm kin}} \ ,
\eeq
where in the very early times limit, $\varphi(t) \sim -\infty$. Note that an additional constant in the field can then be neglected. In addition, the potential being negligible compared to kinetic energy at these moments gets rewritten as
\beq
\rho_{{\rm kin}} \gg V(\varphi) \ \Leftrightarrow \ e^{-\sqrt{6} \times \left(-\sqrt{\frac{1}{6}} \, \ln \rho_{{\rm kin}}\right)} \gg V(- \sqrt{\frac{1}{6}} \, \ln \rho_{{\rm kin}})  \ \Leftrightarrow \ e^{-\sqrt{6}\, \varphi} \gg V(\varphi) \ .
\eeq
Note that any finite scale $V_0$ in the scalar potential can be absorbed by shifting the scalar field in the left-hand side by a constant proportional to $\ln V_0$. As mentioned, such an additional constant to the field can be neglected in these early times; one is then left with the bare exponential. This concludes the proof on this bound on the potential at early times.
\end{proof}

Physically, this statement can be qualitatively understood. When going towards the past, kination indicates that the kinetic energy grows. If the potential also grows towards the past, then there is a risk that it becomes non-negligible. There must therefore be a potential bound, as a function, beyond which the growth in the past is too strong: it turns out to be $e^{-\sqrt{6}\, \varphi}$.

In other words, consider $\dot{\varphi}>0$ at early times with initially $\varphi \sim -\infty$, as in $P_{{\rm kin}}^+$. Take a decreasing potential and the field rolling down the potential. For example, let us take $V = e^{- \lambda \varphi}$. Note that a normalisation constant $V_0>0$ does not play a role here, as it can be reabsorbed as $e^{\ln V_0}$ and then neglected against $\varphi \sim - \infty$. The above result (via the contrapositive implication) is then: if $\lambda >\sqrt{6}$, then the potential is too steep, or backwards, grows too fast in the past, and then dominates the kinetic energy. On the contrary, for  $\lambda <\sqrt{6}$, or any potential subleading it, the converse implication, if true, would mean that neglecting the potential is a valid approximation and kination happens at early times. This point, regarding exponential potentials, was already noticed from the stability study of $P_{{\rm kin}}^+$ in \cite{Andriot:2024jsh}: it was found that the $P_{{\rm kin}}^+$ solution could not be a starting point to a cosmological solution when $\lambda \geq \sqrt{6}$, because it then became a saddle. The result here is stronger, as we do not restrict to exponential potentials.

Considering instead $P_{{\rm kin}}^-$, i.e. $\dot{\varphi}<0$ and $\varphi \sim \infty$, one easily obtains the boundary potential to be $e^{+ \sqrt{6} \varphi}$. In that case, any potential $e^{- \lambda \varphi}$ ($\lambda>0$) is subleading to that limiting potential, and should thus allow for kination: this makes sense as  $e^{- \lambda \varphi}$ is very small in this asymptotics and thus naturally negligible. This explains why $P_{{\rm kin}}^-$ is a valid cosmological starting point (fully unstable fixed point) for any $\lambda>0$, providing an initial kination phase \cite{Andriot:2024jsh}.

A consequence of the (to be proven) converse implication is that any bounded potential (e.g.~a $\cos \varphi$) or even a power law, $V \sim \varphi^n$, allows for initial kination; in particular, this happens in our above examples of exponential or hilltop potentials. On the contrary, a consequence of the lemma (via the contrapositive implication) is that very steep potentials such as exponentials of exponentials, appearing e.g.~in certain regimes of LVS \cite{Balasubramanian:2005zx} or KKLT \cite{Kachru:2003aw} scenarios of string compactifications, cannot provide an initial kination. We also note that $e^{- \lambda \varphi}$ with $\lambda \geq  \sqrt{6}$ does not allow for a cosmological solution with past radiation domination together with today's acceleration \cite{Andriot:2024jsh}. The latter being necessary to a realistic cosmology, an initial kination therefore appears likely in any realistic quintessence model and associated cosmological solutions.\\

So far, we have provided a necessary condition for $V$ to be negligible against kinetic energy at early times, allowing for an early kination phase. However, the evolution $\rho_{{\rm kin}} = \rho_{{\rm kin}0}\, a^{-6}$ is valid later, even during radiation domination, as long as $V$ remains negligible: this is a consequence of \eqref{kincont}. At later times, we have shown that the solution for $(a(t), \varphi(t))$ evolves and interpolates between $P_{{\rm kin}}^{\pm}$ and $P_r$. We can then try to repeat the above proof at a given $t$, not especially small, to get a bound on the potential that ensures it being negligible.

The solution obtained previously admits an $a(t)$ that interpolates between $t^{\frac{1}{3}}$ and $t^{\frac{1}{2}}$. Let us then consider $a \sim t^{n_a}$ and neglect the time variation of $n_a$, i.e.~we take a solution locally in time; we will see in further subsections a way to test this scaling. We also consider $n_a > \frac{1}{3}$, to ensure being away from the early times and the pure $P_{{\rm kin}}^{\pm}$ solution.

With a single, canonical field, we obtain from $\rho_{{\rm kin}}$ that $\dot{\varphi} \sim t^{-3 n_a}$, and ignoring coefficients, $\varphi \sim \varphi_c + t^{-3 n_a +1} \sim \varphi_c + a^{-3 +\frac{1}{n_a}} \sim \varphi_c + \rho_{{\rm kin}}^{\frac{1}{2}-\frac{1}{6n_a}}$ with a constant $\varphi_c$. The limiting case is then, again up to coefficients,
\beq
\rho_{{\rm kin}} = V(\varphi_c + \rho_{{\rm kin}}^{\frac{1}{2}-\frac{1}{6n_a}}) \Rightarrow V(\varphi) = \left(\varphi - \varphi_c \right)^{\frac{6 n_a}{3n_a - 1}} \ .
\eeq
If  $V(\varphi) \geq \left(\varphi - \varphi_c \right)^{\frac{6 n_a}{3n_a - 1}} \geq \left(\varphi - \varphi_c \right)^6$  (with $\frac{1}{3} < n_a \leq \frac{1}{2}$), then kinetic energy is not dominant. On the contrary, if $V(\varphi) < \left(\varphi - \varphi_c \right)^6$, then it can be neglected and the kinetic energy is the main scalar field contribution, consistently with a kination solution.

We can have a similar reasoning to compare and neglect $V$ against $\rho_r$, for e.g.~the kination - radiation phase. The limiting case corresponds to
\beq
\rho_{r0}\, a^{-4} = V(\varphi_c + a^{-3 +\frac{1}{n_a}}) \Rightarrow V(\varphi) \sim \left( \varphi - \varphi_c \right)^{\frac{4 n_a}{3n_a - 1}} \ .
\eeq
We get that $ V(\varphi) \geq \left( \varphi - \varphi_c \right)^{\frac{4 n_a}{3n_a - 1}} \geq \left( \varphi - \varphi_c \right)^4$ does not allow for radiation domination. Considering on the contrary $ V(\varphi) < \left( \varphi - \varphi_c \right)^4$ is consistent with a dominant $\rho_r$.

These results are interesting but a difference with the above proof is that the field value need not be especially large. It may even be dominated by its constant, finite value (as e.g.~in $P_r$). It is therefore harder to say that one potential is subdominant to another one for such intermediate values of $\varphi$: for example, how to then compare our exponential to $\varphi^4$?\\

Coming back to the beginning of this subsubsection, having a negligible potential with some non-zero kinetic energy remains the prime criterion for the kination evolution $\rho_{{\rm kin}} = \rho_{{\rm kin}0}\, a^{-6}$; we have provided with the lemma \eqref{lemmakin} a necessary condition for this. We will make further use of this criterion in the next subsection.

\subsection{Radiation - matter phase}

\subsubsection{Analytic solution $a(t)$}\label{sec:radmatat}

We turn to the radiation - matter phase that lasts between $N_r$ and $N_m$, as defined in \eqref{phases2}. The phase starts with radiation domination and finishes with matter domination, in agreement with the scaling in $a$ of $\rho_r$ and $\rho_m$. We thus expect $P_r$ and $P_m$ to provide asymptotic expressions of a complete analytic solution for this whole phase, at small and large times respectively. We are interested in finding such a complete solution, interpolating between these two solutions.

\begin{figure}[H]
\begin{center}
\begin{subfigure}[H]{0.48\textwidth}
\includegraphics[width=\textwidth]{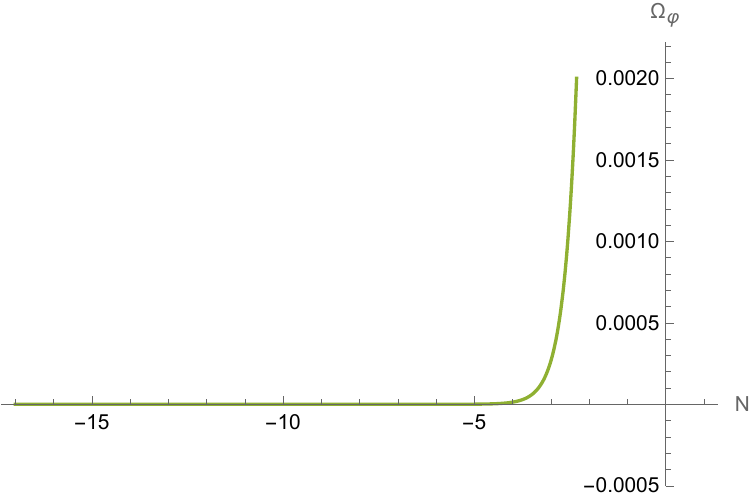}\caption{$\Lambda$CDM}\label{fig:LCDMOp}
\end{subfigure}\quad
\begin{subfigure}[H]{0.48\textwidth}
\includegraphics[width=\textwidth]{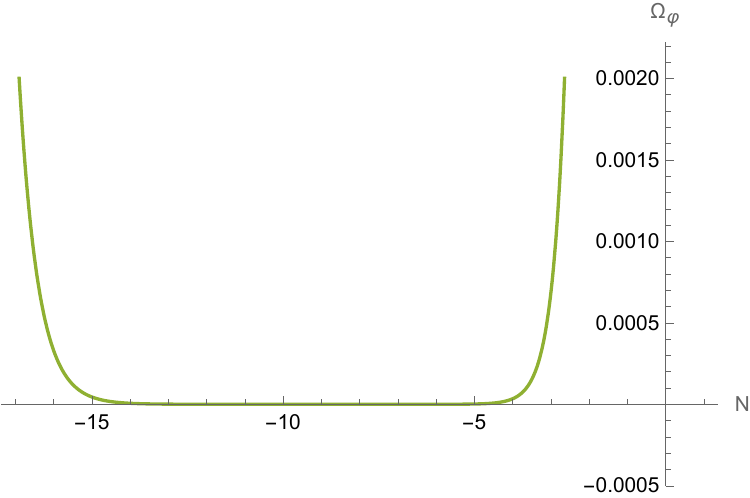}\caption{Quintessence}\label{fig:ExpQuintOp}
\end{subfigure}
\caption{$\Omega_{\varphi}(N)$ during the radiation - matter phase, and beyond. Figure \ref{fig:ExpQuintOp} is obtained from the exponential quintessence example, but a very similar curve can be produced with the hilltop solution.} \label{fig:OpN}
\end{center}
\end{figure}

To that end, we neglect $\Omega_k$ (the scaling in $a$ makes it subdominant) and $\Omega_{\varphi}$: this amounts to set $k=V=0$ and $\varphi^i=\varphi^i_{0rm}$ constant. While these conditions are in agreement with the would-be asymptotic solutions $P_r$ and $P_m$, one may wonder about their legitimacy. One can verify explicitly in our examples, in Figure \ref{fig:LCDMON}, \ref{fig:ExpQuintON} and \ref{fig:HilltopON} or \ref{fig:ONkinV}, that $\Omega_{\varphi}$ appears negligible during the radiation - matter phase. It can also be seen in Table \ref{tab:max} that $\Omega_{\varphi}$ is the smallest of the $\Omega_n$ at $N_r$ and $N_m$: we actually prove it in \eqref{OmOp2} and \eqref{OrOp3}. We also verify in Figure \ref{fig:OpN} that it only gets smaller between $N_r$ and $N_m$, reaching a minimum during this phase. So neglecting $\Omega_{\varphi}$ seems justified, at least to reproduce our example solutions. We will come back to it in more detail in the next subsubsection.

With $V=0$ and $\varphi^i=\varphi^i_{0rm}$, equation $E^i=0$ is satisfied. Thanks to the relation \eqref{relEq}, we are only left to solve $F_1=0$ to find the complete solution. We rewrite this equation, with radiation and matter contributions only, as follows
\beq
\frac{a\, \d a}{\sqrt{\frac{1}{3}(\rho_{r0} + \rho_{m0}\, a) }} =  \d t \ .
\eeq
The solution is straightforward to obtain after an integration by parts, and is given by
\beq
\boxed{\hspace{1em} t-t_* = \frac{2}{\sqrt{3}\, \rho_{m0}^2} \, \left(\rho_{m0}\, a - 2\rho_{r0}  \right) \sqrt{\rho_{m0}\, a + \rho_{r0}} \hspace{1em}}
\eeq
where $t_*$ captures the possible integration constant. This solution has been known for a long time: see \cite{McIntosh, Jacobs}, and \cite{Cohen} with curvature, or \cite[App. B]{Bianchi:2024mlq} for a recent account.

The asymptotic $a(t)$ solutions \eqref{Pnsol} are easily recovered. In the limit of large $a$, for which we neglect the constants, we obtain precisely the $P_m$ solution. In the limit of small $a$, developing the square root to second order, and absorbing the constants in a (shift) redefinition of $t$, we recover the $P_r$ solution.\\

Let us now show that the above solution can completely inverted, to reach an expression of $a(t)$ instead of $t(a)$. To that end, we first rewrite the above solution as follows
\beq
T-2 = (X-2)\sqrt{X+1} \ ,\qquad {\rm where}\quad  T-2= \frac{\sqrt{3}}{2} \frac{\rho_{m0}^2}{\rho_{r0}^{3/2}}\, (t-t_*) \ ,\ X= \frac{\rho_{m0}}{\rho_{r0}} \, a \ ,
\eeq
and $t_*$ has been fixed to give the $-2$, i.e.~giving the same origin $T=t=0$, such that $X(T=0)=0$, i.e.~$a(t=0)=0$. We then take $T\geq0$. Squaring the above solution equation, we reach a 3${}^{{\rm rd}}$ degree polynomial equation
\beq
(X-2)^2 (X+1) = (T-2)^2 \ \Leftrightarrow \ T(4-T) = X^2(3-X)  \ .
\eeq
This equation can be solved following the method of Cardan-Tartaglia, and selecting the real solutions, as we now explain (see \cite{Boyle:2022lcq} for a more general expression in terms of complex roots). The key quantity is
\beq
\Delta = 27\, T(4-T)(4-T(4-T))=27 T(4-T)(T-2)^2 \ .
\eeq
We first get $\Delta > 0 $ for $0<T<4$ and $T\neq 2$. In that case, the method gives three real solutions, namely
\beq
X= 2  \cos \left(\frac{1}{3} \arccos\left( 1- \frac{1}{2} T(4-T) \right) + \frac{2\pi p}{3} \right) + 1 \ ,\ p=0,1,2 \ .
\eeq
Only $p=0,2$ give $X\geq 0$ for $T\in (0,4)$, which is necessary for us. Also, $p=0,2$ give respectively $X=3,0$ at $T=4$: these values will be important to ensure continuity of the complete solution. For $\Delta<0$, i.e.~$T>4$, the (only real) solution is given by
\beq
X= 1+ Y_+ + Y_- \ ,\ Y_\pm= \left(\frac{2- T(4-T) \pm (T-2) \sqrt{T(T-4) } }{2} \right)^{\frac{1}{3}} \ .
\eeq
To verify this, one can first note that $Y_+ Y_- =1$. In addition, with the ansatz $X=1+Y+\frac{1}{Y}$, the above equation gets rewritten into $Y^6 - Y^3 ((T-2)^2-2) +1 =0$. This gives the previous solution for $T\geq 4$. At $T=4$, one obtains $X=3$.

These various real solutions eventually lead us to the complete solution, that we require to be continuous and satisfying $X(T=0)=0$. It is given as follows
\begin{framed}
\begin{itemize}
\vspace{-0.2in}
\item $T\in (0,2)$: $\ X= 2  \cos \left(\frac{1}{3} \arccos\left( 1- \frac{1}{2} T(4-T) \right) + \frac{4\pi}{3} \right) +1 $
  \item $T\in (2,4)$: $\ X= 2  \cos \left(\frac{1}{3} \arccos\left( 1- \frac{1}{2} T(4-T) \right)  \right) + 1$
  \item $T\in (4,\infty)$: $\ X= 1+ \left(\frac{2- T(4-T) + (T-2) \sqrt{T(T-4) } }{2} \right)^{\frac{1}{3}} + \left(\frac{2- T(4-T) - (T-2) \sqrt{T(T-4) } }{2} \right)^{\frac{1}{3}}$
\vspace{-0.2in}
\end{itemize}
\end{framed}
\vspace{-1.2in}
\beq
\label{atsolradmat}
\eeq
\vspace{0.5in}

\begin{figure}[H]
\centering
\includegraphics[width=0.6\textwidth]{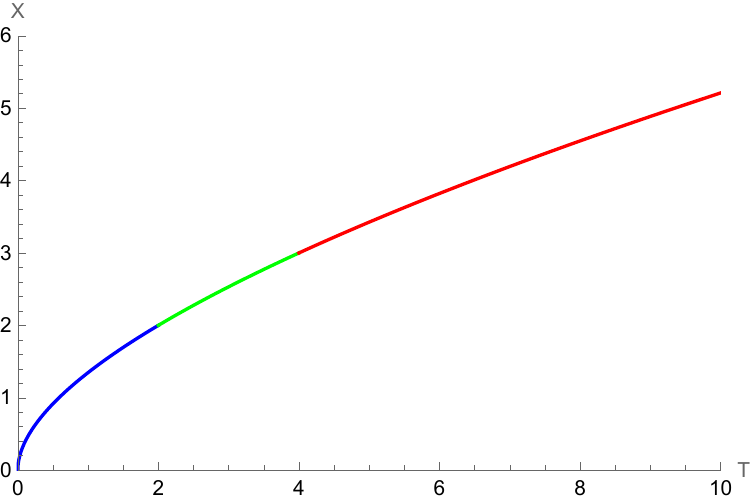}
\caption{$X(T)$, standing up to normalisation for $a(t)$, as the analytic solution for the radiation - matter phase. The three colors of the curve correspond to the three analytic parts of the solution given in \eqref{atsolradmat}. The continuity of the solution is manifest.}\label{fig:XT}
\end{figure}

This complete solution is depicted in Figure \ref{fig:XT}, where we verify its continuity.

From the analytic expressions \eqref{atsolradmat}, the asymptotic behaviours in $t$ or $T$ can be recovered. For large $T$, it is straightforward to recover the $T^{\frac{2}{3}}$ behaviour of $P_m$. For small $T$, we need to use that $\arccos(1-\epsilon) \sim \sqrt{2\epsilon}$ for $\epsilon \ll 1$. With $\cos(x+\frac{\pi}{3})=\frac{1}{2}\cos x - \frac{\sqrt{3}}{2} \sin x$, we conclude that $X\sim \frac{2}{\sqrt{3}} \sqrt{T}$ at $T \sim 0$, reproducing the $T$-scaling of $P_r$.\\

\begin{figure}[H]
\begin{center}
\begin{subfigure}[H]{0.48\textwidth}
\includegraphics[width=\textwidth]{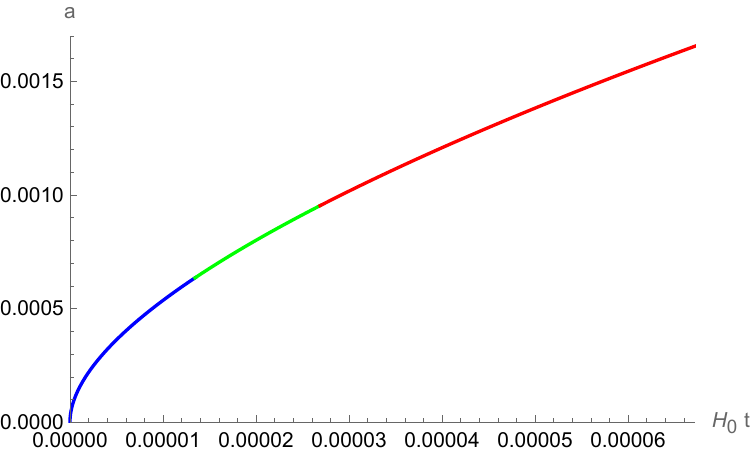}\caption{}\label{fig:atanalytic}
\end{subfigure}\quad
\begin{subfigure}[H]{0.48\textwidth}
\includegraphics[width=\textwidth]{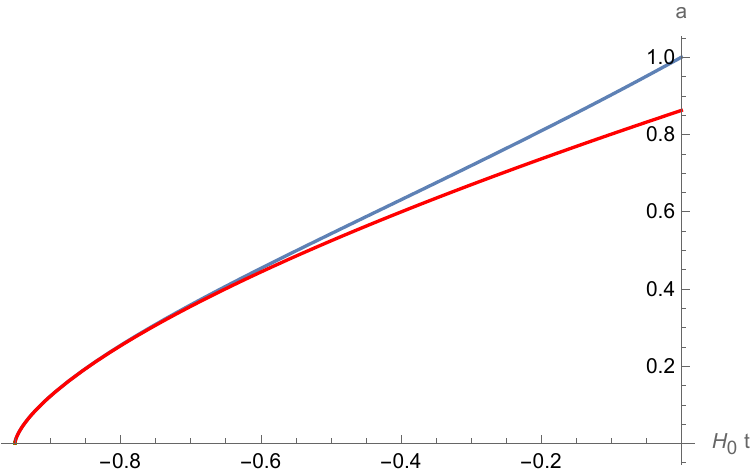}\caption{}\label{fig:atanacomp}
\end{subfigure}
\caption{$a(t)$ in terms of $H_{{\rm today}}\, t$ (denoted $H_0 \, t$ on the graphs) given by the analytic solution for the radiation - matter phase. Figure \ref{fig:atanalytic} gives this curve, analogous to that of Figure \ref{fig:XT}, but here in terms of time: we see that the red solution very quickly becomes the relevant one. In Figure \ref{fig:atanacomp}, where we set $t_{{\rm today}}=0$, we compare this analytic solution (mostly red) to the numerical solution (blue) for $\Lambda$CDM. We verify the good match in the early universe (until dark matter domination), confirming in particular the normalisations discussed in the main text.} \label{fig:analytic}
\end{center}
\end{figure}

The analytic solution \eqref{atsolradmat} should in very good approximation reproduce $a(t)$ during the radiation - matter phase, and in particular, reproduce the $\Lambda$CDM solution from $t=0$ until matter domination; in the later phase of dark energy domination, the analytic solution should differ from that of all models. We can verify this by comparing the numerical solution for $\Lambda$CDM to our analytic one. This comparison first requires to adjust the normalisation of the solution \eqref{atsolradmat} as follows: we rewrite $X$ and $T$ as
\beq
X= \frac{\Omega_{m0}}{\Omega_{r0}}\, \frac{a}{a_0} \ ,\quad T = \frac{3}{2} \sqrt{\Omega_{m0}} \left(\frac{\Omega_{m0}}{\Omega_{r0}} \right)^{\frac{3}{2}} \, H_0\, t \ ,
\eeq
where we introduced an $a_0$ and one has $\Omega_{n}= \Omega_{n0}, H=H_0, a=a_0$ when $t=t_0$. Previously, we typically took $t_0 = t_{{\rm today}}$ and $a_0=1$. Today is however not part of the radiation - matter phase, so we have to use another time $t_0$. We keep as a definition that $a_{{\rm today}}=1$, leaving the e-fold number definition identical: $N= \ln a = \ln \frac{a}{a_0} + \ln a_0$. For $t_0$, we take an appropriate moment during the radiation - matter phase, namely the equality time when $\Omega_r=\Omega_m$. In other words, $a_0 = e^{N_{rm}}$. For all three examples, we read from Table \ref{tab:Neq} the same numerical value $N_{rm} = -8.0548$, giving $a_0= 0.0003176$. We also evaluate numerically for all three examples that $\Omega_{r0}=\Omega_{m0}=0.5000$, a value justified by the fact that $\Omega_{\varphi0}$ is very small in each case:
\bea
\Lambda{\rm CDM}:\quad && \Omega_{\varphi0}=3.483 \cdot 10^{-11} \ ,\nn\\
\text{Exp. Quint.}:\quad && \Omega_{\varphi0}=1.105 \cdot 10^{-10} \ ,\label{Opsmall}\\
\text{Hill. Quint.}:\quad && \Omega_{\varphi0}=7.005 \cdot 10^{-11} \ .\nn
\eea
Finally, we evaluate numerically for all three examples that $H_0/H_{{\rm today}} = 140236$. Using those values for $\Omega_{r0},\Omega_{m0}, a_0, H_0$, we finally obtain analytically a properly normalised expression for $a(t)$ in terms of $H_{{\rm today}}\, t$. This allows the comparison to the numerical solution for $\Lambda$CDM: we depict those curves in Figure \ref{fig:analytic}, and find a good match.

This properly normalised expression also allows to evaluate the duration of the radiation - matter phase. Strictly speaking, the phase starts at $N=N_r$ and ends at $N=N_m$, corresponding with the above radiation - matter equality normalisation, to $X_r= e^{N_r - N_{rm}}$ and $X_m= e^{N_m - N_{rm}}$. In good approximation, the former is $X_r \approx 0$, meaning that the phase can be considered here to start at $T=t=0$. Its duration is thus given by the time $t_m$ corresponding to $X_m$, namely
\beq
H_{{\rm today}}\, t_m= \frac{2}{3}\, \Omega_{m0}^{-\frac{1}{2}}\, \frac{H_{{\rm today}}}{H_0} \,  \left(2+ (X_m - 2) \sqrt{X_m +1}\right) \ ,
\eeq
with $\Omega_{m0} \approx 0.5000$, $H_0/H_{{\rm today}} = 140236$ and $N_{rm} = -8.0548$. The result depends on $N_m$: with Table \ref{tab:max}, we get the following durations for the radiation - matter phase
\bea
\Lambda{\rm CDM}:\quad && H_{{\rm today}}\, t_m \approx 0.0285\ ,\nn\\
\text{Exp. Quint.}:\quad && H_{{\rm today}}\, t_m \approx 0.0200\ ,\label{tradmat}\\
\text{Hill. Quint.}:\quad && H_{{\rm today}}\, t_m \approx 0.0258\ .\nn
\eea
It is worth comparing these values to the numerical estimations of the age of the universe for each model in Section \ref{sec:examples}. In particular, analogously to the computation around \eqref{ageLCDM}, we get for $\Lambda$CDM that $t_m \approx 414 \cdot 10^6\, {\rm yrs}$.

\subsubsection{Analytic solution $\varphi^i(t)$}\label{sec:anafrozenfield}

We have argued that during the radiation - matter phase, the scalar fields contributions are very subdominant compared to radiation and matter: $\rho_{\varphi} \ll \rho_r + \rho_m$. It is then the latter that fix $a(t)$: we obtained above this function analytically by neglecting $\rho_{\varphi}$ and solving $F_1=0$. This can be considered as a ``background solution''. In quintessence models, $\rho_{\varphi}$ is however not strictly vanishing, even though small. The dynamics of the scalar fields are governed by $E^i=0$, in which we then take $H(t)$ entirely fixed by the background solution, as a source term for the fields. There are several motivations for obtaining analytically the solution for $\varphi^i(t)$, that we will develop in the next subsubsection. Let us then tackle here the question of solving $E^i=0$. With canonically normalised fields, it is given by
\beq
\ddot{\varphi}^i + 3 H \dot{\varphi}^i + \del_{\varphi^i}V = 0 \ .
\eeq
Note that the background solution is in good approximation independent of the spatial curvature. In addition, the solution for the scalar fields is dictated only by $E^i=0$, which is independent of the curvature. Therefore we do not need to specify $k$ here.

\begin{figure}[H]
\begin{center}
\begin{subfigure}[H]{0.48\textwidth}
\includegraphics[width=\textwidth]{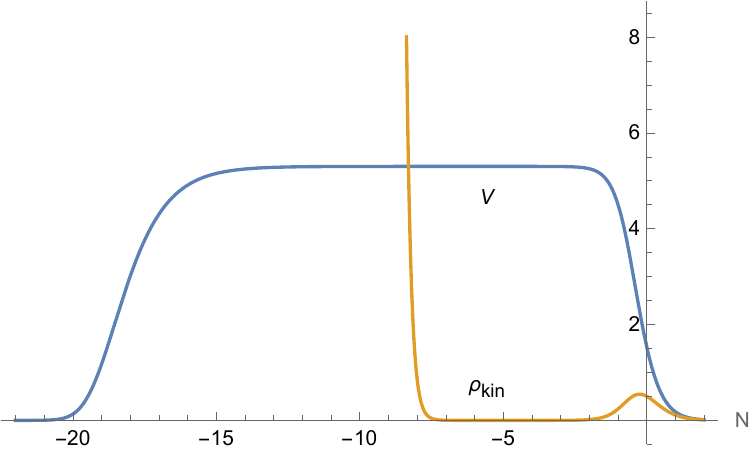}\caption{Exponential}\label{fig:ExpQuintVkin}
\end{subfigure}\quad
\begin{subfigure}[H]{0.48\textwidth}
\includegraphics[width=\textwidth]{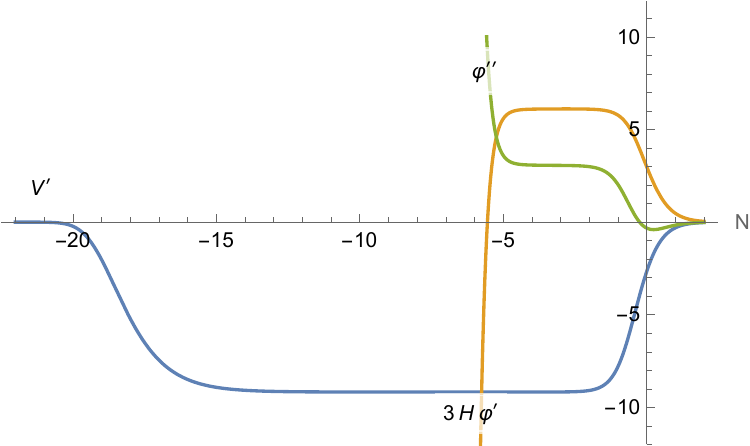}\caption{Exponential}\label{fig:ExpQuintFieldEq}
\end{subfigure}\\
\begin{subfigure}[H]{0.48\textwidth}
\includegraphics[width=\textwidth]{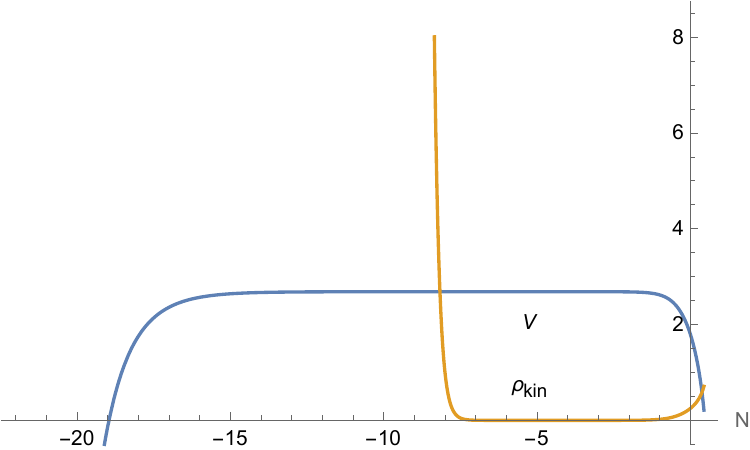}\caption{Hilltop}\label{fig:HilltopVkin}
\end{subfigure}\quad
\begin{subfigure}[H]{0.48\textwidth}
\includegraphics[width=\textwidth]{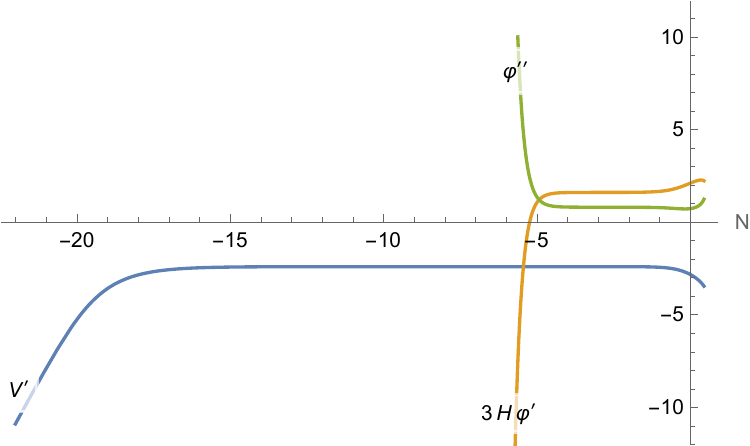}\caption{Hilltop}\label{fig:HilltopFieldEq}
\end{subfigure}
\caption{Evolution of $V$, the kinetic energy $\rho_{{\rm kin}}=\frac{1}{2} \dot{\varphi}^2$, as well as of the different terms in the equation $E^i=0$, here for a single canonical field, namely $\del_{\varphi} V$, $3 H \dot{\varphi}$ and $\ddot{\varphi}$, in terms of $N$. This is depicted for our exponential and hilltop quintessence examples. Strictly speaking, the quantities depicted are those divided by $H_0^2$, due to the formalism used in these examples (see Section \ref{sec:eqreform}). The early times evolutions of $V$ and $\del_{\varphi} V$ are related to the initial condition there being $\dot{\varphi}<0$ in our examples (see footnote \ref{foot:sign}).} \label{fig:VPhi}
\end{center}
\end{figure}

To solve $E^i=0$ and find the analytic solution for $\varphi^i(t)$, helpful hints are given by the numerical solutions. We depict in Figure \ref{fig:VPhi} for our two quintessence examples the evolution of the kinetic energy and the potential, as well as that of the three different terms in $E^i$.

A first observation from Figure \ref{fig:VPhi} is that in a first part of the radiation - matter phase, $\rho_{{\rm kin}} \gg V$, consistently with Figure \ref{fig:ONkinV}. Similarly among the terms of $E^i$, the potential contribution can at first be neglected. This brings us back to the situation discussed for kination: neglecting the potential in $E^i=0$, one gets the continuity equation on $\rho_{{\rm kin}} $ and obtains, as discussed around \eqref{kincont}, the solution
\beq
\rho_{{\rm kin}} = \rho_{{\rm kin}0}\, a^{-6} \ ,\quad \dot{\varphi}^i =\pm \sqrt{2 \, \rho_{{\rm kin}0}}\, c_k^i \, a^{-3} \ ,\quad c_k^i \geq 0 \ ,\ \sum_i (c_k^i)^2 = 1 \ , \label{kinsolradmat}
\eeq
where to get the field expression we took $g_{ij}=\delta_{ij}$.

In this first part of the radiation - matter phase, the fields therefore behave in the same way as in the kination - radiation phase, as long as the potential is negligible (the case in our examples). This is consistent by continuity with the previous phase. Physically, the result is that the kinetic energy continuously decreases to very small values, confirming its subdominance during this phase. Also, $\rho_{\varphi}/\rho_r \sim a^{-2}$ and $\rho_{\varphi}/\rho_m \sim a^{-3}$, and both decrease as long as these expressions are valid. As a consequence $\Omega_{\varphi} \sim \Omega_{{\rm kin}}$ is more and more subdominant. It was already negligible at radiation domination, and it will be even more during this first part of the radiation - matter phase.

Since the kinetic energy becomes very small, the fields $\varphi^i$ become effectively frozen: their value barely changes. The physical explanation to this is the Hubble friction: indeed, the only term affecting $\ddot{\varphi}^i$ via $E^i=0$ is the Hubble friction term, $3H \dot{\varphi}^i$, so it is the one slowing down the field. Recall also that $H$, fixed by the background solution, takes huge values in the early universe, as seen for our examples in Figure \ref{fig:LCDMH}, \ref{fig:ExpQuintHN} or \ref{fig:HilltopHN}. The resulting Hubble friction dramatically slows down, or even freezes, the fields. This is also in line with the field evolution in our solutions, where the freezing of the fields can be seen in Figure \ref{fig:ExpQuintpN} and \ref{fig:HilltopDPhiN}.

The fields become constant, up to small corrections: we denote this constant as in the background solution by $\varphi_{0rm}^i$. We can be more precise and characterise the corrections. The complete solution for the scale factor was given previously as the background solution. For simplicity, we parameterize it here locally as $a(t) = a_{n_a} \, t^{n_a}$ with $\frac{1}{2} \leq n_a \leq \frac{2}{3}$, since the background solution interpolates between $P_r$ and $P_m$; we will later verify this parametrisation. We deduce that $\dot{\varphi}^i =\pm \sqrt{2 \, \rho_{{\rm kin}0}}\, c_k^i \, a_{n_a}^{-3}\ t^{-3 n_a}$, giving locally (i.e.~neglecting $\dot{n}_a$)
\beq
\boxed{\hspace{1em} \varphi^i = \varphi_{0rm}^i \mp c_k^i\, \sqrt{2 \, \rho_{{\rm kin}0}}\ \frac{a_{n_a}^{-3}}{3 n_a - 1}\ t^{-3 n_a + 1} \hspace{1em}} \label{phit1}
\eeq
A more precise expression can in principle be obtained using the complete background solution for $a(t)$. The validity of the above expression will be tested in the next subsubsection. Here, the second term seems to be a correction if one considers a large $t$. However, $t$ is related to a power of $a \sim e^N$ with $N<0$, therefore it rather tends to be small. As we will see and evaluate in the next subsubsection, it is actually the normalisation constants which ensure the second term to be small for any time or $N$ of the period considered.

As a consequence, the potential and its derivatives are also constant in first approximation and given by their value at $\varphi_{0rm}^i$: we denote
\beq
V_{0rm} \equiv V(\varphi_{0rm}^j) \ ,\quad \del_{i}V_{0rm} \equiv \del_{\varphi^i} V (\varphi_{0rm}^j) \ .
\eeq
We verify this explicitly in our examples in Figure \ref{fig:VPhi}. There, $V$ and $V'$ are constant not only during the first part of the radiation - matter phase but actually during the whole of it.\\

The only way to alter these dynamics and the above expressions is to have $\del_{\varphi^i}V$ non-negligible in $E^i=0$; otherwise the solution remains \eqref{kinsolradmat}. Since $\rho_{{\rm kin}}$ is always decreasing, it has to reach the value $V_{0rm} $ at some point, and the potential is then not negligible anymore. Similarly, as $H$ and $|\dot{\varphi}^i|$ are both decreasing, so does $|3H \dot{\varphi}^i|$ and $\ddot{\varphi}^i$, and both terms of $E^i$ must at some point reach the value $|\del_{i}V_{0rm}|$; that third term can then not be neglected anymore. These two moments, when potential contributions become relevant, are clearly identifiable in Figure \ref{fig:VPhi}. The two moments are not the same: the former takes place earlier than the latter. We will come back to the former in the next subsubsection. When it comes to the fields solution however, only the latter matters, namely reaching the value of $|\del_{i}V_{0rm}|$. We then enter the second part of the radiation - matter phase.

In this second part, the dynamics of the scalar fields is not only due to friction but also to the potential force. All three terms in $E^i=0$ are now relevant: we verify in Figure \ref{fig:VPhi} that none can be neglected. This makes the equation more difficult to solve. Interestingly, we also note there that all three terms appear to be constant during most of this second part (and beyond). This is an important hint to find an analytic solution and we will come back to it.

\begin{figure}[H]
\begin{center}
\begin{subfigure}[H]{0.48\textwidth}
\includegraphics[width=\textwidth]{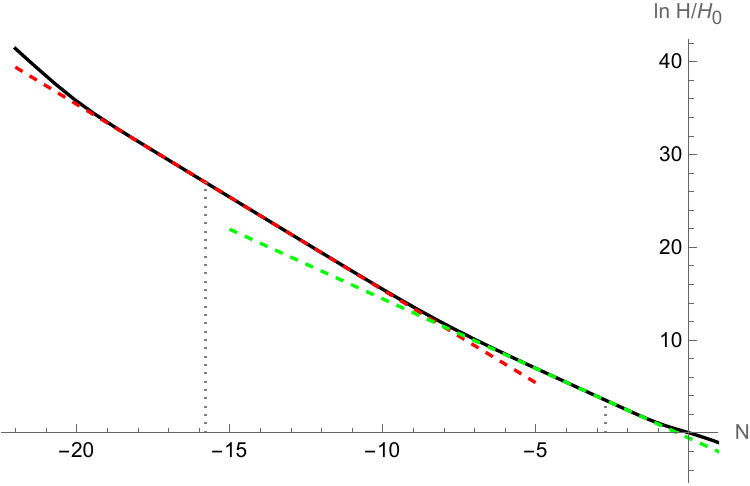}\caption{Exponential}\label{fig:ExpQuintlnHN}
\end{subfigure}\quad
\begin{subfigure}[H]{0.48\textwidth}
\includegraphics[width=\textwidth]{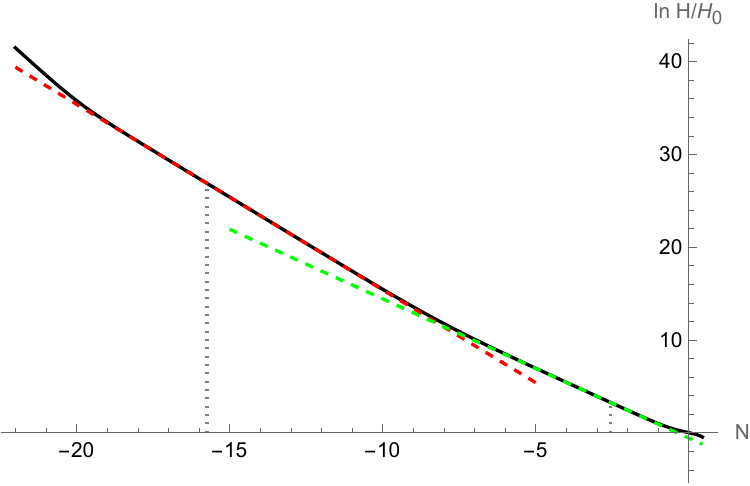}\caption{Hilltop}\label{fig:HilltoplnHN}
\end{subfigure}
\caption{$\ln \frac{H}{H_0} (N)$ (black) for exponential and hilltop quintessence examples. In red, resp. green, are depicted the tangent lines to the curve at $\Omega_{r\, {\rm max}}$, resp. $\Omega_{m\, {\rm max}}$, with slope $-2$, resp. $-\frac32$; the corresponding $N$ values, $N_r$ and $N_m$, given in Table \ref{tab:max}, can be found here following the vertical dotted gray lines. We verify in this way the power law parametrisation of $a(t)$ discussed in the main text. Note also that the two tangent lines consistently intersect at about $N_{rm}$, the radiation - matter equality time.}\label{fig:lnHN}
\end{center}
\end{figure}

To find the solution, we first consider again the local parametrisation $a(t) \sim t^{n_a}$ with $\frac{1}{2} \leq n_a \leq \frac{2}{3}$. Neglecting $\dot{n}_a$, we obtain that $H = n_a\, t^{-1}$. Beyond using this to solve $E^i=0$ as we will do, it also offers a way to verify the parametrisation. Indeed, we then obtain $\ln \frac{H}{n_a} = - \frac{N}{n_a}+{\rm constant}$. We plot the latter in Figure \ref{fig:lnHN} for our two quintessence examples, and verify this local behaviour, linear in $N$.
As a solution ansatz for this second part, we further assume $\dot{\varphi}^i \sim t^{n_{\varphi}}$ with $n_{\varphi}\neq 0$. We then get that $\frac{3 n_a}{n_{\varphi}} \ddot{\varphi}^i = 3 H \dot{\varphi}^i \sim t^{n_{\varphi}-1}$. Each of these two contributions to $E^i=0$ was noticed in Figure \ref{fig:VPhi} to be constant. This is consistent with their combination being equal to $-\del_{i}V_{0rm}$, the constant value of the potential derivative. As a consequence, we refine the ansatz to $n_{\varphi}=1$. This eventually gives us $3 n_a \ddot{\varphi}^i = 3 H \dot{\varphi}^i $. This gives the following rewriting of $E^i=0$, that we can then solve
\beq
\ddot{\varphi}^i = - \frac{\del_{i}V_{0rm}}{3n_a + 1} \ \Rightarrow \ \dot{\varphi}^i = - \frac{\del_{i}V_{0rm}}{3n_a + 1} \ t \ ,\quad \boxed{\hspace{1em} \varphi^i = \varphi_{0rm}^i - \frac{\del_{i}V_{0rm}}{2(3n_a + 1)} \ t^2 \hspace{1em}} \label{phit2}
\eeq
We dropped the possible integration constant in $\dot{\varphi}^i$ consistently with the ansatz.

The above solution to $E^i=0$ in the second part of the radiation - matter phase reproduces the three constant contributions to that equation, noticed numerically. To be consistent with an almost constant field, we also need the second term in $\varphi^i$ to be a small correction, which suggests considering here small times $t$. This is in line with $(\dot{\varphi}^i)^2$ and having a very small kinetic energy, as seems to be the case in Figure \ref{fig:ExpQuintVkin} and \ref{fig:HilltopVkin}. As explained above for the first part, $t$ is related to a power of $e^N$, with $N<0$, which indeed makes it small. We will come back to the evaluation of the correction term in the next subsubsection. Note that the kinetic energy is now increasing: this is due to the potential force. It has therefore reached a minimum during this phase.\\

We have obtained an analytic solution for $\varphi^i(t)$ for the radiation - matter phase, that we separated in two parts. In the next subsection, we will specify better the normalisations of the correction terms in this solution, and make use of this result to compute the field variation during this phase. We will also use our physical understanding of this solution to evaluate the moment at which the potential becomes non-negligible, compared to the kinetic energy.

\subsubsection{Applications: frozen fields ($\Delta \varphi$) and $w_{\varphi}$ transition}\label{sec:anafieldapplication}

Having obtained the analytic solutions for $(a(t), \varphi^i(t))$ during the radiation - matter phase, we now want to make use of it to answer two questions: to what extent can we consider the scalar fields as ``frozen'' during this phase, and when is the transition moment at which $w_{\varphi}$ goes from $+1$ to $-1$?\\

Starting with the scalar field values (considering canonically normalised fields), the first question amounts to compute their variation $\Delta \varphi^i$ during the radiation - matter phase; as mentioned already, this variation seems very small and the fields appear to be almost constant, i.e.~frozen. We can get a first upper bound on this variation with the following reasoning. During this phase, $\Omega_{\varphi}$ is expected to be small. We can consider its maximal value during this phase, $\Omega_{\varphi\, {\rm max}}$. Since $\Omega_{\varphi} < \Omega_{\varphi\, {\rm max}}$ and $V>0$, we obtain $(\dot{\varphi}^i)^2 < H^2\, 6 \Omega_{\varphi\, {\rm max}}$. Taking for illustration $\dot{\varphi}^i \geq 0$ (the result can be adapted to each sign), one gets $\dot{\varphi}^i <  H \, \sqrt{6 \Omega_{\varphi\, {\rm max}}}$. Using $\d N = H \d t$, one deduces
\beq
\boxed{\hspace{1em} \Delta \varphi^i < \sqrt{6 \Omega_{\varphi\, {\rm max}}} \times \Delta N \hspace{1em}} \label{deltaphiOpmax}
\eeq
According to Table \ref{tab:max} and Figure \ref{fig:OpN}, we get in our examples $\Omega_{\varphi\, {\rm max}} \approx \frac{1}{6} \cdot 10^{-2}$. This value, together with roughly $\Delta N \sim 10$ for this phase, give us $\Delta \varphi < 1$ in Planckian units. This first estimate can be greatly improved by restricting to a slightly shorter period. Indeed, as can be seen in Figure \ref{fig:OpN}, $\Omega_{\varphi}$ quickly drops (see also the values in \eqref{Opsmall}). With a shorter period, one then quickly gets a much stronger bound on the variation of the field, which translates into the fact that it appears frozen.

A more accurate estimate for $\Delta \varphi^i$ can be obtained from the analytic solution in \eqref{phit1} and \eqref{phit2}. One unknown in these expressions are the normalisations in the time dependent pieces. A simplifying choice is to pick for $a(t)$ that of $P_r$ for the first part, and that of $P_m$ for the second part. This is justified by Figure \ref{fig:lnHN}, where the complete solution for $a(t)$ seems to be well approximated by these two successive scalings. From these choices, we obtain the following expressions
\begin{framed}
\vspace{-0.2in}
\bea
\hspace{-0.5in} \text{Part 1:}  && \varphi^i = \varphi_{0rm}^i \mp \, c_k^i \, \sqrt{6 \, \frac{\rho_{{\rm kin}0}}{\rho_{r0}} }\ a^{-1} = \varphi_{0rm}^i \mp \, c_k^i \, \sqrt{6}\ e^{N_{{\rm kin}r}-N} \label{PhiN1}\\
\hspace{-0.5in} \text{Part 2:}  && \varphi^i = \varphi_{0rm}^i - \frac{2}{9} \frac{\del_{i}V_{0rm}}{\rho_{m0}} \ a^3 = \varphi_{0rm}^i - \frac{2}{9} \frac{\del_{i}V_{0rm}}{V_{0rm}} e^{4( N_{m\, \Lambda} - N_{m\, q} )} \ e^{3 (N- N_{m\varphi\, \Lambda})} \label{PhiN2}
\eea
\vspace{-0.2in}
\end{framed}
\noindent where for the latter, we used that $V_{0rm} \approx \rho_{\varphi\, m}$, the field density at matter domination, and further that there, $w_{\varphi}\approx -1$, together with the relation to today's density \eqref{rhoratio}.

If we consider the field continuous at the intermediate time between the two parts, and we consider the absolute value of the field displacement for each part, we deduce for one field, with $c_k^i=1$,
\beq
\boxed{\hspace{1em} \Delta \varphi = \sqrt{6}\ e^{N_{{\rm kin}r}-N_r} - \frac{2}{9} \frac{\del_{i}V_{0rm}}{V_{0rm}} e^{4( N_{m\, \Lambda} - N_{m\, q} )} \ e^{3 (N_{m\, q} - N_{m\varphi\, \Lambda})} \hspace{1em}} \label{Deltaphifrozen}
\eeq
where we evaluated for the radiation - matter phase, namely between $N_r$ and $N_{m\, q}$. Taking the data from Section \ref{sec:phases}, and $\frac{\del_{i}V_{0rm}  }{V_{0rm}} \approx -0.8969$ for the hilltop model (numerical evaluation, see also Figure \ref{fig:VPhi}), we obtain (giving each of the two terms above)
\bea
\text{Exp. quint.}:\quad && \Delta \varphi =  0.03618 + 0.0006171 \\
\text{Hill. quint.}:\quad && \Delta \varphi =  0.03725 + 0.0002696
\eea
We also evaluate the correction terms at intermediate values, e.g.~at $N_{rm}$, and those are for both terms much smaller. We see that the biggest contribution comes from the first term, evaluated at $N_r$: this is because the field still evolves a little at the beginning the radiation - matter phase, as can be seen in Figure \ref{fig:ExpQuintpN} and \ref{fig:HilltopDPhiN}, driven by the remaining kination. On the contrary, it appears very frozen at matter domination and beyond. As mentioned at the beginning of this subsubsection, by restricting slightly the time considered, one would obtain a much smaller field displacement.

Let us now compare the above values to numerical ones. To that end, we evaluate numerically for Figure \ref{fig:ExpQuintpN} and \ref{fig:HilltopDPhiN} the moment $N_{\varphi\, {\rm min}}$ at which the field reaches a minimum: this corresponds here to the intermediate time between the two parts. We then evaluate the displacement between $N_r$ or $N_m$ to $N_{\varphi\, {\rm min}}$. We obtain
\bea
\text{Exp. quint.}:\quad && N_{\varphi\, {\rm min}} = -5.4353 \ ,\quad \Delta \varphi =  0.03610 + 0.0006128  \\
\text{Hill. quint.}:\quad && N_{\varphi\, {\rm min}} = -4.7863 \ ,\quad \Delta \varphi =  0.03716 + 0.0002676
\eea
This shows a remarkably good match with our previous analytical results. Note that above, we not even subtracted the evaluation at $N_{\varphi\, {\rm min}}$, and we mentioned it should be very small; this could improve the match of the last digits. This very good match confirms the parametrisation used for $a(t)$, the analytic solutions obtained, and the normalisation choices made, all resulting in the field expressions \eqref{PhiN1} and \eqref{PhiN2}.

We conclude that both analytically and numerically, we see that the field displacement during the radiation - matter phase is small in Planckian units, even more if considered slightly after radiation domination. The high Hubble friction effectively freezes the field in the first part, and the potential force, going against the (still high) friction in the second part, only raises slowly the kinetic energy which remains very small, leading to very little displacement. These statements are also true for multifield cases, since each field $\varphi^i$ is fixed by its own equation $E^i=0$. The important assumption here has been that those are canonically normalised, which is not always possible. We recall however that the decrease of the kinetic energy in the first part remains independent of this assumption, suggesting that the freezing still holds in more general cases. In addition, as argued around Figure \ref{fig:OpN}, $\Omega_{\varphi}$ remains very small until matter domination (on top of being the smallest $\Omega_n$), indicating again a small kinetic energy, so a small field displacement corresponding to a general freezing.\\

We turn to the sharp transition of $w_{\varphi}$ from $+1$ to $-1$, that we observe in our examples in Figure \ref{fig:ExpQuintwN} and \ref{fig:HilltopwN}. From the definition of $w_{\varphi}$, going from $+1$ to $-1$ signals a transition from kinetic energy domination to potential domination. We thus denote the moment of this transition by $N_{{\rm kin}V}$, more precisely defined as the moment for which $w_{\varphi}=0$, meaning when $\rho_{{\rm kin}} = V$. Numerically, we obtain $N_{{\rm kin}V} \approx - 8.299$ for exponential quintessence, and $N_{{\rm kin}V} \approx - 8.158$ for hilltop quintessence.

The above analytic solution and our understanding of it allow us to determine $N_{{\rm kin}V}$. We explained that at first, the potential contributions were negligible, as during kination, and the kinetic energy was dominant but decreasing. Later, as can be seen in Figure \ref{fig:ExpQuintVkin} and \ref{fig:HilltopVkin}, the kinetic energy becomes subdominant to the potential. This evolution is in line with the observed transition.

As discussed in the previous subsubsection, until the transition (and even a little afterwards), the kinetic energy is given by $\rho_{{\rm kin}} = \rho_{{\rm kin}0} \, a^{-6}$. At the transition and after (until matter domination and a little afterwards), the potential is approximately constant: $V = V_{0rm}$. $N_{{\rm kin}V}$ is thus fixed by the equality $V_{0rm} = \rho_{{\rm kin}0} \, a^{-6}$. We can rewrite this in terms of more convenient quantities. We first use, as done above, that at matter domination, the kinetic energy is negligible, giving $V_{0rm} \approx \rho_{\varphi\, m}$. Furthermore, we take at matter domination $w_{\varphi}\approx -1$. We can then relate these quantities to today's dark energy density thanks to \eqref{rhoratio}. We get
\beq
\rho_{{\rm kin}0}\, e^{-6 N_{{\rm kin}V}} = V_{0rm} = e^{4( N_{m\, \Lambda} - N_{m\, q} )} 3 H_0^2 \Omega_{\varphi0} \ \Leftrightarrow \  N_{{\rm kin}V} =  -\frac{2}{3}( N_{m\, \Lambda} - N_{m\, q} ) - \frac{1}{6} \ln \frac{\Omega_{\varphi0} }{\Omega_{{\rm kin}0}} \ .\nn
\eeq
We can now express the latter in terms of equality times, with $\frac{\Omega_{\varphi0} }{\Omega_{{\rm kin}0}} = \frac{\Omega_{\varphi0} }{\Omega_{m0}} \frac{\Omega_{m0}}{\Omega_{r0}} \frac{\Omega_{r0}}{\Omega_{{\rm kin}0}} $, giving
\beq
\boxed{\hspace{1em} N_{{\rm kin}V} =  -\frac{2}{3}( N_{m\, \Lambda} - N_{m\, q} ) + \frac12 N_{m\varphi\, \Lambda} + \frac{1}{6} (2 N_{{\rm kin}r} + N_{rm} ) \hspace{1em}} \label{NkinVequality}
\eeq
We can also use the maxima, with $\frac{\Omega_{\varphi0} }{\Omega_{{\rm kin}0}} = \frac{\Omega_{\varphi0} }{\Omega_{r0}} \frac{\Omega_{r0}}{\Omega_{m0}} \frac{\Omega_{m0}}{\Omega_{{\rm kin}0}}$, giving
\beq
N_{{\rm kin}V} =  -\frac{2}{3}( N_{m\, \Lambda} - N_{m\, q} ) +\frac12 N_r +\frac23 N_{m\, \Lambda} - \frac{1}{6} (N_{rm} + \ln \frac{2}{3} ) \ ,
\eeq
i.e.
\beq
\boxed{\hspace{1em} N_{{\rm kin}V} = - \frac{1}{6}\ln \frac{2}{3} +\frac12 N_r +\frac23 N_{m\, q} - \frac{1}{6} N_{rm} \hspace{1em}} \label{NkinVmax}
\eeq
Both formulas are equivalent. Let us evaluate the first one. We obtain
\bea
\text{Exp. quint.}: && -\frac{2}{3} (N_{m\, \Lambda}-N_{m\, q}) \approx -0.157 \ ,\ \frac{1}{2}  N_{m\varphi\, \Lambda} \approx -0.130\ ,\ \frac{1}{6} \left( N_{rm} + 2 N_{{\rm kin} r} \right) \approx -8.012 \ ,\nn\\
&& \Rightarrow \ N_{{\rm kin}V}  \approx -8.299 \ ,\\
\text{Hill. quint.}: && -\frac{2}{3} (N_{m\, \Lambda}-N_{m\, q}) \approx -0.044 \ ,\ \frac{1}{2}  N_{m\varphi\, \Lambda} \approx -0.130\ ,\ \frac{1}{6} \left( N_{rm} + 2 N_{{\rm kin} r} \right) \approx -7.984 \ ,\nn\\
&& \Rightarrow \ N_{{\rm kin}V}  \approx -8.158 \ .
\eea
We find a perfect match with the numerical values! This is a further confirmation of our analytic solutions.

We note that in the first formula, the first two term are negligible compared to the others: $N_{{\rm kin}V} \approx \frac{1}{6} (2 N_{{\rm kin}r} + N_{rm} )$. Surprisingly, with $N_{rm} \approx -8$ and $N_{{\rm kin}r} \approx -20$ (as chosen by us), we get $2 N_{{\rm kin}r} \approx - 5 N_{rm}$, giving effectively $N_{{\rm kin}V} \approx N_{rm} \approx -8 $! The coincidence between these two quantities is purely numerical due to our choice of $N_{{\rm kin}r}$. An earlier $N_{{\rm kin}r}$ would give a different $N_{{\rm kin}V}$ value. This also shows that $w_{\varphi}=0$ cannot happen later than $N_{rm}$ in the universe history, since $N_{{\rm kin}r} \leq -20$ for BBN to occur. This point is important in case of an observational search of this $w_{\varphi}$ transition. Interestingly, if observed, this transition moment would tell us about  $N_{{\rm kin}r}$, a much earlier event.

\subsection{Matter - dark energy phase}\label{sec:finalphase}

\subsubsection{Analytic solutions: attempts}\label{sec:lastphaseanalytic}

At the end of the previous phase, the kinetic energy, at very small value, was slowly increasing due to the potential force. With such a small speed, the field remained effectively frozen, giving to good approximation a constant potential and a constant first derivative of the potential. This solution was such that the three terms in $E^i=0$ were constant, as seen in Figure \ref{fig:VPhi}. This behaviour remains valid during the start (until $N \sim -1.5$) of the phase of interest here: the matter - dark energy phase, between $N_m < N <0$, as defined in \eqref{phases2}.

To trigger a change in the field dynamics, namely in the three terms of $E^i=0$, there are two options: either the kinetic energy becomes too high, the field cannot be considered constant anymore, so the potential force is not either; or there is a change in $H(t)$. It is unclear to us which one is responsible for the numerically observed change of dynamics, but for sure, the change in $H(t)$ is inevitable; it also is the only change in $\Lambda$CDM.

A decreasing $\rho_m$ in an expanding universe must at some point be of the same order as $\rho_{\varphi}$, if the latter is given by a constant potential (the kinetic energy being at first negligible). For a non-constant potential, which may also decrease, this is less obvious, but in a realistic universe, we know that $\rho_m$ and $\rho_{\varphi}$ become at some point comparable. As a consequence, in $F_1=0$, $\rho_{\varphi}$ cannot be neglected as before, and this enforces a change in $a(t)$. This in turn changes $H(t)$, which as explained, can change the scalar field dynamics.

\begin{figure}[H]
\begin{center}
\begin{subfigure}[H]{0.48\textwidth}
\includegraphics[width=\textwidth]{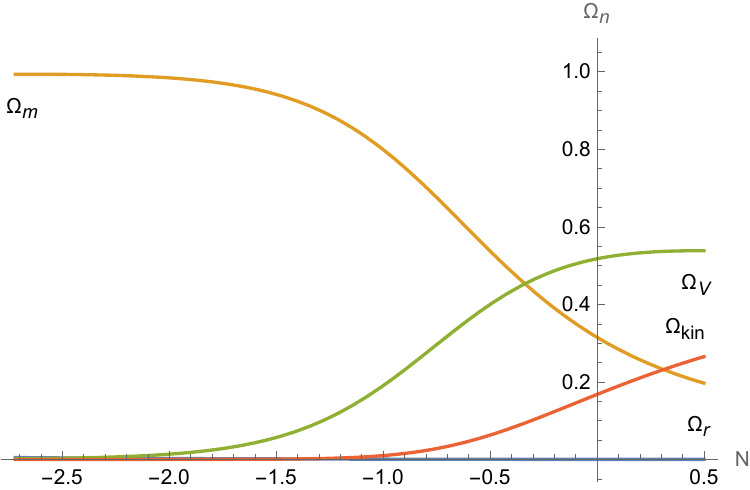}\caption{Exponential}\label{fig:ExpQuintONend}
\end{subfigure}\quad
\begin{subfigure}[H]{0.48\textwidth}
\includegraphics[width=\textwidth]{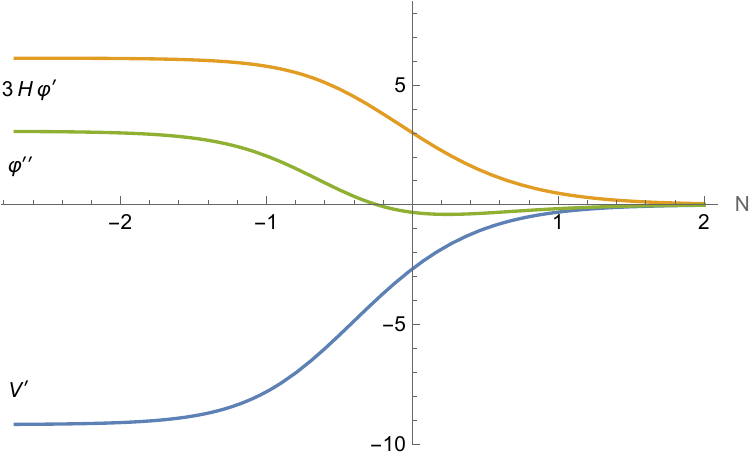}\caption{Exponential}\label{fig:ExpQuintFieldEqZoom}
\end{subfigure}\\
\begin{subfigure}[H]{0.48\textwidth}
\includegraphics[width=\textwidth]{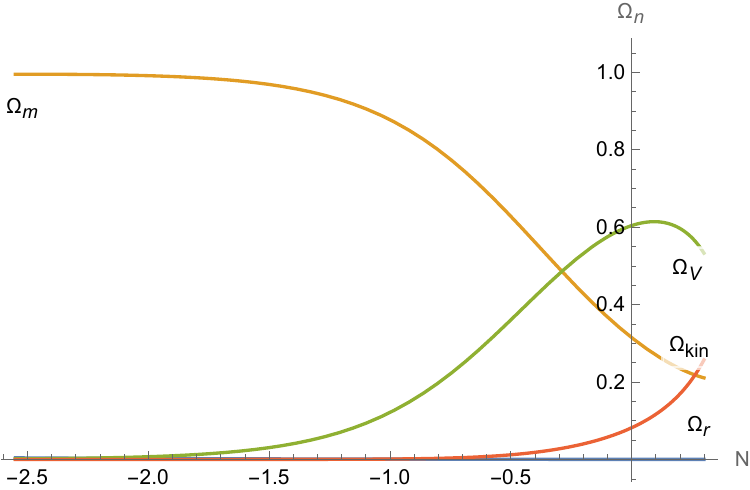}\caption{Hilltop}\label{fig:HilltopONend}
\end{subfigure}\quad
\begin{subfigure}[H]{0.48\textwidth}
\includegraphics[width=\textwidth]{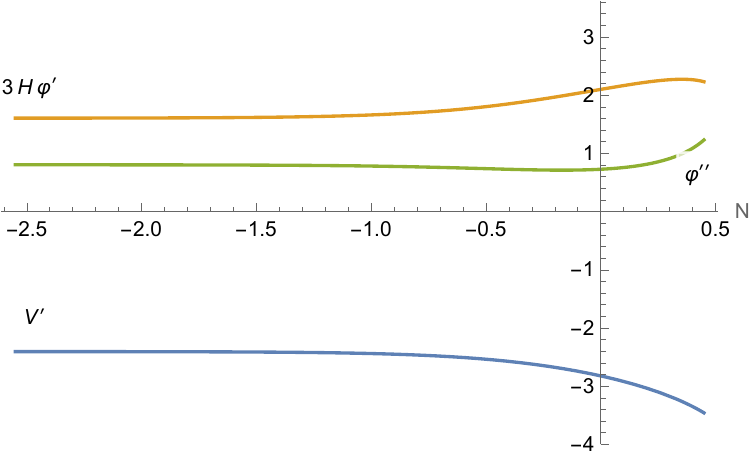}\caption{Hilltop}\label{fig:HilltopFieldEqZoom}
\end{subfigure}
\caption{Evolution of the $\Omega_n$, as well as the different terms in the equation $E^i=0$, for the exponential and hilltop quintessence examples, in the matter - dark energy phase, starting at $N_m$. We refer to Figure \ref{fig:VPhi} for details on the three terms of $E^i=0$.} \label{fig:EndZoom}
\end{center}
\end{figure}

As illustrated by this discussion, the dynamics in this last phase becomes at some point ($N \sim -1.5$) much more complicated to describe, as $\rho_m$, $V$ and then $\rho_{{\rm kin}}$ become all non-negligible. $F_1=0$ and $E^i=0$ become two coupled equations where no term can be neglected, and all are non-constant. We illustrate this in Figure \ref{fig:EndZoom}. Finding an analytic solution is then much more difficult. In addition, the details of model, in particular the potential $V$, become very relevant, while our results were so far model-independent. Indeed, physically, the dynamics are such that the fields are rolling down the potential and gaining speed, while still facing (a not so high) Hubble friction: see Figure \ref{fig:ExpQuintVkin} and \ref{fig:HilltopVkin}. The details of $V$ certainly matter in such dynamics.

There is one case where we can still provide an analytic solution: $\Lambda$CDM. This result is well-known but let us reproduce it here for completeness (see \cite{Boyle:2022lcq} for a general solution in terms of a Jacobi elliptic function, in case of radiation, matter, curvature and cosmological constant). We take that $V=\Lambda = \rho_{\Lambda}>0$ and $\varphi^i$ are constant, so $E^i=0$ is automatically satisfied, and we are left to solve $F_1=0$. We expect to find a solution interpolating between $P_m$ and $P_{\Lambda}$. We can neglect radiation (due to its scaling in $a$), which amounts to set $\rho_{r0}=0$, and we restrict ourselves to the case without curvature, $k=0$. Then, $F_1=0$ gets rewritten as
\beq
\frac{\d a}{\sqrt{\frac{1}{3}\left( \rho_{m0}\, a^{-1} + \rho_{\Lambda} \, a^2 \right)}} = \d t \ \Leftrightarrow \ \frac{\d x}{\sqrt{1+x^2}} = \sqrt{\frac{3}{4}\rho_{\Lambda}}\ \d t \ {\rm with} \ x= \sqrt{\frac{\rho_{\Lambda}}{\rho_{m0}}} \, a^{\frac{3}{2}} \ ,
\eeq
which gets integrated to $x(t)= \sinh \left( \sqrt{\frac{3}{4}\rho_{\Lambda}}\ t\right)$, fixing $x(0)=0$. We recover the well-known solution
\beq
\boxed{\hspace{1em} a(t) = \left( \frac{\rho_{m0}}{\rho_{\Lambda}} \right)^{\frac{1}{3}} \sinh^{\frac{2}{3}} \left( \sqrt{\frac{3}{4}\rho_{\Lambda}}\ t\right) \hspace{1em}}
\eeq
from which we get the asymptotic cases: $a(t)= \left(\frac{3}{4}\, \rho_{m0} \right)^{\frac{1}{3}} \, t^{\frac{2}{3}}$ for $t \sim 0$, as in $P_m$, and $a(t) = \left(\frac{\rho_{m0}}{4\rho_{\Lambda}}\right)^{\frac{1}{3}}\, e^{ \sqrt{\frac{\rho_{\Lambda}}{3}} \ t}$ for $t\sim \infty$, as in $P_{\Lambda}$. Note that the freedom of $\rho_{m0}$ gives the freedom of $a_{\Lambda}$ in $P_{\Lambda}$. Also it is straightforward to verify there that $H=H_0=\sqrt{\frac{\rho_{\Lambda}}{3}} $.

For completeness, let us also recall that one can get from this solution a good approximation of the age of the universe. This is line with remarks in Section \ref{sec:examples}, indicating that most of the universe history in terms of time took place during that phase. Introducing $y=e^{\sqrt{\frac{3}{4}\rho_{\Lambda}}\ t_0} = e^{\frac{3}{2}\sqrt{\Omega_{\Lambda}}\ H_0\, t_0}$, one can solve $a(t_0)=1$ to get
\beq
H_0\, t_0 = \frac{2}{3\sqrt{\Omega_{\Lambda}} } \ln \frac{1+\sqrt{\Omega_{\Lambda}}}{\sqrt{\Omega_{m}}} \approx 0.9510 \ ,\label{ageLCDMapprox}
\eeq
with $\Omega_m=0.315=1-\Omega_{\Lambda}$. This result is very close to the numerical value obtained in Figure \ref{fig:LCDMaT}. Cutting the above solution at the beginning of the matter - dark energy phase, namely at $N=N_m$, meaning removing the corresponding time from \eqref{ageLCDMapprox}, and adding the duration the earlier radiation - matter phase computed in \eqref{tradmat}, we get even closer to the numerical value.\\

\begin{figure}[H]
\begin{center}
\begin{subfigure}[H]{0.48\textwidth}
\includegraphics[width=\textwidth]{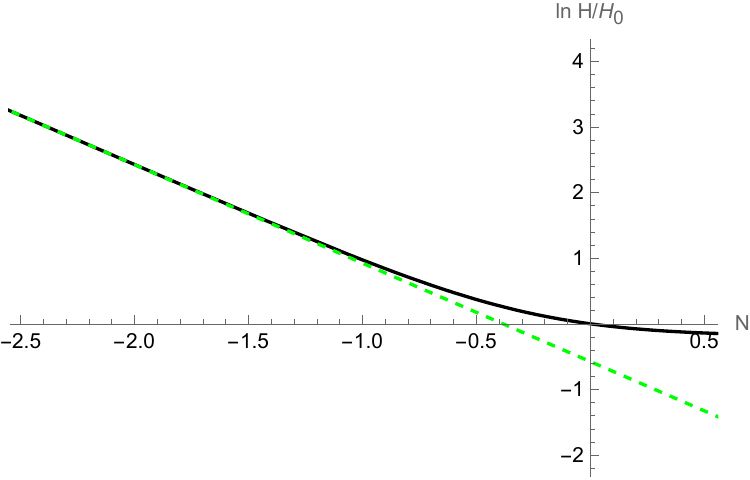}\caption{$\Lambda$CDM}\label{fig:LCDMlnHNend}
\end{subfigure}\\
\begin{subfigure}[H]{0.48\textwidth}
\includegraphics[width=\textwidth]{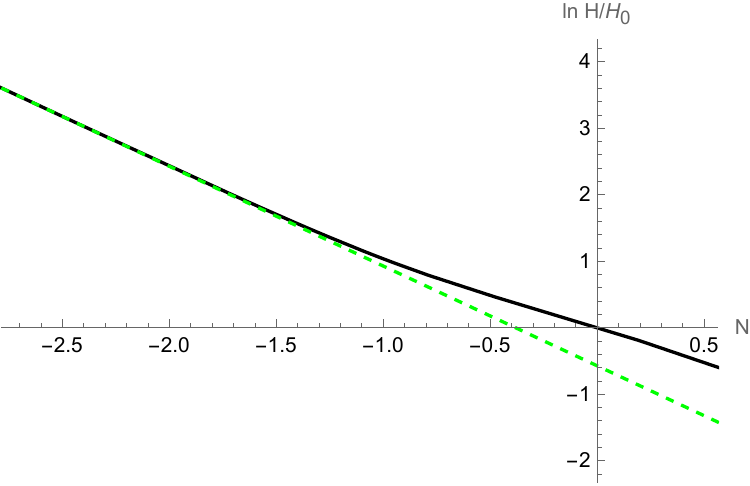}\caption{Exponential}\label{fig:ExpQuintlnHNend}
\end{subfigure}\quad
\begin{subfigure}[H]{0.48\textwidth}
\includegraphics[width=\textwidth]{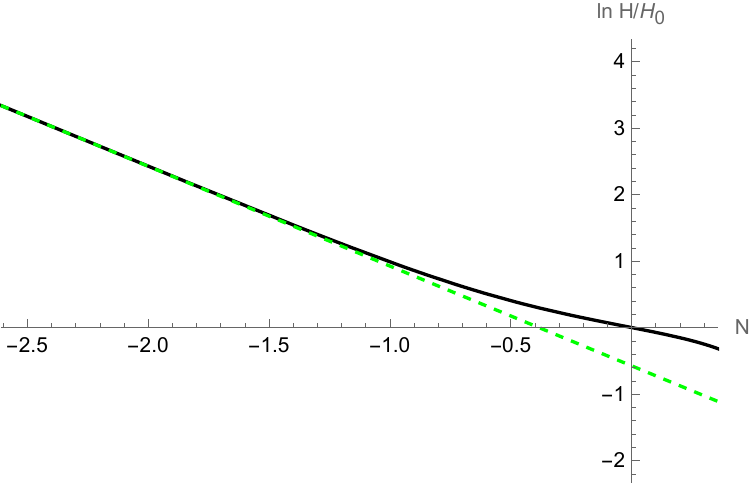}\caption{Hilltop}\label{fig:HilltoplnHNend}
\end{subfigure}
\caption{$\ln \frac{H}{H_0} (N)$ (black) for $\Lambda$CDM, exponential and hilltop quintessence examples, between $N_{m}$ and slightly beyond today. The tangent green line is the same as in Figure \ref{fig:lnHN}. We refer to the main text for more details.}\label{fig:lnHNend}
\end{center}
\end{figure}

Beyond this example, for quintessence models, it is difficult, as explained, to find an analytic solution. A useful parametrisation in the previous phase has been to consider $a(t) \sim t^{n_a}$, giving $H= n_a\, t^{-1}$ and $\ln H = {\rm constant} - \frac{N}{n_a}$. Then, if the plot of $\ln H$ is a line, one can infer the value of $n_a$: see Figure \ref{fig:lnHN}. Focusing here on this last phase, we get the corresponding plots in Figure \ref{fig:lnHNend}. For $\Lambda$CDM and hilltop quintessence, we do not get a line for $\ln H$, which discards this convenient parametrisation.
In the case of exponential quintessence, one is close to getting a line around today's universe, with $n_a \approx 1.04$. Further attempts using this observation are however inconclusive in getting an analytic solution. With the general study of exponential quintessence \cite{Andriot:2024jsh}, we know that in our example, the asymptotic fixed point is $P_{\phi}$, giving a power law with $n_a =\frac{2}{3}$. This does not help in getting an analytic transient solution around today. We refrain from further attempts in getting analytic quintessence solutions in this last phase.

\subsubsection{Integral results: $\Delta \varphi$ and $1+w_{\varphi}$}\label{sec:integral}

We did not obtain an analytic quintessence solution for the matter - dark energy phase, but we may still get interesting results thanks to integrals of relevant quantities over this period. In this subsubsection, we obtain one such results on the field distance $\Delta \varphi$ and one on the integral of $1+w_{\varphi}$.\\

Let us start with $\Delta \varphi$. We consider here the general definition of the field distance during this matter - dark energy phase
\beq
\Delta \varphi = \int_{t(N_m)}^{t_{\rm today}} \sqrt{2 \rho_{\rm kin}} \ \d t \ .
\eeq
In case of canonically normalised fields, we obtain for each field the displacement $\Delta \varphi^i \leq \Delta \varphi$, with an equality for a single field. We leave aside the possible contribution of $V$ to the field distance, captured e.g.~by the recently proposed generalised definitions of \cite{Mohseni:2024njl, Debusschere:2024rmi}. In view of possible corrections to the effective theory that we use, it is important to verify whether the field distance remains sub-Planckian (see e.g.~the refined swampland distance conjecture \cite{Palti:2019pca}), meaning $\Delta \varphi \leq 1$. While its precise value is model and solution dependent, we aim here at getting an upper bound for it.

Similarly to \eqref{deltaphiOpmax}, we first rewrite the field distance as
\beq
\Delta \varphi = \int_{N_m}^{0} \sqrt{6 \Omega_{\rm kin}}\ \d N \ . \label{deltaphiOkinrel}
\eeq
The precise evolution of $\Omega_{\rm kin}$ is model dependent, but it has some universal features. First $\Omega_{\rm kin}$ starts close to zero, since we know it is negligible at and after $N_m$, until a schematic $N=N_k$, where typically $N_k \geq -1.5$. Second, $\Omega_{\rm kin}$ is growing at $N_m$ and after it: in our examples in Figure \ref{fig:EndZoom}, it does so until today, where it reaches its value $\Omega_{\rm kin0}$. This continuous growth from $N_m$ until today can be understood from the expression $\Omega_{\rm kin} = \frac{1}{2} (1 + w_{\varphi}) \Omega_{\varphi}$. By definition of thawing models, and as will be discussed below, $w_{\varphi}$ is typically growing over this period of time. In addition, $\Omega_{\varphi}$ is also growing, as will be clear from \eqref{delOp} with $\Omega_k=0$ that gives $\del_N \Omega_{\varphi} > 0$. This results in a growth of $\Omega_{\rm kin}$ from $N_m$ until today, that will be assumed in the following; note a difference with the evolution of $\rho_{\rm kin}$, in e.g.~Figure \ref{fig:VPhi}.

There is an upper bound to $\Omega_{\rm kin0}$ for a solution that describes an accelerating universe today, as in any realistic quintessence model. Indeed, starting with \eqref{acc}, we obtain (with $k=0$) for an accelerating universe $w_{\varphi} \Omega_{\varphi} < -1/3 - 1/3 \Omega_r < -1/3$, which gives $\Omega_{\rm kin} - \Omega_V < -1/3$. Applied today, we deduce the two following bounds
\beq
\Omega_{\rm kin0}  < \frac{1}{2}\Omega_{\varphi0} - \frac{1}{6} \approx 0.176 \ ,\qquad \Omega_{V0} > \frac{1}{2}\Omega_{\varphi0} + \frac{1}{6} \approx 0.509 \ , \label{Okin0max}
\eeq
where the numerical evaluations are done with the fiducial values \eqref{fidO}. With a growing $\Omega_{\rm kin}$, we deduce the bound $\Omega_{\rm kin} < \frac{1}{2}\Omega_{\varphi0} - \frac{1}{6}$. From this, we get a first upper bound on the field distance
\beq
\boxed{\hspace{1em} \Delta \varphi \leq \Delta N \, \sqrt{3 \Omega_{\varphi0} - 1} \hspace{1em}}\ \approx 1.027 \ \Delta N \ . \label{deltaphiupper1}
\eeq
For $\Delta N$, we can restrict to the range for which $\Omega_{\rm kin}$ is non-negligible, meaning $\Delta N = -N_k$. With $N_k \geq -1.5$, we get an upper bound which is almost the desired Planckian limit.

\begin{figure}[H]
\centering
\includegraphics[width=0.55\textwidth]{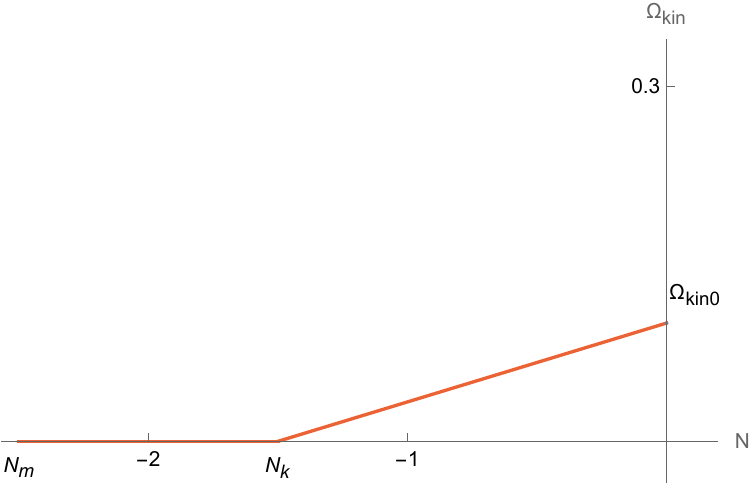}
\caption{Simple parametrisation of $\Omega_{\rm kin}$ given in \eqref{Okinpar}, capturing it initial negligible values, followed by a continuous growth. It can be compared to the evolution of $\Omega_{\rm kin}$ in our examples, depicted in Figure \ref{fig:ExpQuintONend} and \ref{fig:HilltopONend}.}\label{fig:Okinpar}
\end{figure}

We can refine this upper bound as follows. The above computation amounts to take a constant $\Omega_{\rm kin}$ over $N_k < N < 0$, but we argued that it is growing from $0$ to $\Omega_{\rm kin0}$ over this period. To capture the effect of this growth in the integral, let us consider as a very first approximation a linear growth. We take the simple parametrisation depicted in Figure \ref{fig:Okinpar}, which amounts to
\bea
N_m \leq N \leq N_k:\quad && \Omega_{\rm kin} \approx 0 \ ,\nn\\
N_k \leq N \leq 0:\quad && \Omega_{\rm kin} =  \Omega_{\rm kin0} \left( 1- \frac{N}{N_k} \right) \ . \label{Okinpar}
\eea
It is then straightforward to compute the integral \eqref{deltaphiOkinrel}: we obtain
\beq
\Delta \varphi = \frac{2}{3} \Delta N\, \sqrt{6 \Omega_{\rm kin0}} \ , \label{deltaphiOkin0}
\eeq
with $\Delta N= -N_k$. We conclude with the following upper bound, that refines \eqref{deltaphiupper1}
\beq
\boxed{\hspace{1em} \Delta \varphi \leq \frac{2}{3} \Delta N\, \sqrt{3 \Omega_{\varphi0} - 1} \hspace{1em}} \ \approx  0.6848\ \Delta N  \leq 1.027 \ , \label{deltaphiupper2}
\eeq
for $N_k \geq -1.5$. {\sl This evaluation essentially gives a sub-Planckian field distance.} Note that one can obtain a tighter bound depending on $w_{\varphi0}$ by using the expression $\Omega_{\rm kin} = \frac{1}{2} (1 + w_{\varphi}) \Omega_{\varphi}$: this would give $\Delta \varphi \leq \frac{2}{3} \Delta N\, \sqrt{3 \Omega_{\varphi0} + 3 w_{\varphi0} \Omega_{\varphi0}}$, which is smaller than the above since $w_{\varphi0} \Omega_{\varphi0}<-1/3$ in an accelerating universe.

Let us finally assess the validity of these expressions by comparing them to the data of our examples. We see already in Figure \ref{fig:ExpQuintpN} and \ref{fig:HilltopDPhiN} that the field distance between $N_m$ and today are sub-Planckian. We get for them
\bea
\text{Exp. quint.:}&&\quad \Delta \varphi = 0.7082 \ , \\
\text{Hill. quint.:}&&\quad \Delta \varphi = 0.3335 \ .
\eea
We can test the estimate \eqref{deltaphiOkin0} as follows. We take for illustration $N_k$ corresponding to $\Omega_{\rm kin}(N=N_k)=0.005$. We obtain
\bea
\text{Exp. quint.:}&&\ \Omega_{\rm kin0} = 0.1676 \ ,\ N_k=-1.1422 \ ,\quad \frac{2}{3} \Delta N\, \sqrt{6 \Omega_{\rm kin0}} =0.7636 \ , \\
\text{Hill. quint.:}&&\ \Omega_{\rm kin0} = 0.08151 \ ,\ N_k=-0.6860 \ ,\quad \frac{2}{3} \Delta N\, \sqrt{6 \Omega_{\rm kin0}} = 0.3198 \ ,
\eea
from which we see that \eqref{deltaphiOkin0} provides a reasonably good estimate of the actual $\Delta \varphi$. We conclude from this analysis and the upper bounds obtained that under reasonable assumptions, {\sl the field distance during the matter - dark energy phase is sub-Planckian.}\\

\begin{figure}[H]
\begin{center}
\begin{subfigure}[H]{0.48\textwidth}
\includegraphics[width=\textwidth]{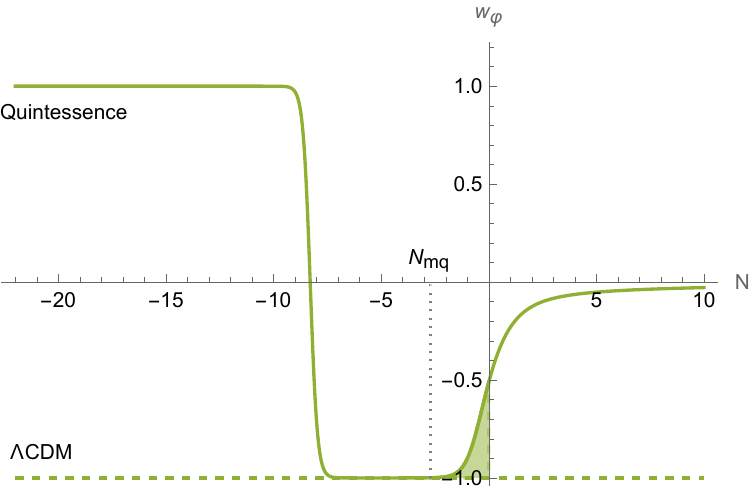}\caption{$w_{\varphi}$}\label{fig:wphiarea}
\end{subfigure}\quad
\begin{subfigure}[H]{0.48\textwidth}
\includegraphics[width=\textwidth]{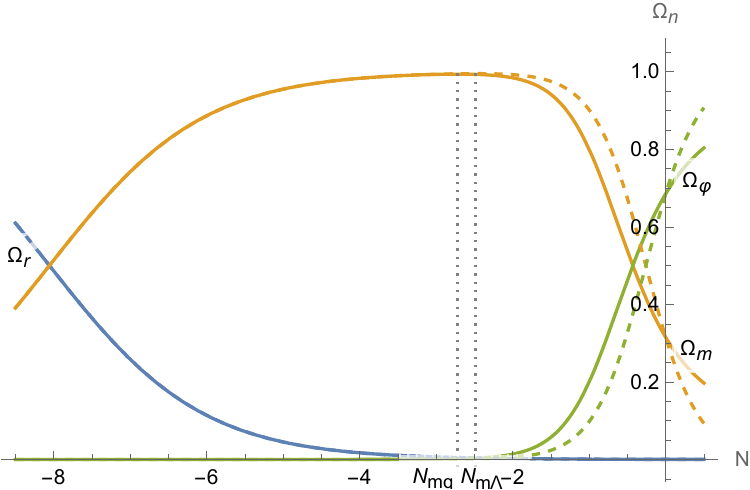}\caption{$\Omega_n$}\label{fig:Ommax}
\end{subfigure}
\caption{$w_{\varphi}$ and $\Omega_n$ for $\Lambda$CDM (dashed) and our exponential quintessence example (plain). In Figure \ref{fig:wphiarea}, we color the area in between the two curves during the matter - dark energy phase. This area is proportional, as shown in \eqref{relareaw}, to the difference between the maxima $N_{m\, \Lambda}$ and $N_{m\, q}$ of $\Omega_m$, indicated in Figure \ref{fig:Ommax} by the vertical dotted gray lines.}\label{fig:difference}
\end{center}
\end{figure}

We turn to a second result obtained thanks to an integral over this last phase. Let us start with the dark energy continuity equation \eqref{rhodot}, that was obtained purely from $E^i=0$. We rewrite it into the following well-known equation
\beq
-\frac{1}{3} \, \frac{\del_N \rho_{\varphi}}{\rho_{\varphi}} = w_{\varphi} +1 \ .
\eeq
The integral of the right-hand side is an interesting quantity as it measures the difference with respect to $\Lambda$CDM, i.e.~with respect to a pure cosmological constant for which $w_{\varphi}=-1$. This difference is also the outcome of the recent measurements by DES \cite{DES:2024tys} and DESI \cite{DESI:2024mwx}, and is therefore relevant. This integral represents an area that we depict in Figure \ref{fig:wphiarea}.

We integrate over the whole matter - dark energy phase. The novelty here is that for the left-hand side, we can use the ratio of $\rho_{\varphi}$ between today and matter domination \eqref{rhoratio}. Using in addition that $w_{\varphi\, m} \approx -1$ (as now justified by our analytical study of the $w_{\varphi}$ transition), we obtain
\beq
\boxed{\hspace{1em} \int_{N_{m\, q}}^0  (w_{\varphi} +1)\, \d N = \frac{4}{3}\, ( N_{m\, \Lambda} - N_{m\, q} ) \hspace{1em}} \label{relareaw}
\eeq
We recall that the right-hand side is the difference in maxima moments of $\Omega_m$ between $\Lambda$CDM and quintessence models, given identical $\Omega_{n0}$ today: we depict this difference in Figure \ref{fig:Ommax} (a similar figure can be found in \cite{Linder:2021syd}, with related discussions). These $N_m$ values are given in Table \ref{tab:max} for our examples. Consistently, the difference vanishes for a cosmological constant. We also recall from the discussion around \eqref{rhoratio} that this difference is positive for quintessence due to Hubble friction; this is also expected from the left-hand side.

The analytic integral relation \eqref{relareaw} between these two quantities is illustrated in Figure \ref{fig:difference}. Around \eqref{relareaequality}, we discuss an analogous relation between a similar integral and the moments of matter - dark energy equality, the difference of which can also be seen in Figure \ref{fig:Ommax}.

We have also investigated the evolution of $\Omega_{\varphi}$. In particular, combining \eqref{rhodot} as well as $f_1=f_2=0$, one can show that
\beq
\del_N \Omega_{\varphi} = 3 w_{\varphi} \Omega_{\varphi} (\Omega_{\varphi} -1) + \Omega_{\varphi} (\Omega_r - \Omega_k) \ , \label{delOp}
\eeq
a result obtained for instance in \cite[(3.13)]{Andriot:2024jsh} for a single field exponential quintessence, and shown here in multifield full generality. Through standard approximations, a similar integration, and making use of the results of Section \ref{sec:phases}, one reaches the same relation \eqref{relareaw}. In the following, we will investigate the implications of this relation, thanks to a specific parametrisation of $w_{\varphi}$.

\subsubsection{$w_0 w_a$ parametrisation and phantom behaviour}\label{sec:phantom}

\begin{figure}[H]
\begin{center}
\begin{subfigure}[H]{0.48\textwidth}
\includegraphics[width=\textwidth]{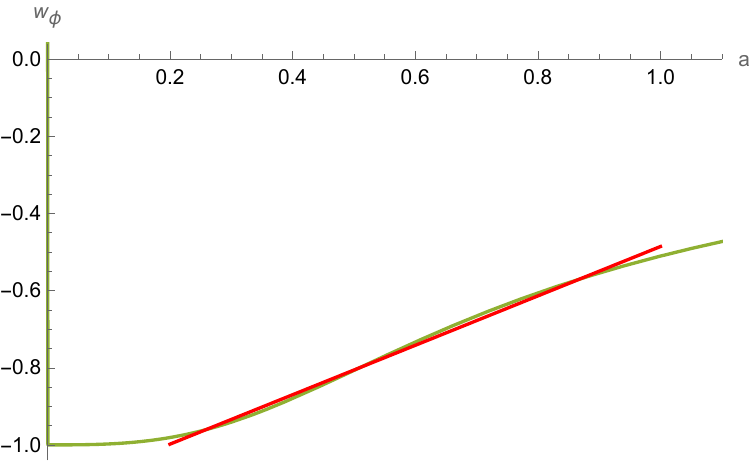}\caption{Exponential\\ $w_0=-0.486 ,\, w_a = -0.640$}\label{fig:ExpQuintwafit}
\end{subfigure}\quad
\begin{subfigure}[H]{0.48\textwidth}
\includegraphics[width=\textwidth]{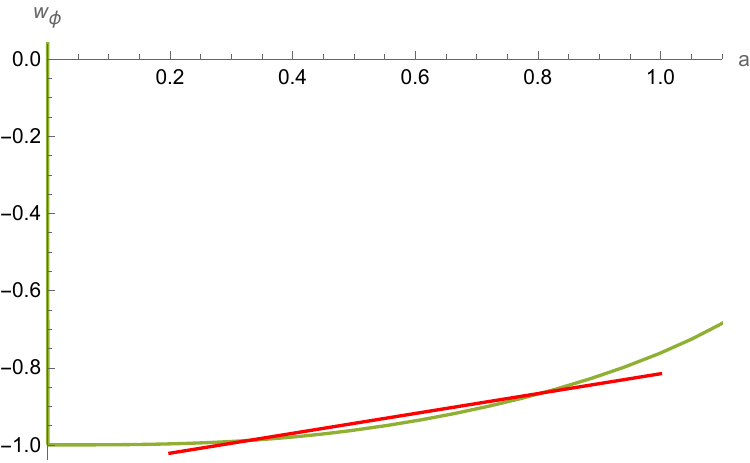}\caption{Hilltop\\ $w_0= -0.816 ,\, w_a = -0.257$}\label{fig:Hilltopwafit}
\end{subfigure}
\caption{$w(a)$ in green, and best fit line in red, corresponding to the $w_0 w_a$ parametrisation between redshifts $z=0$ and $z=4$, for our two quintessence examples.}\label{fig:wafit}
\end{center}
\end{figure}

One analytic parametrisation of $w_{\varphi}$ in the matter - dark energy phase is the Chevallier-Polarski-Linder (CPL) or $w_0 w_a$ parametrisation. It proposes a linear behaviour in $a$ around today ($a=1$)
\beq
w_{\varphi} = w_0 + w_a (1-a) \ . \label{w0wa}
\eeq
It has been used recently in the DES and DESI observations \cite{DES:2024tys, DESI:2024mwx} to parameterize a varying $w_{\varphi}$ between redshift $z=4$ ($a=0.2$, $N\approx -1.61$) and today at $z=0$. Having a concrete model with a solution displaying an evolving $w_{\varphi}$, one can consider the $w_0 w_a$ linear parametrisation as a best fit to the actual curve. We do so in our two examples (as in \cite{Andriot:2024jsh}) to extract values of $w_0, w_a$, and compare those to the observational values. We give our results in Figure \ref{fig:wafit}: our particular examples are not compatible with the values of DES or DESI.
In quintessence models and with our analytic solutions, one starts around $N_m$ with $w_{\varphi}\approx -1$. With this as an anchor point, if the line aims at a small $|w_0|$, then the slope $|w_a|$ would be large; on the contrary, we can also have a large $|w_0|$ and a small $|w_a|$. Figure \ref{fig:wafit} illustrates these two possibilities.

\begin{figure}[H]
\begin{center}
\begin{subfigure}[H]{0.48\textwidth}
\includegraphics[width=\textwidth]{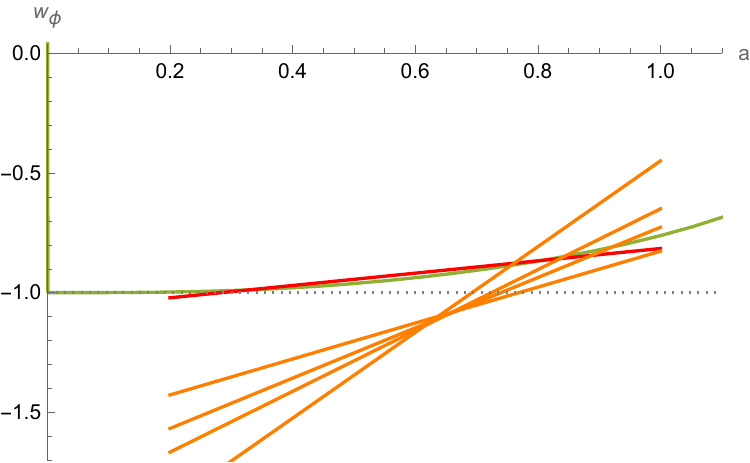}\caption{}\label{fig:wpDESIcentral}
\end{subfigure}\quad
\begin{subfigure}[H]{0.48\textwidth}
\includegraphics[width=\textwidth]{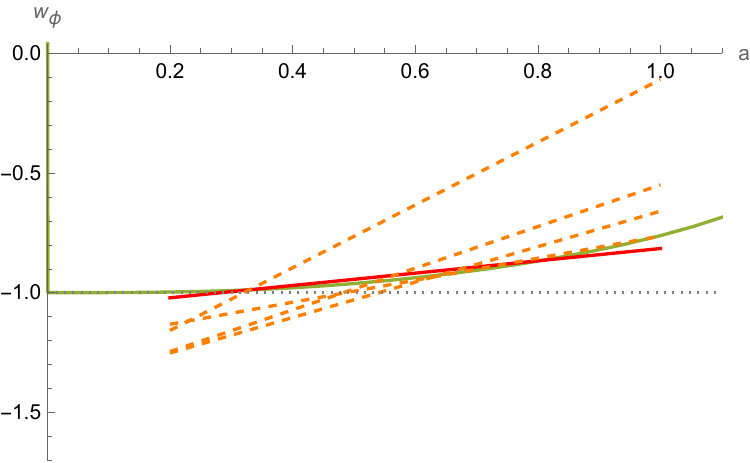}\caption{}\label{fig:wpDESItop}
\end{subfigure}
\caption{$w_0 w_a$ parametrisation of $w(a)$ between redshifts $z=0$ and $z=4$. In orange are the DESI observational results \cite{DESI:2024mwx} for different data sets: the DESI+CMB data (top line with highest $w_0$) and the three DESI + SNe data (lower three lines). These are superposed to the green $w(a)$ and its red best fit for our hilltop example, given in Figure \ref{fig:Hilltopwafit}, to help the comparison. The $w_{\varphi}=-1$ line, corresponding to the phantom behaviour threshold, is depicted in dashed gray. In Figure \ref{fig:wpDESIcentral} are used DESI central values, and in Figure \ref{fig:wpDESItop} are used the error bars limiting values which are favorable (highest $w_0$ and smallest $-w_a$).}\label{fig:wpDESI}
\end{center}
\end{figure}

The DESI values are difficult to match in general, because the central values of $|w_0|$ and $|w_a|$ are both large.\footnote{Planck 2018 results \cite{Planck:2018vyg} discuss as well possible values for $w_0, w_a$ using several data sets. Two data sets are presented in \cite[Tab. 6]{Planck:2018vyg} with equally good fits. One is highlighted in the main text, because it gives more precise results and can be compatible with no phantom behaviour: it provides central values close $\Lambda$CDM, namely $w_0 =-0.957, w_a=-0.29$. Allowing however some phantom behaviour, we can also consider the other data set results, giving $w_0=-0.76, w_a = -0.72$. Interestingly, the latter are close to the DESI values.} Such values (and parameters) imply that $w_{\varphi} < -1$ at $N_m$, contrary to the above. Values such that $w_{\varphi} < -1$ are referred to as a phantom behaviour. As discussed in Section \ref{sec:setting}, a quintessence model admits by definition $-1 \leq w_{\varphi} \leq 1$ so a priori, it cannot lead to a phantom behaviour, hence the difficulties encountered in reproducing DESI results. We illustrate this situation in Figure \ref{fig:wpDESI} comparing DESI results to our hilltop example.
Let us point-out that a quintessence model can lead to some phantom behaviour as an artefact of the $w_0 w_a$ parametrisation (see \cite{Wolf:2023uno} for related ideas with the hilltop potential). Indeed, as mentioned, one has $w_{\varphi} \approx -1$ at matter domination and for some time afterwards, and then $w_{\varphi}$ starts growing. Therefore, $\del_a w_{\varphi}$ was initially vanishing and then becomes positive, from which we deduce that $\del_a^2 w_{\varphi}>0 $. This shows that the initial growth is convex.\footnote{Note that for a function $f$, $\del_N f = e^N \, \del_a f$, meaning that growing in $a$ is equivalent to growing in $N$. However, one has $\del_N^2 f = e^{2N}\,\del_a^2 f + e^N \, \del_a f$. This means that convex in $N$, as observed in our two examples, does not imply convex in $a$. A counter-example to such an equivalence is indeed provided in Figure \ref{fig:ExpQuintwafit} by the exponential quintessence.} The best fit line of a convex curve necessarily crosses it. Therefore, this fitting line must initially be below the curve, at least if the latter stays convex. For a curve which is at $-1$ for some time before rising, then the best fit line must go below $-1$: this generates a (fake) phantom behaviour. This parametrisation artefact is visible in Figure \ref{fig:Hilltopwafit}, where the red line goes below $-1$ at the first moments considered; it can be made even more manifest by increasing the potential parameter $k$ in that model.

Such an initial value $w_{\varphi}=-1$ followed by a convex growth can thus produce a (fake) phantom behaviour, as a result of the parametrisation. However, as made explicit in Figure \ref{fig:wpDESI}, the DESI results lead to a much more drastic phantom behaviour, which does not seem to be due to a parametrisation artefact. As mentioned above, they are difficult to reproduce with a quintessence model. The results which would be the less difficult to accommodate are those obtained with the DESI+CMB data alone (without SNe) when pushed to the boundary of its error bars; finding a corresponding quintessence curve remains challenging.

Let us add that a DESI ``model-agnostic'' reconstruction of $w_{\varphi}(z)$ \cite[Fig. 1]{DESI:2024aqx} shows again large portion of phantom behaviour in the past (although DESI + Planck data alone is not displayed); equivalently, $\rho_{\varphi}/\rho_{\varphi0}$ (corresponding in that work to $f_{DE}$) is shown to grow with time towards the past. The latter is not possible within quintessence because of Hubble friction, as indicated by \eqref{rhodotE}. With \eqref{rhodot}, this is equivalent to say that one must have $w_{\varphi} \geq -1$. Insisting on having such a phantom behaviour, that goes beyond an artefact of the $w_0 w_a$ parametrisation, would thus be a major theoretical challenge, requiring for example non-standard (negative) kinetic energy, non-minimal couplings or modified gravity (see e.g.~\cite[Sec. 7.4.1]{Planck:2018vyg}).

A possible way-out can be seen in the fact that error bars quickly become huge for $w_{\varphi}(z)$ in \cite{DESI:2024aqx} once it becomes phantom. This is related to the fact that the measurement is challenging in the past, especially when it is not dark energy dominated. Therefore one option would be to discard the data (and period of time) indicating a phantom behaviour, and wait for more precise measurements in that earlier phase. Doing so may allow a match to a quintessence model with the most recent period of time. Further interesting remarks on the reality of such a phantom regime and the observations can be found in \cite[App. A]{DESI:2024kob}.

\subsubsection{Bounds on $w_0, w_a$}\label{sec:w0wabounds}

\begin{figure}[H]
\centering
\includegraphics[width=0.55\textwidth]{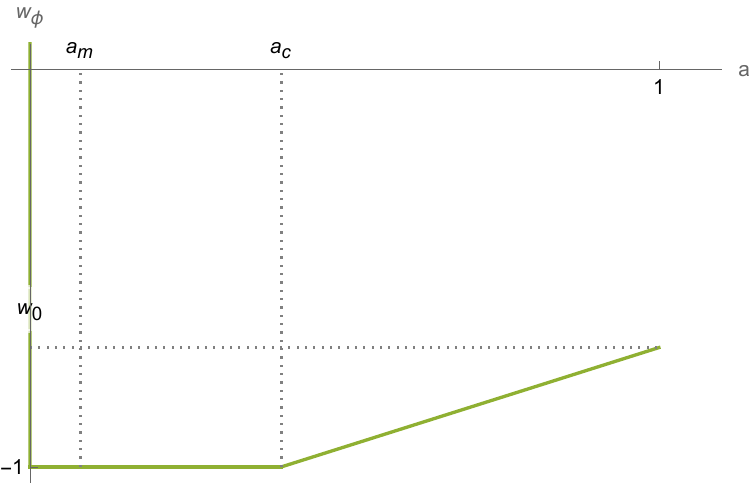}
\caption{Parametrisation of $w_{\varphi}$ considered in the main text, with $a_m=e^{N_m}$ and $a_c=e^{N_c}$: $w_{\varphi}=-1$ for $N_m \leq N \leq N_c$ and $w_{\varphi} = w_0 + w_a (1-a)$ for $N_c \leq N \leq 0$.}\label{fig:wphiparam}
\end{figure}

With the analytic expression \eqref{w0wa} of $w_{\varphi}$, we can evaluate the integral in the relation \eqref{relareaw}. As suggested in the previous discussion, the $w_0 w_a$ parametrisation need not apply to $w_{\varphi}$ for the whole matter - dark energy phase, or not even up to $z=4$ ($N \approx -1.61$). As e.g.~in our examples in Figure \ref{fig:wafit}, we would rather consider \eqref{w0wa} to capture the growth of $w_{\varphi}$ starting at a certain $N_c$ until today at $N=0$. Before that, for $N_m \leq N \leq N_c$, we consider $w_{\varphi}=-1$. We summarize this parametrisation of $w_{\varphi}$ in Figure \ref{fig:wphiparam}.
With this parametrisation of $w_{\varphi}$, the relation \eqref{relareaw} becomes
\beq
-N_c\left( 1 + w_0 +w_a \right) - w_a(1-e^{N_c}) = \frac{4}{3}( N_{m\, \Lambda} - N_{m\, q} )  \ .
\eeq
As a first check of this expression, we note that $N_c=0$ consistently enforces $N_{m\, \Lambda} = N_{m\, q}$, i.e.~going back to $\Lambda$CDM with $w_{\varphi}=-1$. For quintessence however, we rather expect $N_c<0$, giving us
\beq
 1 + w_0 +w_a =\frac{1}{-N_c} \left( w_a (1-e^{N_c}) +\frac{4}{3}( N_{m\, \Lambda} - N_{m\, q} ) \right) \ ,
\eeq
from which we deduce
\beq
\boxed{\hspace{1em}  w_0 +w_a < -1 \ \Leftrightarrow \ w_a < -\frac{4}{3} \frac{ N_{m\, \Lambda} - N_{m\, q} }{1-e^{N_c}} \ \ (< 0) \hspace{1em}} \label{waupperbound}
\eeq
We recall that we expect $w_a<0$ since we have a growing $w_{\varphi}$ (from $-1$ to $w_{\varphi0}$). Having $w_0 +w_a < -1$ thus gives an upper bound to $|w_a|$ thanks to \eqref{waupperbound}.

In addition, with the above parametrisation, we have at $N_c$ that $w_{\varphi} = -1$. This gives $w_a (1-e^{N_c}) = -(1+w_0)$. From this we deduce the alternative condition
\beq
\boxed{\hspace{1em}  w_0 +w_a < -1 \ \Leftrightarrow \ w_0 > -1 + \frac{4}{3} \left( N_{m\, \Lambda} - N_{m\, q} \right) \ \ (> -1)  \hspace{1em}} \label{w0lowerbound}
\eeq
consistent with the fact that we must have $ w_0 > -1$. This gives a lower bound to $w_0$.\\

The inequality $w_0 +w_a < -1 $ is worth being commented on. There is at first sight no reason for the $w_{\varphi}$ parameters to obey it, but it is consistent with the latest observations \cite{DESI:2024mwx}, and was discussed in \cite{Shlivko:2024llw}. We can prove here the following lemma
\beq
\text{{\bf Lemma.}} \quad \quad \boxed{\hspace{1em}  \exists\ a > 0 \ {\rm s.t.} \ w_{\varphi} < -1 \quad \Leftrightarrow \quad w_0 +w_a < -1 \hspace{1em}} \label{lemmaw}
\eeq
assuming the $w_0 w_a$ parametrisation to hold as well as a growing $w_{\varphi}$ (meaning $w_a<0$). For completeness, let us mention that a related implication (not equivalence) was discussed in \cite{Vagnozzi:2018jhn}.
\begin{proof}
The $w_0 w_a$ parametrisation gives that $ w_{\varphi}(a) < -1$ at a certain $a>0$ is equivalent to $1+w_0 +w_a <a \, w_a$. Since $w_a<0$, we conclude $1+w_0 +w_a <0$, i.e.~prove the implication. Converse-wise, if $w_0 +w_a < -1$, we can simply take $a=(1+w_0 +w_a)/(2w_a) >0$. We then get that $1+w_0 +w_a <a \, w_a$, from which we deduce $ w_{\varphi}(a) < -1$ and prove the converse.
\end{proof}
Given the solutions discussed in this work, it is clear that $w_a<0$, because in the last phase, $w_{\varphi}$ grows from $-1$ to some value $w_{\varphi0}>-1$. To use the above equivalence, the main question is therefore whether the moment at which $w_{\varphi} < -1$ is within the range of validity of $w_0 w_a$ parametrisation. Both for DESI observational results, as well as for our hilltop example where the phantom behaviour appears as an artefact of the parametrisation, this condition is considered to be true. Then, $w_0 +w_a < -1$ holds.

Considering this inequality true, we get an upper bound on $w_a$ \eqref{waupperbound} and a lower bound on $w_0$ \eqref{w0lowerbound}. Let us evaluate them in some examples for illustration. To evaluate the former bound, we need to fix $N_c$. In agreement with the discussion of the previous subsubsection, and with the above parametrisation, $N_c$ is the moment at which could appear the phantom behaviour. Let us take here from the DESI central values in Figure \ref{fig:wpDESIcentral} this value to correspond to $a_c=0.75$, i.e.~$N_c \approx -0.289$. Furthermore, using for illustration the values for $N_{m\, \Lambda} - N_{m\, q}$ obtained in our two examples, we get the following bounds
\bea
N_{m\, \Lambda} - N_{m\, q} \approx 0.236 \ \ \text{(Exp. quint.)}:\quad && w_a < -1.259 \ ,\ w_0 > -0.685\\
N_{m\, \Lambda} - N_{m\, q} \approx 0.066 \ \ \text{(Hill. quint.)}:\quad && w_a < -0.352 \ ,\ w_0 >- 0.912
\eea
These bound values seem competitive with the DESI results, and could thus be interesting in view of future measurements.

\section{Outlook: string theory, cosmology, and observational targets}\label{sec:outlook}

In this work, we have studied (possibly realistic) cosmological solutions from (multi)field quintessence models. We have provided analytic expressions for these solutions, as well as characterisations of their features, in a model-independent manner. A summary of our results can be found in Section \ref{sec:intro}. We make here a few remarks on the implications of these results, and suggest four observational targets related to the general features of these quintessence solutions.

\begin{itemize}
  \item {\bf Remarks for string theory model building}
\end{itemize}

String phenomenology is the branch that aims at relating string theory to phenomenologically valid models. In the last decades, a lot of activity in this field had to do with the problem of moduli stabilisation: see for instance \cite{McAllister:2023vgy}, as well as the approach described e.g.~in \cite{Andriot:2022yyj}. The problem is about massless scalar fields, a.k.a.~moduli. Those are ubiquitous in 4d effective theories from string theory, because of the extra dimensions (see e.g.~\cite{VanRiet:2023pnx}). Massless or light scalar fields could lead to long range fifth forces or varying coupling constants, that are so far unobserved. The idea is then to stabilise them by generating a potential with a vacuum, or at least giving them a (heavy) mass.

Here, we have studied in depth the freezing of scalar fields, due to the high Hubble friction, occurring during the radiation - matter phase, and a little after that. In case of canonically normalized fields, we have obtained an analytic expression for the field variation \eqref{Deltaphifrozen} that generically gives $\Delta \varphi^i \lesssim 10^{-2}$. We have further argued that beyond canonical normalisation, we still expect a high degree of freezing because kinetic energy is maintained very small during this phase. This observation could lead to a {\sl change of paradigm in string phenomenology}: indeed, the freezing leads effectively (at least at the classical level) to the same effect as moduli stabilisation. The benefit is that there is no need anymore of a vacuum, which was highly demanding from model building; the fields now stay almost constant on a potential slope. This phenomenon was used in \cite{Agrawal:2018rcg} to argue in favor of having a de Sitter maximum in a scalar potential, instead of a minimum, to reproduce the observed cosmology: fields would be maintained frozen around the maximum, until recently when they would start rolling.

The difficulties due to varying scalar fields, mentioned above, would nevertheless reappear in the recent universe, when indeed fields are not frozen anymore and roll down the potential. In Section \ref{sec:finalphase}, we proposed this to start at $N_c \approx -1.5$ e-folds, when today is $N=0$. We showed there that the resulting (multi)field distance generically remains sub-Planckian, $\Delta \varphi \leq 1$. It then becomes very model dependent to know whether the fields variation in this last phase would lead to observational issues or not. If that is the case, screening mechanisms (see e.g.~\cite{SevillanoMunoz:2024ayh}) could become crucial. To answer these questions, knowing the coupling to matter, if any, becomes necessary. Getting that information from string model building is a non-trivial task, but we believe it is now an urgent one.

\vspace{0.3in}

\begin{itemize}
  \item {\bf Remarks for cosmology}
\end{itemize}

It is not obvious what the output of the present analysis is with respect to the Hubble tension. Indeed, we do not predict any value for $H_0$, as our formalism rather considers $H/H_0$ or a dependence in $H_0 \, t$. As mentioned already, we noticed a shorter value for $H_0\, t_{\rm age}$ in our quintessence examples, which would give a shorter age of the universe for a given $H_0$, or a smaller $H_0$ for a given age.

Still, we make the following related remark. Solutions to the Hubble tension have been proposed under the name of early dark energy scenarios (see e.g.~\cite{Kamionkowski:2022pkx, Poulin:2023lkg, Niedermann:2023ssr}), suggesting a rise of the dark energy contribution during the radiation - matter phase. We would like to stress that in the models considered in this work, such a rise is not possible: Hubble friction and the resulting freezing of fields generically prevents from such a change in $\Omega_{\varphi}$ (which rather reaches a minimum during this phase). Different scenarios would then be needed, involving e.g.~non-trivial couplings to matter, or else. It could also be interesting to see if the early kination phase (for which $\Omega_{\varphi}$ is large), and the specific evolution and later transition of $w_{\varphi}$ that we described in detail, could be of any help in these matters.\\

Another comment has to do the electroweak phase transition in the early universe, during which the Higgs field is supposed to roll-down its potential (while the potential evolves thermally to generate a new minimum) \cite{LISACosmologyWorkingGroup:2022jok, Salvio:2024upo}. This ``rolling-down'' classical picture is probably only valid for a second order phase transition; a first order one may rather happen by a non-perturbative creation of bubbles. In any case, this phase transition may be considered to happen during radiation domination: if that is the case, the freezing described above could affect the Higgs scalar field, at least for a small enough scalar potential. This would prevent the naive rolling-down and thus possibly forbid a (second order) phase transition.

There are several ways out. First, if we consider recombination to have taken place at $N=-\ln(1+z)=-7.00$ ($z=1100$, $T= 0.259 {\rm eV}$, and $T \propto 1+z$), we get the electroweak phase transition ($T=100 {\rm GeV}$, $z= 4.25 \cdot 10^{14}$) to have happened at $N=-33.68$. This is much earlier than the start of radiation domination considered in this work ($N\approx -20$, $z= 4.85 \cdot 10^8$, $T= 114 {\rm keV}$), even though we mentioned that this start could easily be adjusted by further tuning of the initial conditions. A way out to the above obstruction is then to claim that the electroweak phase transition happens during the Higgs kination phase instead of the radiation domination phase; this seems however unlikely with respect to the rest of the standard model constituents. Another way out could simply be that the scalar potential contribution cannot be neglected against the friction, avoiding the freezing; this might be checkable. It remains an interesting idea to set constraints on this phase transition (e.g.~its order) using the Hubble friction and freezing argument.

One may ask a similar question for pions $\pi^0$, appearing after the QCD phase transition ($T=150 {\rm MeV}$, $z=6.38 \cdot 10^{11}$, $N=-27.18$). However, pions are not required to evolve in their scalar potential; they rather quickly decay.\\

Finally, while spatial curvature could be neglected during most of the universe history that we described, it could in principle affect the latest universe, and we discarded it for simplicity in our study of the matter - dark energy phase. It would interesting to see if it can affect the results we obtained there. For example, we found in \cite{Andriot:2024jsh} that when taken off the dark energy budget, a non-zero $\Omega_k$ could quantitatively impact various values and bounds we obtained for the recent universe; the same may hold here.

\begin{itemize}
  \item {\bf Observational targets}
\end{itemize}

Last but not least, inspired by the features identified and characterised in our study of quintessence models and their realistic solutions, we propose here a list of four observational targets. The main motivation is to distinguish the universe evolution based on a quintessence model versus that of $\Lambda$CDM.

\begin{enumerate}
  \item The first obvious target, already tackled by current and coming experiments, is the recent evolution of $w_{\varphi}$, away from $-1$. As stressed in our study, the quintessence models considered generically give $w_{\varphi} \approx -1$ at matter domination ($N\approx -2.5$, $z\approx 11$), but also before that, and after that until $N=N_c$. We suggested that $N_c \geq -1.5$, giving $z_c \leq 3.5$, depending on the model. The fact that $w_{\varphi} \approx -1$ for so long makes it difficult to distinguish the resulting evolution from that of $\Lambda$CDM. We combined this feature together with the $w_0 w_a$ parametrisation into a schematic parametrisation described in Figure \ref{fig:wphiparam}, that could avoid the phantom behaviour issues.

      A result of interest here is the evaluation of the integral $\int (w_{\varphi}+1) \,\d N$ over this phase, as given in terms of the difference $N_{m\, \Lambda} - N_{m\, q}$ in \eqref{relareawintro} or \eqref{relareaw}. We come back to that quantity in the next item; here we suggest that it could be used to normalise the integral. In addition, this relation allowed us to provide bounds on $w_0$ and $w_a$ in \eqref{waupperbound} and \eqref{w0lowerbound}, that could serve as priors for these parameters. Note that these bounds are valid, even if the phantom behaviour mentioned e.g.~in \eqref{w0waintro} is cut-off by the use of the parametrisation of Figure \ref{fig:wphiparam}, introducing $N_c$.

  \item We introduced the quantity $N_{m\, \Lambda} - N_{m\, q}$; it appears in several expressions obtained in this work. This quantity describes the difference between the moment of maximum $\Omega_m$, i.e.~matter domination, as obtained with a $\Lambda$CDM or with a quintessence evolution. This offers a second interesting observational target, as we now explain. $N_{m\, \Lambda}$ and $N_{m\, q}$ should not be computed by running two distinct universe histories with either models on a given data set, because they are based on the same today values $\Omega_{n0}$. Rather, one should run the evolution on a data set with a quintessence model (or any such generic evolution), and first obtain the $\Omega_{n0}$: those define $N_{m\, \Lambda}$ with the formulas of Section \ref{sec:phases}; a foreseeable difficulty though is the need of $\Omega_{r0}$. Then, with the resulting evolution of $\Omega_m$, one should determine precisely the maximum moment $N_{m\, q}$ (around $z\approx 11$). As a result, one obtains the difference $N_{m\, \Lambda} - N_{m\, q}$, which is another way to detect quintessence, and distinguish it from $\Lambda$CDM. Its relation \eqref{relareawintro} to $\int (w_{\varphi}+1) \,\d N$ over the last phase is illustrated in Figure \ref{fig:difference}. It would be useful to evaluate both sides of that relation.

  \item Another shift, due to the difference between the $\Lambda$CDM and quintessence evolution, is that of the matter - dark energy equality moment, $N_{m\varphi}$ (broadly between $z \approx 0.2 - 0.5$). This offers a third observational target. To get this difference, one can again run on a data set a (generic) quintessence evolution. Determining the $\Omega_{n0}$ first provides analytically $N_{m\varphi\, \Lambda}$ as indicated in Section \ref{sec:phases}. With respect to $N_{m\, \Lambda}$, the advantage here is that $N_{m\varphi\, \Lambda}$ does not depend on radiation. Then, using the evolution of $\Omega_m$ and $\Omega_{\varphi}$, one may determine precisely the equality moment, giving $N_{m\varphi\, q}$. We have not emphasized much the resulting difference  $N_{m\varphi\, \Lambda} - N_{m\varphi\, q}$ but it appears in the ratio of dark energy densities \eqref{rhoratioequality}. As done in Section \ref{sec:integral}, we can relate it to $\int (w_{\varphi}+1) \,\d N$, however not over the whole matter - dark energy phase, but from the equality time
      \beq
      \int_{N_{m\varphi\, q}}^{0} (w_{\varphi}+1) \,\d N = N_{m\varphi\, \Lambda} - N_{m\varphi\, q} \ . \label{relareaequality}
      \eeq
      This dependence on the equality time in the left-hand side is not very convenient, because the value of $w_{\varphi}$ at this moment is not clear and very model dependent. This relation may still prove useful, e.g.~as a normalisation. In any case, determining this difference in equality times is again a way to detect quintessence and distinguish it from $\Lambda$CDM. In addition, it could be an {\sl accessible target, given the low redshifts involved}.

  \item A fourth observational target is the transition of $w_{\varphi}$ from $+1$ to $-1$. This is more ambitious on several levels. To start with, we are not entirely sure that this transition happened since it relies on an earlier kination phase (with negligible potential) which as discussed, could be avoided in an early universe history. In addition, it requires a measurement of $w_{\varphi}$ during radiation domination, when $\Omega_{\varphi}$ is very subdominant. However, the fact the transition from $+1$ to $-1$ is fairly sharp may help in its detection. Beyond a clear distinction from $\Lambda$CDM, an important motivation for such an observation is that it teaches us something about a much earlier period of the universe. Indeed, the moment of this transition, $N_{{\rm kin}V}$ is related as in \eqref{NkinVequalityintro} to the much earlier equality time $N_{{\rm kin}r}$, or as in \eqref{NkinVmax}, to the maximum of radiation $N_r$. A drawback of this relation is that we do not know when this transition occurred. As argued, if we take $N_{{\rm kin}r} \leq -20$, we get $N_{{\rm kin}V} \lesssim -8$, i.e.~$z \gtrsim 3000$. But it could also have happened much earlier. While very challenging, this observation would have far reaching implications. It would be interesting to think of the effect of this transition in terms of primordial gravitational waves, or as an imprint on the CMB, if any.

\end{enumerate}

\vspace{0.4in}

\subsection*{Acknowledgements}

We would like to thank G.~B\'elanger, J.~Larena, A.~Moradinezhad, V.~Poulin, F.~Revello, P.~Salati and F.~Tonioni for insightful discussions.

\newpage

\addcontentsline{toc}{section}{References}

\providecommand{\href}[2]{#2}\begingroup\raggedright\endgroup

\end{document}